\documentclass[12pt]{article}
\usepackage{amsmath}
\usepackage{graphicx}
\usepackage{enumerate}
\usepackage{natbib}
\usepackage{url} 
\usepackage[colorlinks, citecolor={blue}]{hyperref}

\newcommand{\blind}{1}

\usepackage{amsmath,amsfonts,amssymb, graphicx}
\usepackage{natbib}
\usepackage[utf8]{inputenc}
\usepackage{mathtools}
\usepackage{lmodern}
\usepackage{anyfontsize}

\usepackage{libertine}
\usepackage{float}
\usepackage{bbm}
\usepackage{subfigure}
\usepackage{algorithm}
\usepackage{algpseudocode}
\usepackage{mathrsfs}
\usepackage{url} 
\usepackage{tikz}
\usepackage{enumerate}
\usepackage{enumitem}

\usepackage{xcolor}

\newtheorem{remark}{Remark}

\newtheorem{theorem}{Theorem}[section]
\newtheorem{lemma}[theorem]{Lemma}

\newtheorem{alg}{Algorithm}

\begin{document}

\def\spacingset#1{\renewcommand{\baselinestretch}%
{#1}\small\normalsize} \spacingset{1}


\if1\blind
{
  \title{\bf Modeling Neural Switching via Drift-Diffusion Models}
  \author{Nicholas Marco \thanks{
    The authors gratefully acknowledge from NIH awards R01 DC013096 and R01 DC016363.}\hspace{.2cm}\\
    Department of Statistical Science, Duke University\\
    Jennifer M. Groh \\
    Department of Neurobiology, Department of Psychology \& Neuroscience,\\ Department of Biomedical Engineering, Department of Computer Science,\\ Duke University\\
    and\\
    Surya T. Tokdar\\
    Department of Statistical Science, Duke University}
  \maketitle
} \fi

\if0\blind
{
  \bigskip
  \bigskip
  \bigskip
  \begin{center}
    {\LARGE\bf Modeling Neural Switching via Drift-Diffusion Models}
\end{center}
  \medskip
} \fi

\bigskip
\begin{abstract}
Neural encoding is a field in neuroscience that focuses on characterizing how information from stimuli is encoded in the spiking activity of neurons. When more than one stimulus is present, a theory known as multiplexing posits that neurons temporally switch between encoding various stimuli, creating a fluctuating firing pattern. Here, we propose a new statistical framework to analyze rate fluctuations and discern whether neurons employ multiplexing as a means of encoding multiple stimuli. We adopt a mechanistic approach to modeling multiplexing by constructing a non-Markovian endogenous state-space model. Specifically, we posit that multiplexing arises from competition between the stimuli, which are modeled as latent drift-diffusion processes. 
We propose a new MCMC algorithm for conducting posterior inference on similar types of state-space models, where typical state-space MCMC methods fail due to strong dependence between the parameters. In addition, we develop alternative models that represent a wide class of alternative encoding theories and perform model comparison using WAIC to determine whether the data suggest the occurrence multiplexing over alternative theories of neural encoding. Using the proposed framework, we provide evidence of multiplexing within the inferior colliculus and novel insight into the switching dynamics.
\end{abstract}

\noindent%
\vfill

\newpage
\spacingset{1.9}
\section{Introduction}

\begin{figure}[t]
    \centering
    \includegraphics[width = 0.90\textwidth]{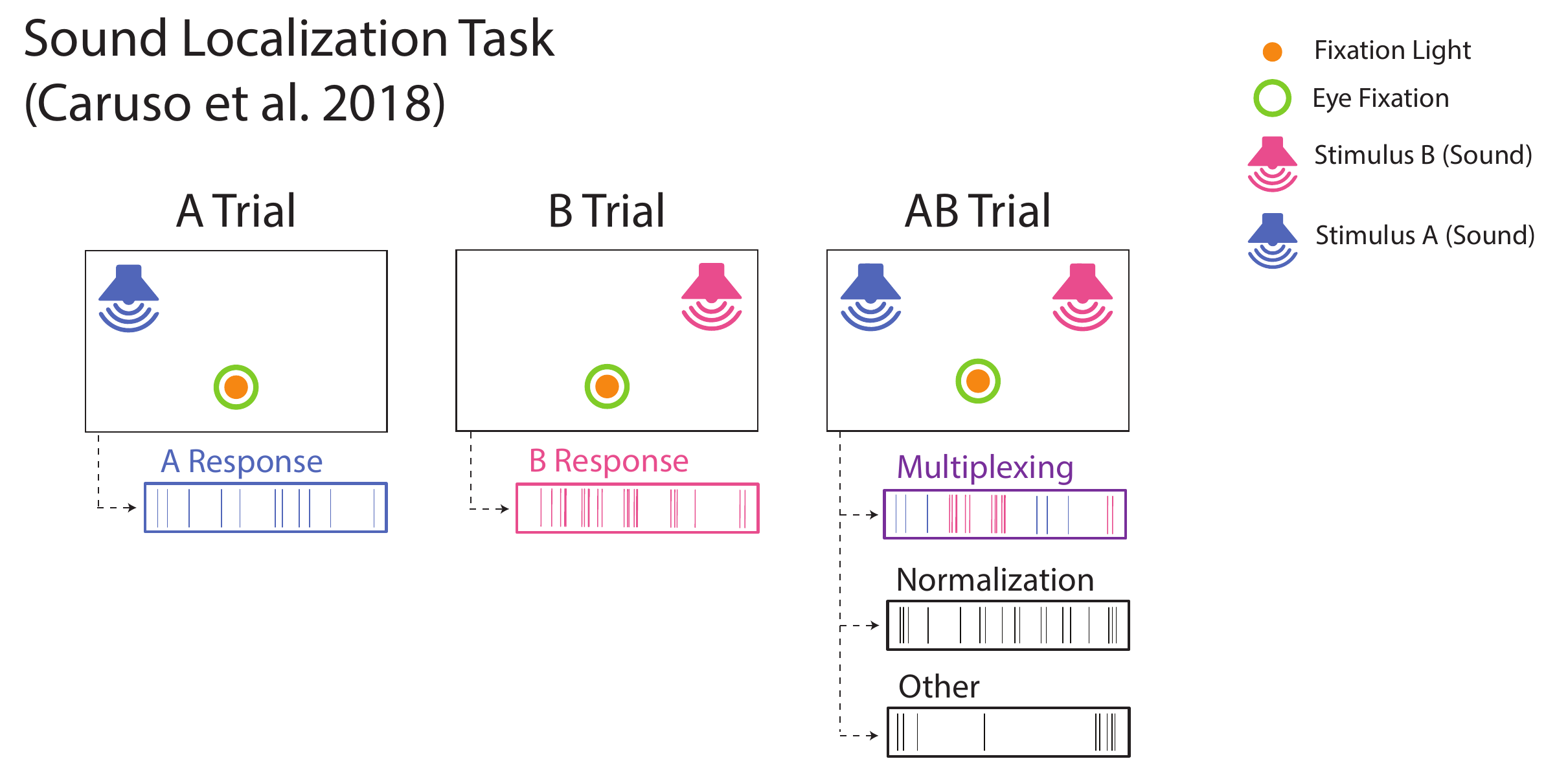}
    \caption{Diagram of the stimuli used in \cite{caruso2018single}, along with synthetic spike trains illustrating possible neural encodings for the dual-stimuli trials ($AB$ condition). Our scientific objective is to determine if multiplexing is occurring in the $AB$ condition trials and, if so, at what timescale.}
    \label{fig: caruso_task}
\end{figure}

In neuroscience research, a theory known as multiplexing posits that when presented with multiple stimuli together, individual neurons can switch over time between encoding each member of the stimulus ensemble, causing a fluctuating pattern of firing rates \citep{caruso2018single, mohl2020sensitivity, jun2022coordinated, schmehl2024multiple, groh2024signal}. Multiplexing is a scalable encoding scheme that offers a clear explanation of how information about each distinct stimulus is preserved. This theory is typically examined with extracellular recordings of neurons from various brain regions and under various triplets of conditions, such as the \textit{triplet} shown in Figure \ref{fig: caruso_task}. A triplet of conditions involves spike train recordings from a single neuron across three sets of trials: trials recorded under the $A$ condition (only $A$ stimulus), trials conducted under the $B$ condition (only $B$ stimulus), and trials under the $AB$ condition ($A$ and $B$ stimuli concurrently). Using the spike trains obtained under the $A$ condition and the $B$ condition as benchmarks, our aim is to determine whether the neuron utilizes multiplexing to encode the two stimuli and, if so, to infer the timescale at which the switching occurs.

Here, we introduce a new statistical framework to test and analyze rate fluctuations of multiplexing neurons by proposing a state-space model for point process data, which facilitates detailed and localized inference on which stimulus is being encoded by each spike emitted by a neuron. Our state-space model specification deviates from existing statistical literature in that the state changes are continuous-time, non-Markovian, and endogenous -- they are determined by the interactions between latent continuous processes, which also determine the emission of spikes. Specifically, for single stimulus trials, we adopt the integrate-and-fire model \citep{burkitt2006review} which posits that spikes are generated as a result of a latent drift-diffusion process hitting a boundary. We propose that multiplexing emerges in the dual-stimuli trials as a result of competition between the latent drift-diffusion processes associated with each stimulus. Under this competition framework, a spike is emitted whenever the first diffusion process hits the boundary. Therefore, competition among the single-stimuli drift diffusion processes controls both the transition probabilities of the proposed state-space model and the emission of a spike under the dual-stimuli trials, leading to the notion of an endogenous state-space model.

Statistical inference for the proposed state-space model is challenging due to observing only the hitting times of the continuous-time process, coupled with the model's non-Markovian and endogenous nature. Here, we propose a Bayesian approach as it enables us to obtain posterior probabilities of the latent states and obtain posterior predictive distributions for key scientific quantities of interest, such as the rate of switching or the time encoding each stimulus. However, the non-Markovian and endogenous nature of the proposed state-space model result in poor sampling performance when using typical Markov chain Monte Carlo (MCMC) methods for state-space models. 
To facilitate scalable and efficient posterior inference, we provide a novel MCMC algorithm for our proposed model, which is generalizable to general state-space models with similar dependence structures. 


A important question is whether such an intricate state-space model is at all necessary to explain the observations. While a simpler model for multiplexing could be devised, the proposed state-space model adopts a mechanistic approach, allowing its parameters to be directly mapped to biophysical mechanisms that generate spikes. Moreover, the state-space model implies a neural motif that could lead to this type of multiplexing behavior. This mechanistic construction probabilistically defines a falsifiable model for multiplexing, which in turn allows us to test whether the data support the posited biophysical mechanisms and neural motif. 

To statistically assess the necessity of our multiplexing-specific model to explain the observed data, we pitted it against a simpler and more general point process model that encapsulates alternative encoding theories (normalization, subadditivity, winner-take-all, etc.) with some level of abstraction. 
Using the proposed models, we conduct model comparison using the widely applicable information criterion (WAIC) \citep{watanabe2010asymptotic, watanabe2013widely} to determine whether the recorded neural activity is consistent with our mechanistic model for multiplexing. Through simulation studies, we show that WAIC is highly informative and reliable in model selection, while being computationally fast to compute. 

Using the proposed statistical framework, we analyze spike train data collected from neurons in the inferior colliculus (IC) of two macaque monkeys during sound localization tasks \citep{caruso2018single}. With this more granular framework, we were able to provide novel insight into the timescale at which switching occurs and obtain posterior probabilities of which stimulus was encoded in an individual spike. 
We will start by reviewing previous statistical evidence of multiplexing, 
showing that our proposed framework is related to previous frameworks, but is (1) more flexible, (2) provides more compelling evidence of multiplexing, and (3) provides novel insight into the switching behavior of multiplexing neurons. 
\section{Scientific Background}
\label{sec: scientific_background}
\subsection{Statistical Evidence of Multiplexing}
\label{sec: stat_evidence_multiplexing}

\begin{figure}[t]
    \centering
    \includegraphics[width = 0.99\textwidth]{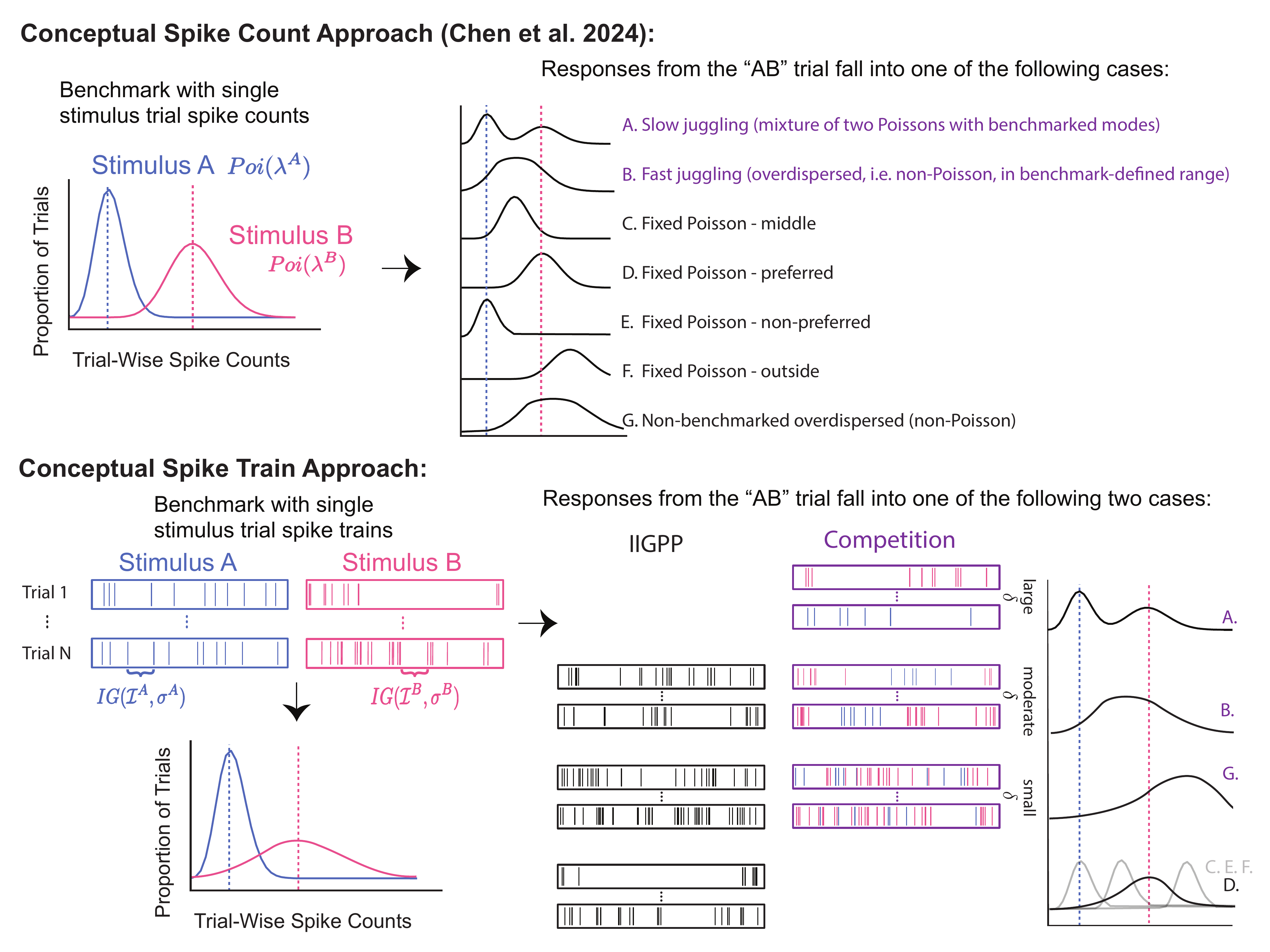}
    \caption{Illustrative differences in the spike count approach used in \cite{chen2024} and the proposed spike train approach. The visualization of the spike train approach demonstrates how the IIGPP (non-multiplexing) and Competition (multiplexing) models can result in identical spike count distributions; highlighting the necessity for a more granular approach.
    }
    \label{fig: spike_train_spike_count}
\end{figure}

 \cite{caruso2018single} were the first to propose and investigate multiplexing as a theory of neural encoding under exposure to multiple stimuli, with various refinements carried out in later works. In a recent work, \cite{chen2024} proposed a statistical framework, known as SCAMPI, to determine whether trial-wise spike count data supported the occurrence of multiplexing activity. As illustrated in Figure \ref{fig: spike_train_spike_count}, the framework assumes that the A condition and B condition spike counts are Poisson distributed and aims to categorize the $AB$ response as one of seven different responses, with purple representing the categories in which multiplexing is believed to possibly occur. However, a spike count analysis relies solely on the aggregate number of spikes within a trial; requiring a more granular level of analysis to determine whether multiplexing occurs on a time scale shorter than the duration of the trial. This is apparent as both the inhomogeneous inverse Gaussian point process (IIGPP) model (non-multiplexing model) and the competition model (multiplexing model) can generate spike count distributions that would be classified as \textsc{Fast Juggling} (B) and \textsc{Non-Benchmarked Overdispersed} (G). 

To achieve a more granular level of analysis, \cite{glynn2021analyzing} developed the dynamic admixture point process (DAPP) model. DAPP uses spike train data to model within-trial fluctuations in neuronal firing rates under the $AB$ condition. Although DAPP is capable of modeling fluctuating firing patterns, it is not a model specific to multiplexing; making it challenging to ascertain whether a neuron is multiplexing and providing minimal information about the timescale of any potential switching. We address this gap by constructing a mechanistic statistical model for multiplexing, allowing us to gain insight into the timescale and characteristics of multiplexing, as well as flexible alternative statistical models that represent alternative encoding theories with some level of abstraction. With these models, we are able to provide insight into how often multiplexing occurs and, if so, insight into the temporal dynamics of the switching process. Unlike DAPP and SCAMPI, our proposed models do not rely on Poisson distributional assumptions. Instead, we leverage the more flexible integrate-and-fire framework as the foundation of our statistical models. 


\subsection{Integrate-and-Fire Models}
\label{sec: IF Models}


Integrate-and-fire models \citep{burkitt2006review} are mathematical models that characterize the temporal dynamics of the membrane potential of a neuron through a set of differential equations with varying degrees of computational complexity and biophysical realism. 
The general form of a leaky integrate-and-fire model can be expressed as the following stochastic differential equation:
\begin{equation}
    \text{d}V(t) = \left[-g(t)\left(V(t) - V_0\right) + I(t)\right] \text{d}t + \sigma \text{d}W(t),
    \label{eq: LIF}
\end{equation}
where $V(t)$ denotes the membrane voltage, $V_0$ denotes the resting potential, $g(t)$ represents the membrane conductance,  $I(t)$ represents the input current, and  $W(t)$ denotes a Wiener process \citep{paninski2009statistical}. The process starts at the resting potential, $V_0$, and the generation of an action potential, or spike, occurs when the membrane voltage crosses some voltage threshold, $V_{th}$. Once the membrane voltage crosses the threshold, the membrane voltage returns to the resting potential, $V_0$.

What is relevant to our study is that the leaky integrate-and-fire model \eqref{eq: LIF} can be used to directly model extracelluar recordings of spike trains. Given a diffusion process representing the voltage process, the interspike interval times (ISIs), or the time between consecutive action potentials, can be modeled as first passage times. Although the leaky integrate-and-fire model induces a distribution on the ISIs, the distribution cannot be expressed in an analytic form. However, a simpler version of integrate-and-fire models, known as the perfect integrator model, leads to ISI distributions that have an analytic form. The perfect integrator model can be expressed as the following stochastic differential equation: $\text{d}V(t) = I \text{d}t + \sigma \text{d}W(t)$.
Assuming the perfect integrator model results in ISIs that follow an inverse Gaussian distribution, characterized by a mean parameter $\mu = \frac{V_{th}-V_0}{I}$ and a shape parameter $\lambda = (\frac{V_{th}-V_0}{\sigma})^2$ \citep{folks1978inverse}.
 More formally, let $\mathcal{S}_i := \{S_{i1}, \dots, S_{in_i}\}$ be the $i^{th}$ spike train, where $S_{ij}$ is defined as the time of the $j^{th}$ spike in the $i^{th}$ spike train ($i = 1, \dots, N$). Then, the $i^{th}$ ISI can be defined as $X_{ij} := S_{ij} - S_{i(j-1)}$ for $i = 1, \dots, N$ and $j = 1, \dots, n_i$, where $S_{i0} := 0$. Assuming the perfect integrator model, we have $X_{ij} \sim IG(\frac{V_{th}-V_0}{I}, (\frac{V_{th}-V_0}{\sigma})^2 )$.

\vspace{-1em}
\begin{remark}[Identifiability]
    Since we are working with extracellular recordings of neurons, we only have information on the hitting times (spikes) of the diffusion process. As specified, the model is currently unidentifiable. Since the resting potential ($V_0$) and the threshold potential ($V_{th}$) are similar for most neurons and are not of direct scientific interest for our case study, we assume that $V_{th} - V_0 = 1$. This assumption provides us with an identifiable model, enabling us to capture the firing rate through the input current ($I$) and capture the ISI variability through $\sigma$.
\end{remark}
\vspace{-1em}
\begin{remark}[Time-Inhomogeneity] To allow time-inhomogeneous firing rates, we allow the input current of the voltage process to change once a spike occurs. Specifically, letting $V_{ij}(t)$ be the voltage process associated with $X_{ij}$, we assume that $\text{d}V_{ij}(t) = \mathcal{I}(S_{i(j-1)}) \text{d}t + \sigma \text{d}W(t)$, where $S_{ij} := \sum_{k = 1}^j X_{ik}$ for $1 \le j \le n_i$. Thus, the pdf of $X_{ij}$ is 
\begin{equation}
    f(x_{ij}\mid \mathcal{I}(\cdot), \sigma, s_{i(j-1)}) = \frac{1}{\sigma\sqrt{2\pi x_{ij}^3}}\exp\left( -\frac{\left(1 - \mathcal{I}(s_{i(j-1)})x_{ij}\right)^2}{2\sigma^2x_{ij}}\right),
    \label{eq: ISI_IG_simp}
\end{equation}
for $i = 1, \dots, N$ and $j = 1, \dots, n_i$. Although not as flexible as allowing the input current to change between spikes, as in Equation \ref{eq: LIF}, this construction allows us to capture time-inhomogeneous firing rates while maintaining a computationally tractable model.
\end{remark}
\vspace{-1em}

Compared to using a typical Poisson process \citep{kass2014analysis}, this type of inverse Gaussian point process is more flexible; allowing us to capture ISI variability. 
In addition, modeling neural spike trains with this type of inverse Gaussian point process is well supported from a biophysical standpoint, as it can be formulated through an integrate-and-fire framework. We will refer to this general point process model \eqref{eq: ISI_IG_simp} as the inhomogeneous inverse Gaussian point process (IIGPP) model throughout the manuscript, and it will serve as the model for the spike trains observed under a single stimulus, as well as our alternative model to multiplexing.

\section{Statistical Models for Spike Trains}
\label{sec: stat_model_spike_trains}

\subsection{A Competition Model for Two Stimuli}
\label{sec: Competition_Model}

Multiplexing theory posits that neurons temporally switch between encoding multiple stimuli, causing a fluctuating spiking pattern \citep{groh2024signal}. A natural model for multiplexing is a state-space model in which hidden states identify the particular stimulus being encoded at a given time. Could one construct such a state-space model which interweaves perfect integrator models for single-stimulus responses? We offer an affirmative answer by proposing that transitions between hidden states arise from competition between the latent diffusion processes associated with each single stimulus. 
%
By letting competition drive the switching behavior, we offer a mechanistic explanation of multiplexing as a statistical phenomenon. In addition, we furnish a neural motif that could lead to this particular type of neural behavior. 

\begin{figure}[h!]
    \centering
    \includegraphics[width = 0.99\textwidth]{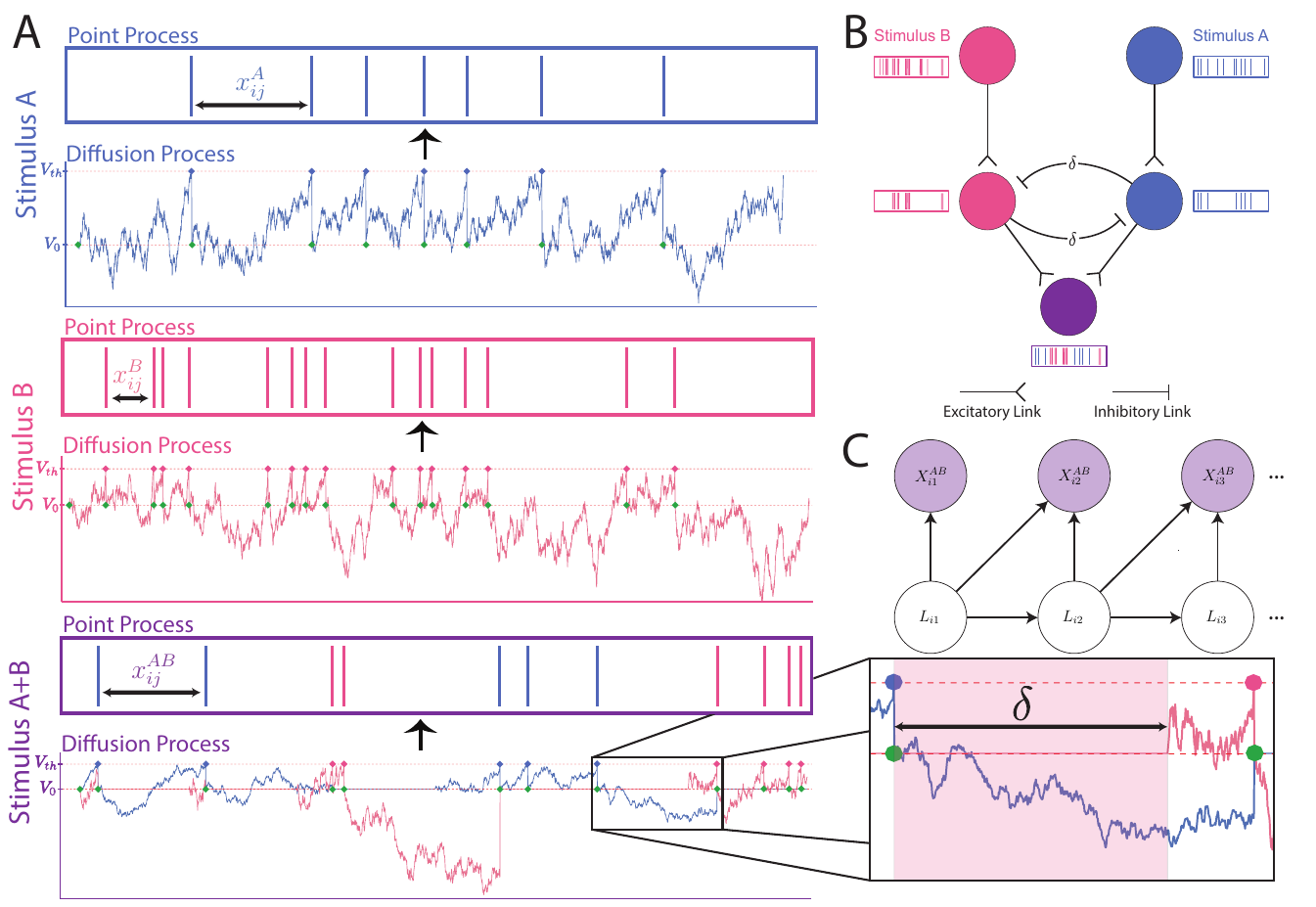}
    \caption{(Subfigure A) Illustration of how the latent drift diffusion processes relate to the observed spike trains. The start of the drift diffusion process is denoted by a green circle, while the hitting time is denoted by a red or blue circle depending on which stimulus the spike is encoding. (Subfigure B) Potential neural motif that could lead to the competition framework proposed in this manuscript. (Subfigure C) Directed acyclic graph (DAG) characterizing the dependence between the labels ($L_{ij}$) and the observed ISIs ($X_{ij}^{AB}$).}
    \label{fig: neural_circuit}
\end{figure}

In our experimental setup, we observe spike trains under three different conditions ($A, B, AB$), consisting of combinations of two different stimuli (stimulus $A$ and stimulus $B$). In this manuscript, we will let $\mathscr{H} \in \{A, B, AB\}$ denote one of the three conditions and let $\mathscr{S} \in \{A, B\}$ denote one of the two stimuli. Let $\mathcal{S}^{\mathscr{H}}_i = (S^{\mathscr{H}}_{ij}, 1 \le j \le n^{\mathscr{H}}_i)$ be the $i^{th}$ spike train generated under condition $\mathscr{H}$ for $i = 1,\dots, N^{\mathscr{H}}$ and $\mathscr{H} = A, B, AB$. Let $X_{ij}^{\mathscr{H}} := S^{\mathscr{H}}_{ij} - S^{\mathscr{H}}_{i(j-1)}$ be the corresponding ISIs and $\mathcal{T} := [0, T]$ be the experimental time window of interest. From Equation \ref{eq: ISI_IG_simp}, the probability density function of $X_{ij}^{\mathscr{S}}$ given $(S^{\mathscr{S}}_{ik}=s_k, 0\le k < j)$ is
\begin{equation}
    f^\mathscr{S}_j(x \mid s_{j-1}) = \frac{1}{\sigma^{\mathscr{S}}\sqrt{2\pi x^3}}\exp\left\{ -\frac{\left(1 - \mathcal{I}^{\mathscr{S}}(s_{j-1})x\right)^2}{2(\sigma^{\mathscr{S}})^2x}\right\},
    \label{eq: ISI_single_pdf}
\end{equation}
for $i = 1, \dots, N^{\mathscr{S}}$, $j = 1, \dots, n_i^{\mathscr{S}}$, and ${\mathscr{S}}  = A, B$. Given the two voltage processes for stimulus $A$ and stimulus $B$, we can model the $AB$ process as a competition between the two processes: an action potential is generated once the $A$ voltage process or the $B$ voltage process reaches the voltage threshold. Once either of the voltage processes reaches the voltage threshold, \textit{both} the $A$ and $B$ voltage processes are reset to their resting potential, $V_0$. To control how often the dual-stimuli process switches between encodings, we will add inhibition through the introduction of a time delay for one of the processes. Inhibition, which can be visualized in Figure \ref{fig: neural_circuit}, will control the overall firing rate and the probability of switching.  Lastly, we will let $L_{ij} \in \{A, B\}$, which we refer to as labels, be latent variables that represent which process generated the the $j^{th}$ spike in the $i^{th}$ spike train under the $AB$ condition for $i = 1, \dots, N^{AB}$ and $j = 1, \dots, n^{AB}_i$. These labels represent the states in the proposed state-space model and will allow us to infer what stimulus is being encoded in an individual spike.

The proposed competition process shares similarities with accumulator models, which are often used to model multicategory decision tasks \citep{ratcliff2008diffusion, paulon2021bayesian}. Accumulator models typically use diffusion processes to model the accumulation of evidence over time for different possible decisions; the chosen decision corresponds to the diffusion process that reaches its boundary first. Since the decisions are known from the experiment, the goal is to conduct inference on the latent diffusion processes. However, we go one step further by also inferring which diffusion process reached the voltage threshold to generate the spike. 


From a computational perspective, an important aspect of the above modeling framework is that it can be fully characterized through the distribution of the ISIs. The joint probability density function of the labels and ISIs corresponding to the first spike in the spike trains can be expressed as
\begin{equation}
f^{AB}_1\left(x_{i1}^{AB}, L_{i1} = \mathscr{S}\mid\boldsymbol{\theta}\right) = f^{\mathscr{S}}_1\left(x^{AB}_{i1}\right)\left[1 - F^{^{\mathscr{S}^C}}_1\left(x^{AB}_{i1}\right) \right]
\label{eq: joint_pdf_first}
\end{equation}
for $i = 1, \dots, N^{AB}$, where $f^{\mathscr{S}}_j(x_{ij}) := f^{\mathscr{S}}_j(x_{ij}\mid  s^{\mathscr{S}}_{i(j-1)})$, $\mathscr{S}^C := \{A, B\} \setminus \mathscr{S}$, $F^{\mathscr{S}}_j(\cdot)$ is the cumulative density function of $X_{ij}^\mathscr{S}$, and $\boldsymbol{\theta}:= \{\mathcal{I}^A(\cdot), \mathcal{I}^B(\cdot), \sigma^A, \sigma^B, \delta\}$. Similarly, the joint probability density function of the labels and the rest of the ISIs can be expressed in a similar functional form, but with the inclusion of inhibition through a time delay for one of the processes. Specifically, we have 
\begin{equation}
\begin{split}
\label{eq: joint_pdf_rest}
f^{AB}_j\left(x_{ij}^{AB}, L_{ij} = \mathscr{S}\mid\boldsymbol{\theta}, l_{i(j-1)}, s^{AB}_{i(j-1)}\right)  =   f^{\mathscr{S}}_j\left(x^{AB}_{ij} - \delta \mathbbm{1}\left\{l_{i(j-1)} = \mathscr{S}^C\right\}\right)\\ 
\times \left[1 - F^{{\mathscr{S}^C}}_j\left(x^{AB}_{ij} - \delta \mathbbm{1}\left\{l_{i(j-1)} = \mathscr{S}\right\}\right) \right],
\end{split}
\end{equation}
for $i = 1, \dots, N^{AB}$ and $j = 2, \dots, n_i^{AB}-1$. Thus, if the ISI is smaller than the time delay, then the label of the $j^{th}$ spike must be equal to the value of the previous label (i.e. $l_{ij} = l_{i(j-1)}$). For the last spike, we have to account for not observing any spikes after the last spike ($s_{in_{i}^{AB}}^{AB}$) in the time window $\mathcal{T}$. Letting $\tilde{n}_i$ be defined as the last spike in the $i^{th}$ dual-stimuli spike train (i.e. $x_{\tilde{n}_i}^{AB} := x^{AB}_{in_{i}^{\mathcal{AB}}}$; 
 $x_{\tilde{n}_i +1}^{AB} := x^{AB}_{i(n_{i}^{\mathcal{AB}}+1)}$), we have
\begin{equation}
\label{eq: joint_pdf_last_spike}
\begin{split}
f^{AB}_{\tilde{n}_i}&\left(x^{AB}_{\tilde{n}_i}, X^{AB}_{\tilde{n}_i + 1} > T - s^{AB}_{\tilde{n}_i}, L_{\tilde{n}_i} = \mathscr{S}\mid\boldsymbol{\theta}, l_{\tilde{n}_i - 1}, s^{AB}_{\tilde{n}_i -1}\right) = \\
 & f^{\mathscr{S}}_{\tilde{n}_i}\left(x^{AB}_{\tilde{n}_i} - \delta \mathbbm{1}\left\{l_{\tilde{n}_i - 1} = \mathscr{S}^C\right\}\right)\left[1 - F^{{\mathscr{S}^C}}_{\tilde{n}_i}\left(x^{AB}_{\tilde{n}_i} - \delta \mathbbm{1}\left\{l_{\tilde{n}_i - 1} = \mathscr{S}\right\}\right) \right]\\ 
&\times \left[1 - F^{{\mathscr{S}^C}}_{\tilde{n}_i + 1} \left(T - s_{\tilde{n}_i}^{AB} - \delta\right) \right] \left[1 - F^{{\mathscr{S}}}_{\tilde{n}_i + 1}\left(T - s_{\tilde{n}_i}^{AB}\right) \right],    
\end{split}
\end{equation}
for $i = 1, \dots, N^{AB}$.
Using Equation \ref{eq: joint_pdf_rest}, the conditional density of ISIs recorded under the $AB$ can be expressed as 
\begin{equation}
    \label{eq: llik_AB}
    f_{X^{AB}}\left(x_{ij}^{AB}\mid l_{i(j-1)}, l_{ij}, \boldsymbol{\theta}, s^{AB}_{i(j-1)}
    \right) = \frac{f^{AB}_j\left(x_{ij}^{AB}, l_{ij}\mid\boldsymbol{\theta}, l_{i(j-1)},  s^{AB}_{i(j-1)}\right)}{\int_0^\infty f^{AB}_j\left(x, l_{ij}\mid\boldsymbol{\theta}, l_{i(j-1)}, s^{AB}_{i(j-1)}\right)\text{d}x},
\end{equation}
for $i = 1, \dots, N^{AB}$ and $j = 2, \dots, n_i^{AB} - 1$. Similarly, expressions for $f_{X^{AB}}\left(x_{i1}^{AB}\mid l_{i1}, \boldsymbol{\theta}\right)$ and $f_{X^{AB}}\left(x_{\tilde{n}_i}^{AB}, X^{AB}_{\tilde{n}_i + 1} > T - s_{\tilde{n}_i}^{AB}\mid \allowbreak l_{\tilde{n}_i}, l_{\tilde{n}_i - 1}, \boldsymbol{\theta}, s^{AB}_{\tilde{n}_i}\right)$ can be derived from Equations \ref{eq: joint_pdf_first} and \ref{eq: joint_pdf_last_spike}. Importantly, the competition framework leads to a tilted density \eqref{eq: llik_AB}, where $\delta$ affects the conditional distribution of the ISIs observed under the $AB$ condition.

As illustrated in Figure \ref{fig: spike_train_spike_count}, different values of $\delta$ under this competition framework lead to spike count distributions under the $AB$ condition that coincide with different categories identified in \citet{chen2024}. Specifically, a model where $\delta$ is large results in spike counts that correspond to \textsc{Slow Juggling}. When $\delta$ is large, the probability of switching encodings within the trial is small, leading to spike counts that have a mixture distribution. 
The category referred to as \textsc{Fast Juggling} in \cite{chen2024} corresponds to models where $\delta$ is moderate. When $\delta$ is moderate, neurons switch encodings within a trial with clear blocks of time where the neuron only encodes one stimulus. Due to tilting, the expected ISI time conditional on the label is shorter than the corresponding expected ISI time under the single-stimulus condition. Lastly, the modeling framework extends the theory of multiplexing by allowing some subadditive processes to be considered multiplexing. Specifically, as $\delta$ approaches zero, the model posits constant switching, resulting in a subadditive effect in the expected spike counts and firing rate. 
In general, this framework posits that the spiking behavior and the probability of switching encodings are dependent on (1) the distribution of ISIs under single-stimulus conditions and (2) the penalty for switching between encodings, $\delta$.

\vspace{-1em}
\begin{remark}[Scientific Implications]
    The competition framework implies the neural circuit in Subfigure 3B. In this neural circuit, we have two neurons at the top level, each encoding their respective stimulus-specific signal. Each of these top-level neurons has an excitatory link to the next layer of neurons, which have mutual inhibitory links between the second layer of neurons. In this second layer, if one neuron fires, then it inhibits the other neuron from firing, which is controlled by the $\delta$ parameter in our model. Lastly, these second-layer neurons are connected to a final neuron through an excitatory link, which would have a fluctuating firing rate; encoding for both the $A$ stimulus and the $B$ stimulus over time. 
\end{remark}
\vspace{-1em}

\subsection{Model Specification}
\label{sec: Model Specification}
Using the integrate-and-fire framework specified in Section \ref{sec: IF Models} to model the spike trains observed under condition $A$ and condition $B$, and utilizing the competition framework specified in Section \ref{sec: Competition_Model} to model the spike trains observed under the $AB$ condition, we can specify a statistical model for the spike trains observed from the ``triplets'' of conditions. 
Let $\mathcal{T} = [0, T]$ be the experimental time window of interest. To allow for a time-inhomogeneous point process, we will utilize B-splines \citep{eilers1996flexible}, which will allow for a flexible yet computationally efficient model. Let $\mathbf{b}(t):= [b_1(t), \dots, b_P(t)]^\intercal \in \mathbb{R}^P$ be the set of basis functions, without including an intercept term, evaluated at $t \in \mathcal{T}$. The input current, which depends on the time of the previous spike, can be specified as $\mathcal{I}^\mathscr{S}(s) = I^\mathscr{S} \exp\left(\left(\boldsymbol{\phi}^\mathscr{S}\right)^\intercal \mathbf{b}(s) \right)$, where $\boldsymbol{\phi}^\mathscr{S} \in \mathbb{R}^P$ are the spline coefficients that model the time-heterogeneity of the firing rate under stimulus $\mathscr{S} \in \{A,B\}$.
Thus, the probability density function of $X^\mathscr{S}_{ij}$ given $(S^{\mathscr{S}}_{ik}=s_k, 0\le k < j)$ is
\begin{footnotesize}
    \begin{equation}
    \label{Inverse_Gaussian_spline}
    f^{\mathscr{S}}_j(x \mid \boldsymbol{\theta}, s_{j-1}) = \frac{1}{\sigma^{\mathscr{S}}\sqrt{2\pi x^3}}\exp\left( -\frac{\left(1 - I^\mathscr{S} \exp\left\{\left(\boldsymbol{\phi}^\mathscr{S}\right)^\intercal \mathbf{b}\left(s_{j-1}\right) \right\}x\right)^2}{2(\sigma^{\mathscr{S}})^2x}\right),
\end{equation}
\end{footnotesize}
for $i = 1, \dots, N^\mathscr{S}$, $j = 1,\dots, n_i^{\mathscr{S}}$, and $\mathscr{S} = A, B$. Here, we let $\boldsymbol{\theta}$ denote the set of parameters $\{I^A, I^B, \sigma^A, \sigma^B, \delta, \boldsymbol{\phi}^A, \boldsymbol{\phi}^B\}$. The likelihood function of $\boldsymbol{\theta}$ given observing the spike train $\mathcal{S}_i^\mathscr{S}$ in the experimental window $\mathcal{T}$ can be expressed as
\begin{footnotesize}
    \begin{equation}
    \label{eq: likelihood_IIGPP}
    \mathcal{L}^{\mathscr{S}}(\boldsymbol{\theta} \mid \mathcal{S}_i^{\mathscr{S}} ) = \left(\prod_{j =1}^{n_i^{\mathscr{S}}} f^{\mathscr{S}}_j(x^{\mathscr{S}}_{ij}\mid \boldsymbol{\theta}, s^{\mathscr{S}}_{i(j-1)})\right) \left[1 - F^{\mathscr{S}}_{n_i^{\mathscr{S}} + 1}(X_{i(n_i^{\mathscr{S}} + 1)}^\mathscr{S} > T - s^{\mathscr{S}}_{in_i^{\mathscr{S}}}\mid \boldsymbol{\theta}, s^{\mathscr{S}}_{in_i^{\mathscr{S}}})\right],
\end{equation}
\end{footnotesize}
where the last term accounts for the probability of not observing another spike in the rest of the experimental time window. Using these specified probabilistic models for the single-stimulus response, we can use Equations \ref{eq: joint_pdf_first}, \ref{eq: joint_pdf_rest}, and \ref{eq: joint_pdf_last_spike} to specify the joint distribution of the latent labels and the observed spike trains under the $AB$ condition.

\section{Estimation and Model Comparison}
\label{sec: posterior_inference_model_comparison}
We carry out estimation under a Bayesian paradigm, which is a fairly typical choice in the literature for statistical inference on state-space models \citep{chib1996calculating}. Among other things, Bayesian estimation offers probabilistic expressions to infer the latent states and also quantify the uncertainty associated with such inference. It also offers a natural quantification of the fit of the model to the data by utilizing the framework of posterior predictive evaluations, as will be made clear shortly. 
Importantly, with suitably chosen prior distributions, we are able to make use of filtering algorithms to carry out the computational task in linear time. 


\subsection{Prior Specification}
\label{sec: Prior_specification}
A hierarchical shrinkage prior is placed on the $\boldsymbol{\phi}^\mathscr{S}$ parameters to promote shrinkage of the basis coefficients toward zero. Specifically, we will assume that $\boldsymbol{\phi}^\mathscr{S} \sim \mathcal{N}_P(\mathbf{0}, \tau^\mathscr{S}\mathbf{I})$ and $\sqrt{\tau^\mathscr{S}} \sim t^+(\nu, \gamma)$
for $\mathscr{S} = A, B$, where $t^+(\nu, \gamma)$ denotes the half-t distribution with $\nu$ degrees of freedom and scale parameter $\gamma$. Compared to using an inverse-gamma prior on $\tau^\mathscr{S}$, the half-Cauchy distribution has been shown to be more flexible, to have better behavior near the shrinkage point, and less sensitive to the choice of hyperparameter $\gamma$ \citep{gelman2006prior, polson2012half}. The use of a half-t prior also allows for conditionally conjugate sampling schemes \citep{Wand2011}, leading to simple and efficient sampling algorithms. 
Through simulations, we found that setting $\gamma = 2$ and $\nu = 0.25$ led to good results in both time-homogeneous and time-inhomogeneous settings; however, the results were relatively robust to the choice of $\gamma$ and $\nu$. 

Weakly informative priors are utilized for the remaining parameters in $\boldsymbol{\theta}$, such that $I^\mathscr{S} \sim IG(\alpha_I, \beta_I)$, $\sigma^\mathscr{S} \sim IG(\alpha_\sigma, \beta_\sigma)$, and  $\delta \sim \Gamma(\alpha_\delta, \beta_\delta)$ for $\mathscr{S} = A, B$, where $\Gamma(\alpha, \beta)$ denotes the Gamma distribution with shape $\alpha$ and rate $\beta$. In practice, we set $\alpha_I = 40$, $\beta_I = 1$, $\alpha_\sigma = \sqrt{40}$, $\beta_\sigma = 1$, $\alpha_\delta =  0.01$, $\beta_\delta = 0.1$. Thus, we have a fully specified Bayesian framework to model the spike trains generated from the triplet of conditions.

\subsection{Posterior Inference}
\label{sec: posterior_inference}
Posterior inference is performed using an efficient Markov chain Monte Carlo (MCMC) algorithm to obtain samples from the posterior distribution. Sampling from the conditional posterior distributions of the sets of parameters $\{I^A, I^B, \sigma^A, \sigma^B \}$ and $\{\boldsymbol{\phi}^A, \boldsymbol{\phi}^B \}$ is carried out using Hamiltonian Monte Carlo (HMC) \citep{neal2011mcmc}. The parameters $I^A, I^B, \sigma^A,$ and $\sigma^B$ are transformed to remove positivity constraints, and HMC is carried out in the transformed space. 
While the number of leapfrog steps is user-specified, the step size and mass matrix used in the HMC sampling scheme are adaptively learned in the initial warm-up blocks of the algorithm. 


Although $\delta$ is a continuous variable with support on $\mathbb{R}^+$, HMC is not a suitable method to obtain samples from the posterior distribution due to the complicated posterior geometry. Furthermore, the dependence between $\delta$ and the labels, $L_{ij}$, necessitates a joint sampler to achieve efficient sampling with good mixing. To construct an efficient MCMC sampling scheme, we will start by specifying an efficient MCMC algorithm for sampling from the posterior distribution of the labels.
While the simplest MCMC scheme would consist of local updates, updating each $L_{ij}$ at a time, the complicated dependency among the labels and the addition of a time delay make this MCMC scheme unsuitable. 
Therefore, an MCMC scheme that simultaneously updates all the labels from a spike train is needed to have an efficient sampler. From Figure \ref{fig: neural_circuit}, it is apparent that our model does not assume the Markov property, as dependence exists between the previous label and the current ISI. However, forward filtering backward sampling (FFBS) \citep{baum1970maximization, baum1972inequality, chib1996calculating}, a commonly used MCMC algorithm for inference on Markovian state-space models, can be adapted to sample from the posterior distribution of the labels. 

The FFBS algorithm starts with the forward filtering step, where the objective is to calculate the label probabilities, marginalizing out the dependence on the previous labels. 
While $P(L_{i1} = \mathscr{S} \mid x^{AB}_{i1}, \boldsymbol{\theta})$ can be calculated from Equation \ref{eq: joint_pdf_first}, we recursively calculate $P(L_{ij} = \mathscr{S} \mid \{x^{AB}_{ik}\}_{k=1}^j, \boldsymbol{\theta})$ given that
{\footnotesize\begin{align}
    \label{eq: forward_filtration}
    P(L_{ij} =  \mathscr{S} \mid \{x^{AB}_{ik}\}_{k=1}^j, \boldsymbol{\theta}) & \propto  \sum_{\mathscr{S}' = \{ A, B\}}  \bigg[P\left(L_{i(j-1)} = \mathscr{S}'\mid \{x^{AB}_{ik}\}_{k=1}^{j-1}, \boldsymbol{\theta}\right) \\
    \nonumber & \times f^{AB}_{j}\left(x_{ij}^{AB}, L_{ij} = \mathscr{S}\mid L_{i(j-1)} = \mathscr{S}', \boldsymbol{\theta}, s^{AB}_{i(j-1)}\right)\bigg],
\end{align}}
for $j = 2, \dots, n_i^{AB} -1$ and $i = 1, \dots, N^{AB}$. Similarly, the probability of the last label in each spike train ($P(L_{in_i^{AB}} = \mathscr{S} \mid \mathcal{S}_i^{AB}, \boldsymbol{\theta})$) can be calculated using Equation \ref{eq: joint_pdf_last_spike}, which accounts for not observing another spike in the experimental time domain.

Upon calculating all the marginal probabilities, samples can be drawn from the posterior distribution using the backward sampling step. Specifically, a sample, denoted $\{\tilde{l}_{ij}\}_{j=1}^{\tilde{n}_i}$, from $P(\mathbf{L}_i = \mathbf{l}_i\mid \mathcal{S}_i^{AB}, \boldsymbol{\theta})$ can be drawn using Algorithm \ref{alg1}.
\begin{alg}
\label{alg1}
Let $P(L_{ij} = \mathscr{S} \mid \{x^{AB}_{ik}\}_{k=1}^j, \boldsymbol{\theta})$ and $P(L_{\tilde{n}_i} =  \mathscr{S} \mid \mathcal{S}_i^{AB}, \boldsymbol{\theta})$ be calculated in the forward filtering step (Equation \ref{eq: forward_filtration}) for $i = 1,\dots, N^{AB}$ and $j = 1, \dots, n_i^{AB}-1$. For $i = 1, \dots, N^{AB}$ repeat the following:
\begin{enumerate}
    \item Set $\tilde{l}_{\tilde{n}_i} = \mathscr{S}$ with probability $P(L_{\tilde{n}_i} = \mathscr{S} \mid \mathcal{S}_i^{AB}, \boldsymbol{\theta})$ for $\mathscr{S} \in \{A, B\}$,
    \label{BS1}
    \item For $ j = n_i^{AB} - 1, \dots, 1$, set $\tilde{l}_{ij} = \mathscr{S}$ with probability  
    \begin{footnotesize}
        $P(L_{ij} = \mathscr{S} \mid \{x^{AB}_{ik}\}_{k=1}^{j+1}, \{\tilde{l}_{ik}\}_{k=j+1}^{n_i^{AB}}, \boldsymbol{\theta}) \propto P(L_{ij} = \mathscr{S} \mid \{x^{AB}_{ik}\}_{k=1}^j, \boldsymbol{\theta})f^{AB}_{j+1}\left(x_{i(j+1)}^{AB}, \tilde{l}_{i(j+1)} \mid L_{ij} = \mathscr{S}, \boldsymbol{\theta}, s^{AB}_{i(j-1)}\right)$.
    \end{footnotesize}
    \label{BS2}
\end{enumerate}
\end{alg}
\vspace{-1em}

 Although the FFBS algorithm provides an efficient way to sample from $P(\mathbf{L}_i = \mathbf{l}_i \mid \mathcal{S}_i^{AB}, \boldsymbol{\theta})$, updating the labels and $\delta$ separately leads to poor mixing results, as detailed in the Supplementary Materials. Therefore, we propose a method to efficiently sample from $f(\delta, \{\mathbf{l}_i\}_{i = 1}^{N^{AB}} \mid \{\mathcal{S}_i^{AB}\}_{i =1}^{N^{AB}}, \boldsymbol{\theta}_{-\delta}, \alpha_\delta, \beta_\delta)$, where $\boldsymbol{\theta}_{-\delta} := \boldsymbol{\theta} \setminus \{\delta\}$. Our method requires a user-defined proposal distribution for $\delta$, $q_\delta$, such that $q_\delta$ has support on $\mathbb{R}^+$ and $q_\delta(\delta) > 0$ whenever $f(\delta|\{\mathcal{S}_i^{AB}\}_{i=1}^{N^{AB}}, \boldsymbol{\theta}_{-\delta}) > 0$. Using this, we can efficiently draw samples from $f(\delta, \{\mathbf{l}_i\}_{i = 1}^{N^{AB}} \mid \allowbreak \{\mathcal{S}_i^{AB}\}_{i =1}^{N^{AB}}, \boldsymbol{\theta}_{-\delta}, \alpha_\delta, \beta_\delta)$ using Algorithm \ref{alg2}.

\begin{alg}
\label{alg2}
Given $\{\mathbf{x}_i^{AB}\}_{i =1}^{N^{AB}}$, $\mathcal{T}$, $q_\delta$, $\alpha_\delta$, $\beta_\delta$, and $\boldsymbol{\theta}_{-\delta}$, generate a Markov chain $\Big(\Big\{\Big(\delta^{s}, \{\mathbf{l}_i^{s}\}_{i = 1}^{N^{AB}}\Big) : s \in \mathbb{N}\Big\}\Big)$ as follows: \begin{enumerate}
    \item Start with some $\delta^0$ and $\{\mathbf{l}_i^{0}\}_{i = 1}^{N^{AB}}$.
    \label{joint sampler 1}
    \item For $s = 1, 2, \dots$ repeat the following:
    \label{joint sampler 2}
        \begin{enumerate}
            \item Set $\hat{\delta}_0 = \delta^{s-1}$ and randomly generate $\hat{\delta}_1, \dots, \hat{\delta}_{M_\delta}$ from $q_\delta(\delta)$ ($M_\delta \ge 1$).
            \item Letting $\hat{\boldsymbol{\theta}}_m := \{I^A, I^B, \sigma^A, \sigma^B, \hat{\delta}_m, \boldsymbol{\phi}^A, \boldsymbol{\phi}^B\}$, run the forward filtering step to obtain the marginal likelihood via the normalizing constants 
                \begin{equation}
                \label{eq: marginal_likelihood}
                \begin{split}
                \mathcal{L}_{Comp}^{AB}\left(\hat{\boldsymbol{\theta}}_m \mid \mathcal{S}_i^{AB}\right) = f\left(x_{i1}^{AB} \mid \hat{\boldsymbol{\theta}}_m \right)\left(\prod_{j=1}^{n_i^{AB} - 1} f\left( x_{ij}^{AB} \mid \{x_{ik}^{AB}\}_{k=1}^{j-1}, \hat{\boldsymbol{\theta}}_m\right)\right)\\
                     \times f\left( x^{AB}_{\tilde{n}_i}, X^{AB}_{\tilde{n}_i +1} > T - s_{\tilde{n}_i}^{AB} \mid \{x_{ik}^{AB}\}_{k=1}^{n_i^{AB}-1}, \hat{\boldsymbol{\theta}}_m \right).
                \end{split}
            \end{equation}
            
            \item Set $\delta^s = \hat{\delta}_m$ with probability proportional to {\footnotesize$\frac{f\left(\hat{\delta}_m\mid \left\{\mathcal{S}_i^{AB}\right\}_{i=1}^{N^{AB}}, \boldsymbol{\theta}_{-\delta}, \alpha_\delta, \beta_\delta\right)}{q_\delta(\hat{\delta}_m)}$}.
            \item Conditionally on $\delta^s$, obtain $\{\mathbf{l}_i^s\}_{i=1}^{N^{AB}}$ through backward sampling (Algorithm \ref{alg1}).
        \end{enumerate}
\end{enumerate}
\end{alg}

Algorithm \ref{alg2} starts by drawing $M_\delta$ independent samples from $q_\delta(\delta)$ and then running the forward filtering step of FFBS $M_\delta + 1$ times to obtain the marginal likelihood under the different proposed values of $\delta$. We can then calculate the acceptance probability for each proposed $\delta$, which is equivalent to Barker's proposal \citep{barker1965monte} when $M_\delta = 1$ and, more generally, to the acceptance probability obtained in ensemble MCMC \citep{neal2011mcmc_ensemble} when the ensemble states are independent and identically distributed under the ensemble base measure. Conditionally on the chosen $\delta$, independent samples from the posterior distribution of the labels are drawn by running the backwards sampling step. The efficiency of the algorithm can be largely attributed to the ability to marginalize out dependency of the states when sampling $\delta$, the parameter that affects the transition probability between states.

\begin{lemma}
    The Markov chain $\left(\left\{\left(\delta^{s}, \{\mathbf{l}_i^{s}\}_{i = 1}^{N^{AB}}\right) : s \in \mathbb{N}\right\}\right)$ generated from Algorithm \ref{alg2} satisfies detailed balance with respect to $f\left(\delta, \{\mathbf{l}_i\}_{i = 1}^{N^{AB}} \mid \left\{\mathcal{S}_i^{AB}\right\}_{i=1}^{N^{AB}}, \boldsymbol{\theta}_{-\delta}, \alpha_\delta, \beta_\delta\right)$.
    \label{lemma: detailed_balance}
\end{lemma}

From Lemma \ref{lemma: detailed_balance}, we show that the constructed Markov chain generated from Algorithm \ref{alg2} is a reversible Markov chain, and therefore the Markov chain will converge to its stationary distribution, $f\big(\delta, \{\mathbf{l}_i\}_{i = 1}^{N^{AB}} \mid \allowbreak \left\{\mathcal{S}_i^{AB}\right\}_{i=1}^{N^{AB}}, \boldsymbol{\theta}_{-\delta}, \alpha_\delta, \beta_\delta\big)$. The proof of Lemma \ref{lemma: detailed_balance} can be found in the Supplementary Materials.

The efficiency of the proposed sampling strategy is highly dependent on the choice of the proposal distribution, $q_\delta(\delta)$. Following \citet{gaasemyr2003adaptive}, we choose an adaptive approach consisting of a mixture of distributions such that
$q_\delta(\delta\mid \alpha_\delta, \beta_\delta) = \alpha \pi_\delta(\delta\mid \alpha_\delta, \beta_\delta) + (1 - \alpha)f_\delta(\delta\mid\boldsymbol{\psi}_\delta),$
for some $\alpha \in (0,1)$, where $\pi_\delta(\cdot\mid \alpha_\delta, \beta_\delta)$ is the prior density function of $\delta$ and $\boldsymbol{\psi}_\delta$ is a set of parameters. In our implementation, we choose $f_\delta(\delta|\boldsymbol{\psi}_\delta)$ to be the probability density function of a log-normal distribution with $\boldsymbol{\psi}_\delta = (\mu, \sigma^2)$, where $\boldsymbol{\psi}_\delta$ is adaptively learned. 
This approach allows us to generate a proportion of our proposed moves from a distribution that is adapted to have geometry similar to the posterior distribution, leading to high acceptance rates. The remaining proposals are drawn from the prior distribution, which is generally assumed to be relatively diffuse, thereby permitting large moves in the parameter space.

Using the methods discussed in this section, we have a scalable and efficient MCMC sampling scheme that facilitates posterior inference. The Supplementary Materials include detailed versions of the algorithms, a comparison between the proposed sampling strategy and other sampling methods, and in-depth information on how we adapt the Markov chain to achieve an efficient algorithm.

\subsection{Model Comparison}
\label{sec: Model Comparison}
In order to obtain evidence of multiplexing,
we adopt a Bayesian model comparison framework in which we use the widely applicable information criterion (WAIC) \citep{watanabe2010asymptotic, watanabe2013widely} to carry out model selection between the multiplexing state-space model and an alternative model. The alternative model, referred to as the inhomogeneous inverse Gaussian point process (IIGPP) model, represents a wide class of alternative theories of neural encoding with a higher level of abstraction. As illustrated in Table \ref{tab: two_models}, the competition model and the IIGPP model have the same structure under the single-stimulus conditions; however, the IIGPP model assumes that the ISIs observed under the dual-stimuli condition are inverse Gaussian distributed with parameters specific to the dual-stimuli condition.

\begin{table}
\caption{\label{tab: two_models}Specification of the likelihood functions for each of the models for each condition.}
\resizebox{\columnwidth}{!}{%
\begin{tabular}{l|c|c} 
   $\mathscr{H}$ & IIGPP Model & Competition Model\\
    \hline
    $A$ & $\prod_{i=1}^{N^A}\mathcal{L}^A\left(I^A, \sigma^A, \phi^A \mid \mathcal{S}_i^A\right)$ [Eq \ref{eq: likelihood_IIGPP}] & $\prod_{i=1}^{N^A}\mathcal{L}^A\left( I^A, \sigma^A, \phi^A \mid \mathcal{S}_i^A\right)$[Eq \ref{eq: likelihood_IIGPP}]\\
    $B$ & $\prod_{i=1}^{N^B}\mathcal{L}^B\left( I^B, \sigma^B, \phi^B \mid \mathcal{S}_i^B\right)$ [Eq \ref{eq: likelihood_IIGPP}] & $\prod_{i=1}^{N^B}\mathcal{L}^B\left( I^B, \sigma^B, \phi^B \mid \mathcal{S}_i^B\right)$ [Eq \ref{eq: likelihood_IIGPP}]\\
    $AB$ & $\prod_{i=1}^{N^{AB}}\mathcal{L}_{IIGPP}^{AB}\left( I^{AB}, \sigma^{AB}, \phi^{AB} \mid \mathcal{S}_i^{AB}\right)$ [Eq \ref{eq: likelihood_IIGPP}] & $\prod_{i=1}^{N^{AB}} \mathcal{L}^{AB}_{comp}\left(\boldsymbol{\theta} \mid \mathcal{S}_i^{AB} \right)$ [Eq \ref{eq: marginal_likelihood}]\\
\end{tabular}}
\end{table}

WAIC has been shown to be asymptotically equivalent to leave-one-out cross-validation in Bayesian estimation \citep{watanabe2010asymptotic} and is relatively easy to calculate for most statistical models. 
WAIC is calculated by first calculating the log pointwise predictive density (lppd) and then adjusting for the effective number of parameters ($p$). Following \citet{gelman2014understanding}, the computed log pointwise predictive density is defined as
\begin{equation}
    \label{lppd}
    \text{lppd}_{\mathcal{M}} = \sum_{\mathscr{H} \in \{A, B, AB\}}\sum_{i=1}^{N^{\mathscr{H}}}\log \left(\frac{1}{S} \sum_{s = 1}^S \mathcal{L}^{\mathscr{H}}_{\mathcal{M}}\left(\boldsymbol{\theta}^s \mid \mathcal{S}_i^{\mathscr{H}}\right) \right),
\end{equation}
and the computed effective number of parameters is defined as
\begin{equation}
    \label{eff_num_param}
    p_{\mathcal{M}} = \sum_{\mathscr{H} \in \{A, B, AB\}}\sum_{i=1}^{N^{\mathscr{H}}}\text{Var}_{s=1}^S \left(\log \mathcal{L}^{\mathscr{H}}_{\mathcal{M}}\left(\boldsymbol{\theta}^s \mid \mathcal{S}_i^{\mathscr{H}}\right) \right),
\end{equation}
where $\boldsymbol{\theta}^s$ denotes the $s^{th}$ posterior draw of $\boldsymbol{\theta}$,  $\text{Var}_{s=1}^S(x_s) := \frac{1}{S-1}\sum_{s=1}^L(x_s - \bar{x})^2$, and $\mathcal{L}^\mathscr{H}_{\mathcal{M}}$ is the likelihood function under condition $\mathscr{H}$ for the corresponding model $\mathcal{M} \in \{IIGPP, comp\}$, as specified in Table \ref{tab: two_models}. Using these quantities, WAIC is defined as $\text{WAIC}_\mathcal{M} = -2(\text{lppd}_\mathcal{M} - p_{\mathcal{M}})$. Using this definition of WAIC, the preferable model is the one with the smallest WAIC. 

Through comprehensive simulation studies, WAIC was shown to be highly informative in model selection; reliably distinguishing between data generated from the competition model and the IIGPP model. In addition to determining whether a neuron is multiplexing, we are also scientifically interested in determining whether a neuron employs a winner-take-all scheme. Under a winner-take-all encoding scheme, the neuron will encode only the $A$ stimulus or the $B$ stimulus for all trials (i.e. no switching encodings between trials). Fundamentally, a winner-take-all model can be considered a nested model within the class of IIGPP models, with parameters equal to one of the single-stimulus set of parameters (i.e. $\sigma^{AB} = \sigma^{\mathscr{S}}$, $I^{AB} =I^{\mathscr{S}}$, and $\phi^{AB} = \phi^{\mathscr{S}}$ for winner-take-all ($\mathscr{S}$) schemes). Similarly, we consistently identified the generating model through the use of WAIC, when considering this wider class of generating models (competition, IIGPP, and winner-take-all).

While it is important to have good performance in cases where the spike trains are generated from the correct models, it is also crucial to understand how WAIC performs under model misspecification. To gain insight into how WAIC performs under model misspecification, we conducted two simulation studies focusing on two different types of model misspecification. Under our modeling assumptions, we assume that all $AB$ condition trials in a triplet come from either an IIGPP model or a competition model. Thus, in the first simulation study, we studied the performance of WAIC when only a proportion of the $AB$ condition spike trains were generated from the competition model, and the rest were generated from the IIGPP model. The simulations showed that WAIC generally chose the IIGPP model when 30\% or fewer spike trains were generated from the multiplexing model, and generally chose the competition model when 75\% or more were generated from the competition model. Crucially, this simulation study illustrated that WAIC behaved in a predictable manner; largely choosing the model from which most of the spike trains were generated.

In the second simulation study, we looked at how WAIC behaves when spike trains were generated from a simple hidden Markov model (HMM) in which a neuron continued to encode stimulus $\mathscr{S}$ in the next spike with probability $p_s$ and switched to encoding $\mathscr{S}^C$ with probability $1-p_s$. A key assumption in our modeling is that the IIGPP model is flexible enough to represent a suitable class of alternatives. However, the IIGPP model does not assume a fluctuating spiking rate, making it unclear whether WAIC would suggest the competition model for any type of fluctuating spiking pattern. From this simulation study, we found that WAIC reliably chose the IIGPP model, except when $p_s$ was close to 1. When $p_s$ was close to 1, the neuron would likely encode only one stimulus in each trial while switching between trials, which is similar to \textsc{Slow Switching} in our competition model. Overall, this simulation study illustrated that WAIC is sensitive to the state transition probabilities; correctly suggesting the alternative model (IIGPP) when the switching was sufficiently different from the posited transition probabilities under the competition framework. Additional details for all of the simulation studies conducted in this section can be found in Section 3 of the Supplementary Materials.

\section{Simulation Study: Recovery of Scientific Quantities of Interest}
\label{sec: Sim_Study}
To assess whether statistical inference on our state-space model for multiplexing is tractable, we carried out a simulation study in which we studied the inferential accuracy under various sample sizes and the mixing properties of the MCMC. The ability to recover parameters and posterior predictive distributions under a small sample size is paramount in single-electrode recording studies, as the time-intensive recording process often leaves us with triplets that contain only a few samples per condition. 
In addition to recovering the parameters controlling the firing rate, ISI variability, and level of inhibition, we are also interested in accurately recovering posterior predictive distributions of scientific interest. Specifically, we are interested in inferring the posterior predictive distribution of the number of switches, the time spent encoding stimulus $A$, and the number of spikes generated in a trial under the $AB$ condition. We assessed our ability to recover the model parameters and posterior predictive distributions of interest by calculating the relative squared error (RSE) and the relative Wasserstein distance (RWD), respectively, across different quantities of observed spike trains ($N^A = N^B = N^{AB} = 5, 10, 25, 50$). For each sample size, 100 data sets were generated from the competition framework. 

\begin{figure}[t]
    \centering
    \includegraphics[width = \textwidth]{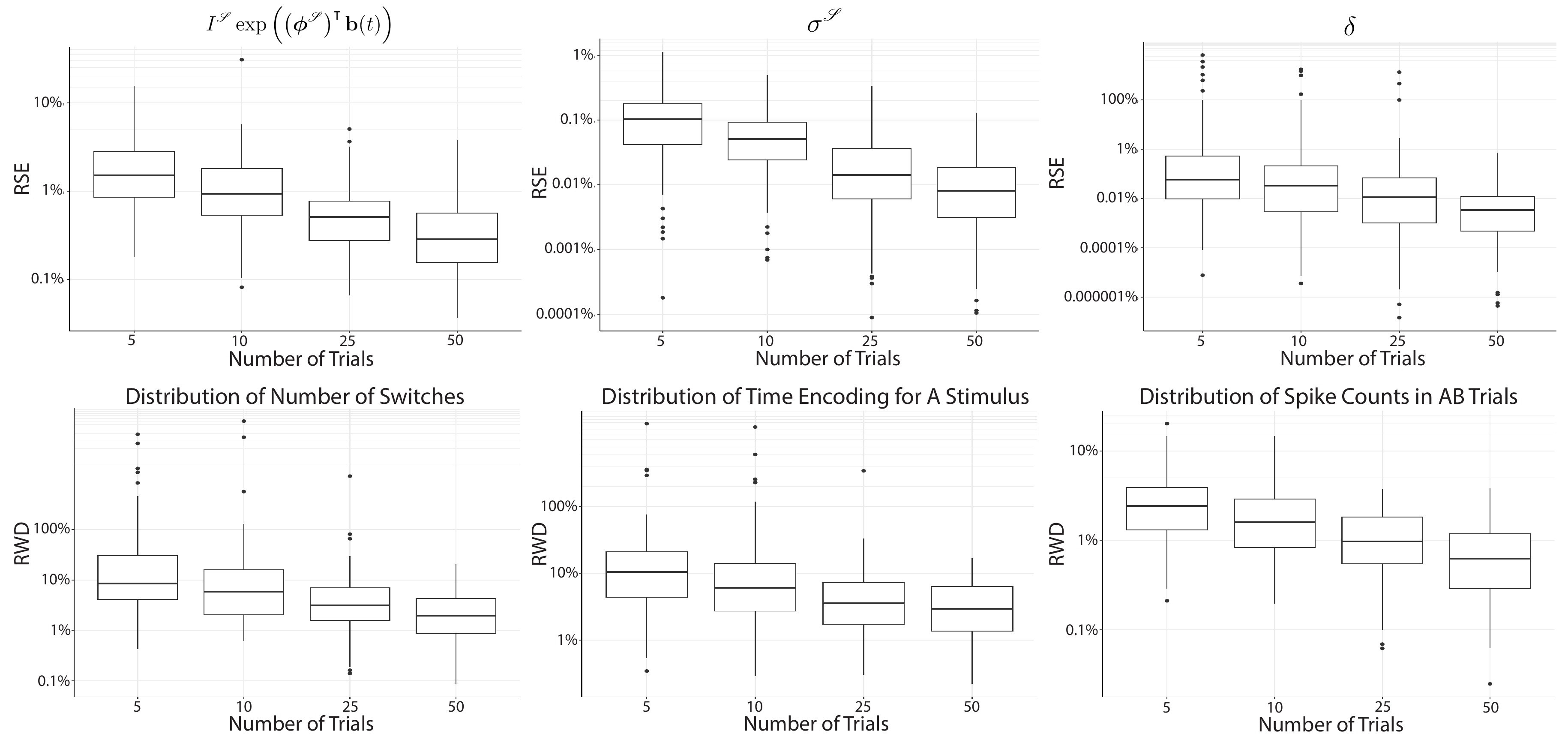}
    \caption{Performance metrics for the recovery of model parameters and posterior predictive distributions of interest, quantified by the relative squared error (RSE) and relative Wasserstein distance (RWD).}
    \label{fig: recovery_sim}
\end{figure}

Unlike typical MCMC methods for state-space models, our proposed sampling strategy was able to avoid getting stuck in local modes and do an effective job of exploring the posterior distribution. Specifically, we found that the frequentist coverage of the credible intervals for our model parameters was close to the nominal 95\%, especially as the sample size increased. As shown in Figure \ref{fig: recovery_sim}, we were able to accurately recover the time-inhomogeneous input current and $\sigma^{\mathscr{S}}$ with as little as five trials per condition. Although the median RSE for $\delta$ was relatively small, we can see that some of our estimates of $\delta$ were far from the truth. This often occurred when $\delta$ was relatively large and a switch did not occur in any of the observed trials, leading to an overestimation of $\delta$ and subsequently a high RSE value. As more trials were added, the probability of not observing a switch in any of the trials decreased, leading to better recovery of $\delta$. 
Detailed information on how the simulation study was conducted, the coverage probabilities, and how RSE and RWD were defined can be found in Section 4 of the Supplementary Materials.

\section{Case Study: Inferior Colliculus Neural Recordings under Auditory Stimuli}
\label{sec: Case_Study}

To gain insight into whether the brain employs multiplexing to encode information from multiple stimuli, we analyzed neural spike trains recorded from the inferior colliculus (IC) of macaque monkeys during sound localization tasks, which was originally collected and analyzed by \citet{caruso2018single}. The IC is an early station along the auditory pathway \citep{adams1979ascending, moore1963ascending} that is believed to encode sound location via the level of neural activity \citep{groh2003monotonic, mcalpine2003sound, grothe2010mechanisms}. Given that individual neurons can only output one level of neural activity at a time, it is not clear how the brain is able to encode and preserve information about multiple auditory stimuli. Considering this, \citet{caruso2018single} focused on understanding how the brain retains information from various auditory stimuli originating from different locations and investigated whether the brain employs multiplexing to preserve information from these distinct stimuli.

The data set collected by \citet{caruso2018single} contains 2225 triplets recorded from two different female macaques, consisting of 166 different neurons and a varying number of trials for each of the three conditions. As in \citet{caruso2018single}, we will require that each condition in the triplet contains at least five successful trials, leaving us with 1231 triplets. We will add two additional requirements for inclusion in the analysis: (1) single-stimulus spike trains can be represented by a time-inhomogeneous inverse Gaussian point process (Equation \ref{eq: likelihood_IIGPP}), and (2) distinguishably different distributions of spike trains for the $A$ and $B$ conditions. The first condition will exclude cases where the single-stimulus spike trains appear to be generated from a mixture of distributions or multiplexing itself, and is carried out by calculating posterior p-values \citep{meng1994posterior, gelman1996posterior}; allowing us to calculate the discrepancy between the observed spike trains and the spike trains generated from the fitted model that assumes a time-inhomogeneous inverse Gaussian point process. The second condition removes cases where the $A$ and $B$ conditions are indistinguishable from each other, and is carried out by comparing the pointwise predictive distribution under a joint model of the $A$ and $B$ conditions with the pointwise predictive distribution under separate models for the $A$ and $B$ conditions. After excluding triplets that do not meet these requirements, we have 571 triplets, which is more than the 159 triplets analyzed using DAPP \citep{glynn2021analyzing} and the 442 triplets analyzed using SCAMPI \citep{chen2024}.


In this analysis, we consider three main encoding scenarios: multiplexing, winner-take-all \citep{chen2017mechanisms}, and alternative encoding strategies represented by the IIGPP model. Similarly to \citet{chen2024}, we will differentiate between two types of multiplexing: \textsc{Slow Juggling} and \textsc{Fast Juggling}. \textsc{Slow Juggling} consists of triplets in which the neuron switches between encodings from trial to trial but rarely switches within a trial, while \textsc{Fast Juggling} consists of triplets in which the neuron often switches between encodings within a trial. 
Using WAIC, we will categorize the triplets into one of the following categories: (1) \textsc{IIGPP}, (2) \textsc{Winner-Take-All (Preferred)} -- denoting winner-take-all scenarios where the neuron under the $AB$ condition encodes the stimulus that elicits the higher overall firing rate between the two stimuli, (3) \textsc{Winner-Take-All (Non-Preferred)} -- denoting winner-take-all scenarios where the neuron under the $AB$ condition encodes the stimulus that elicits the lower overall firing rate between the two stimuli, (4) \textsc{Slow Juggling} -- denoting scenarios where WAIC suggest the competition model and the posterior predictive mean number of switches in an $AB$ trial is less that 0.5, and (5) \textsc{Fast Juggling}  -- denoting scenarios where WAIC suggest the competition model and the posterior predictive mean number of switches in an $AB$ trial is greater than or equal to 0.5. 

\begin{figure}[t]
    \centering
    \includegraphics[width = 0.99\textwidth]{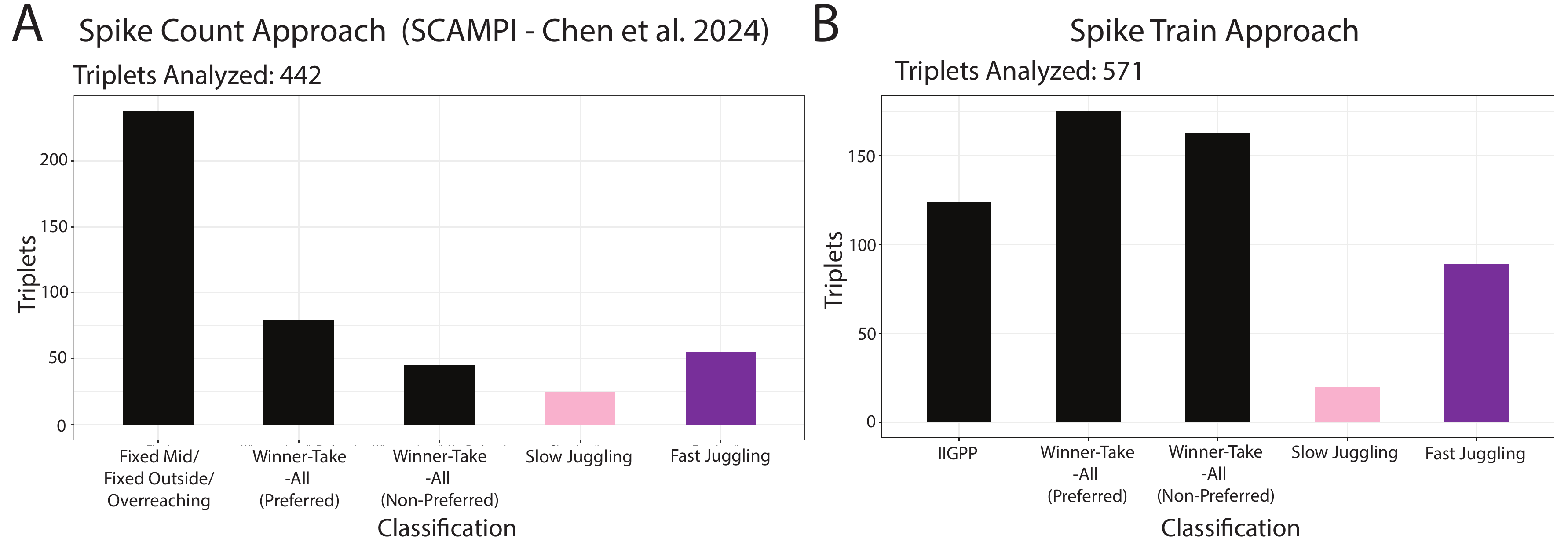}
    \caption{(Subfigure A) Classification results obtained by using SCAMPI \citep{chen2024}. (Subfigure B) Results obtained from the proposed spike train approach.}
    \label{fig: population_inference_caruso}
\end{figure}

\begin{figure}[!ht]
    \centering
    \includegraphics[width = 0.99\textwidth]{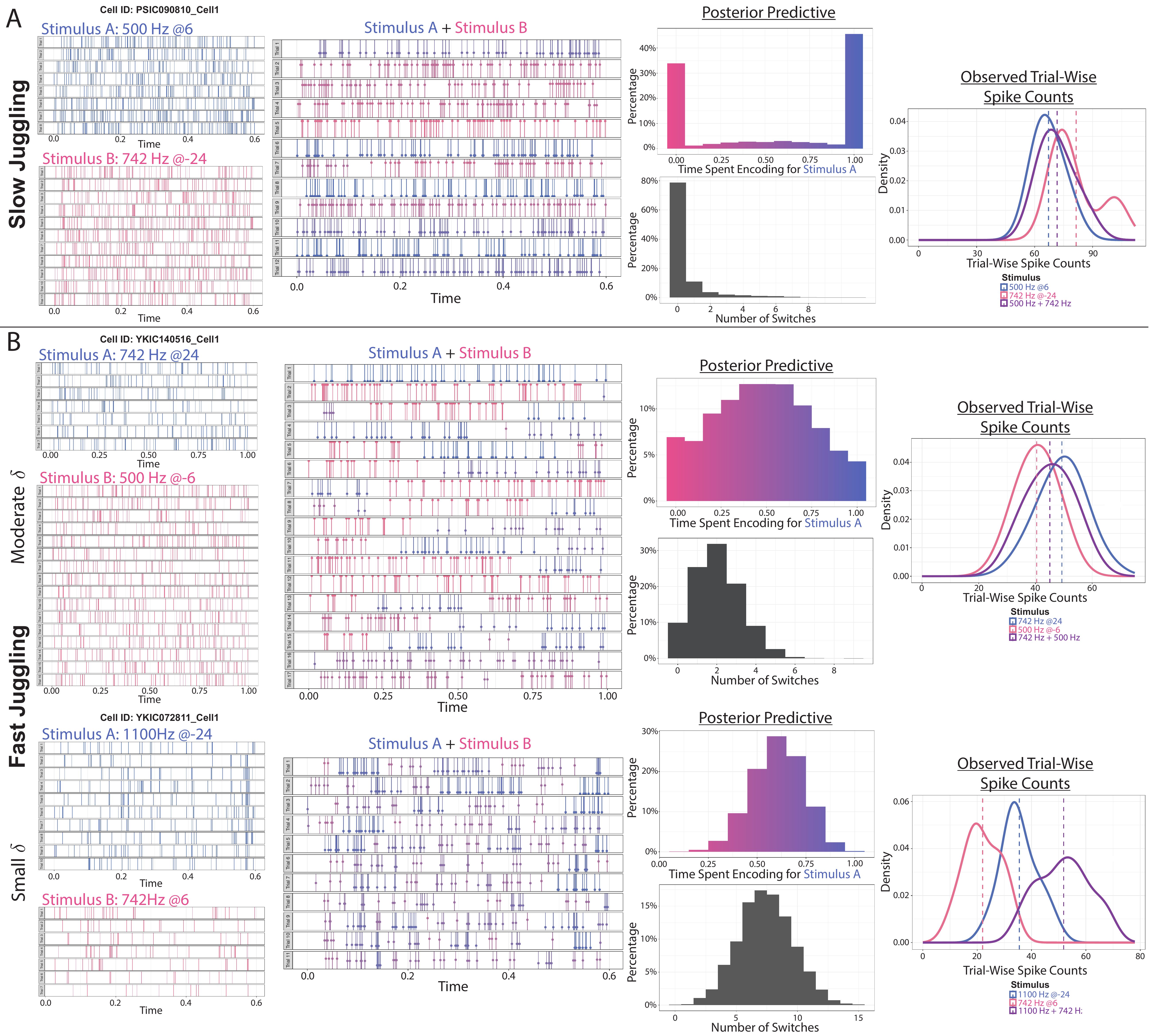}
    \caption{(Subfigure A) Results from a \textsc{Slow Juggling} triplet.  (Subfigure B) Results from two \textsc{Fast Juggling} triplets; one having a moderate $\delta$, and the other having a small $\delta$ leading to an almost additive process. The circles (and color) on the $AB$ condition spikes  denote the posterior probability of encoding each stimulus.}
    \label{fig: individual_ST_inference_caruso}
\end{figure}

As shown in Subfigure B of Figure \ref{fig: population_inference_caruso}, 19.1\% (109) of the triplets exhibited behavior consistent with multiplexing. Among the 109 triplets, 18.3\% (20) of the triplets had neurons that exhibited slow switching behavior and 81.7\% (89) of the triplets had neurons that displayed fast switching behavior. Similarly, the analysis conducted by \cite{chen2024} found that 18.1\% (80) of the analyzed triplets had spike count characteristics consistent with multiplexing; however, a closer examination revealed a notable proportion of triplets in which the two statistical frameworks showed discrepancies. This underscores the importance of a more granular spike train analysis, as time-inhomogeneous firing rates and ISI variability can lead to spike count distributions suggesting incorrect coding schemes. This can be seen in Figure \ref{fig: individual_ST_inference_caruso} where the \textsc{Slow Juggling} triplet and \textsc{Fast Juggling} (moderate $\delta$) triplet have similar spike count distributions, yet notably different spike train behavior. 

Due to the mechanistic nature of our competition model, we are able to gain novel insight into spike-level information that cannot be obtained from spike count analyses or even in the more granular DAPP analysis. This is illustrated in Figure \ref{fig: individual_ST_inference_caruso}, which shows three different triplets in which WAIC suggested the occurrence of multiplexing. Subfigure A illustrates an example of slow switching behavior, where the neuron was unlikely to switch encodings within a trial but frequently switched encodings between trials, as evident in the posterior predictive distributions for the time encoding stimulus $A$ and the number of switches within a trial.
Subfigure B contains a visualization of two triplets in which the neuron displays characteristics of fast switching; one with a moderate $\delta$ and the other with a small $\delta$. In the moderate $\delta$ case, we commonly observe switching between the encodings within the trials; however, we still observe entire trials in which the neuron encodes only one for one stimulus. 
In the small $\delta$ case, we can see an almost additive effect in the spike counts. 
Although we may not necessarily be able to recover which spikes encode which stimulus, WAIC still suggested the competition framework over the IIGPP model. Through simulation studies, WAIC was shown to be very informative in model selection, even if $\delta$ was zero (Section 3.1 of the Supplementary Materials), suggesting that these triplets display multiplexing characteristics despite being unable to decode the stimulus being encoded in individual spikes. In fact, we were unable to recover the spike-level encodings when analyzing data simulated from the competition model where $\delta = 0$ (Section 5.5 of the Supplementary Materials); suggesting that the spike-level encodings cannot be recovered in general when $\delta$ is close to zero.

Overall, this model allows for novel insight into spike-level dynamics and provides compelling evidence of multiplexing. 
In addition, our framework allows us to view some forms of subadditivity \citep{goris2024response} as a special case of multiplexing (small $\delta$); broadening the previous thought that the firing rate of a multiplexing neuron should lie between the single-stimulus firing rates of the neuron \citep{groh2024signal}.  Additional details on this case study, including a detailed description of the experimental setup and inclusion criteria, can be found in Section 5 of the Supplementary Materials.
 
\section{Discussion}
\label{sec: Discussion}

In this manuscript, we specified a mechanistic state-space model for multiplexing -- a neural encoding theory that posits that individual neurons can temporally switch between encoding different stimuli over time \citep{groh2024signal}. The state-space model differs from most state-space models in that it is non-Markovian and endogenous, requiring novel MCMC sampling schemes to achieve efficient posterior inference. To determine whether the data offered adequate evidence toward multiplexing, we conducted model comparison using WAIC. 
Using the proposed statistical framework, we analyzed spike train data from the inferior colliculus (IC); providing novel insight into the temporal dynamics of multiplexing and the potential mechanisms by which multiplexing is regulated. Crucially, the probabilistic models developed in this manuscript can be extended to models for a population of neurons, thereby providing insight into how neurons coordinate to encode stimuli. 

Although our model is relatively flexible compared to typical statistical models for spike trains, we do not assume any second-order stochastic variation. Specifically, we assume that the model parameters do not vary over the trials in the same triplet. Despite DAPP's fairly restrictive assumptions and non-mechanistic nature -- leading to difficulties in determining whether a neuron is multiplexing --  the primary purpose of DAPP is to capture second-order stochasticity. Specifically, \citet{glynn2021analyzing} suggested that there is a non-negligible amount of second-order stochastic variation in the IC data set. By extending the proposed framework to allow for second-order stochasiticity, we would potentially be able to analyze more triplets and gain insight into the extent of second-order stochastic variation present in neural spike trains.

As discussed in Section \ref{sec: posterior_inference}, the non-Markovian and endogenous nature of the proposed state-space model for multiplexing led to poor sampling performance when using typical MCMC schemes for state-space models. Specifically, we observed that typical MCMC schemes were sensitive to the initial state of the Markov chain and often got stuck in local modes. Therefore, we constructed an efficient algorithm (Algorithm \ref{alg2}) to sample from the conditional posterior distribution of the labels and $\delta$. Although this algorithm was constructed for the proposed model, this sampling scheme is generalizable state-space models with similar dependence structures; allowing for variables that affect the transition probabilities to be sampled with the dependence of the states marginalized out. In addition, Algorithm \ref{alg2} can be parallelizable on multiple levels. By exploiting the independence of the spike trains, the forward filtering and backwards sampling steps for different spike trains can be computed in parallel, thereby allowing the marginal likelihood to be computed through parallel operations. Furthermore, calculating the marginal likelihoods of $\hat{\boldsymbol{\theta}}_1,\dots, \hat{\boldsymbol{\theta}}_{M_\delta}$ can be computed in parallel, allowing the acceptance probabilities to be calculated in parallel. These properties make this algorithm particularly suitable for applications where the evaluation of the likelihood function may be computationally expensive. 

As shown through extensive simulation studies, we found that WAIC was highly informative in model selection for these types of models. However, when considering using WAIC for point process data using state-space models, one must consider whether to (1) calculate WAIC using the posterior predictive distributions of the individual spikes ($x_{ij}$) or the posterior predictive distributions of the spike trains ($\mathcal{S}_i$) and whether to (2) use the marginal likelihood, where the states are marginalized out, or use the conditional likelihood. Here, computing WAIC based on the posterior predictive distributions of individual spikes is similar to performing leave-one-spike-out cross-validation, whereas calculating WAIC from the posterior predictive distributions of spike trains is akin to leave-one-spike-train-out cross-validation. A key assumption in \cite{watanabe2010asymptotic} is that the observations are independent with respect to the true distribution, which generally does not coincide with the posited statistical model and does not depend on the model parameters. For general time-inhomogeneous point processes, the individual spikes in a spike train are not independent; indicating that WAIC should be calculated using the posterior predictive distributions of the spike trains ($\mathcal{S}_i$) to ensure that the theoretical properties of WAIC are preserved. 

When considering state-space models, a key consideration is whether the states should be marginalized out when calculating WAIC. 
For state-space models, we found that using WAIC calculated with the states marginalized out was more informative; differentiating between models that had different structural transition schemes.  Specifically, conditional WAIC did not sufficiently penalize data that make improbable switches between states under a specified state-space model. Since we specified a mechanistic state-space model where the transition probabilities are controlled by competition between the two stimuli, it is crucial in this application that the model selection takes this into account. Consequently, we employed the marginal likelihood to compute WAIC, and advocate for its preference over the conditional likelihood in scenarios where transition probabilities have a structural/mechanistic nature. 

Although we found evidence of multiplexing in the IC of two macaque monkeys, further analyses of spike trains recorded across various stimuli, different brain regions, and various species are necessary to determine when the brain uses multiplexing and whether multiplexing facilitates more efficient neural computation and representation. The R package \textsc{NeuralComp} is available on GitHub to fit the proposed models and perform the discussed Bayesian analysis of neural spike train data.

\section{Supplementary Materials}

\begin{description}
    \item[Supplementary Materials:] Contains proofs for all lemmas in the manuscript, algorithms for calculating WAIC, supplemental simulation studies, as well as additional details for the case studies and simulation studies presented in the main manuscript.
    \item[\textsc{NeuralComp}:] The R package to fit the IIGPP and competition models can be found on GitHub at \url{https://github.com/ndmarco/NeuralComp}.
    \item[Simulation Studies:] The R scripts for the simulation studies can be found on GitHub at \\
    \url{https://github.com/ndmarco/NeuralComp_Sim_Study}.
    \item[Case Study:] The R scripts for the case study can be found on GitHub at \\
    \url{https://github.com/ndmarco/NeuralComp_Case_Study}.
\end{description}
\vspace{-1em}
\section{Disclosure}
The authors report that there are no competing interests to declare.

\bibliographystyle{abbrvnat}
\bibliography{Neural_Switching}

\end{document}


\def\spacingset#1{\renewcommand{\baselinestretch}%
{#1}\small\normalsize} \spacingset{1}


\if1\blind
{
  \title{\bf Supplementary Materials for ``Modeling Neural Switching via Drift-Diffusion Models''}
  \author{Nicholas Marco \thanks{
    The authors gratefully acknowledge from NIH awards R01 DC013096 and R01 DC016363.}\hspace{.2cm}\\
    Department of Statistical Science, Duke University\\
    Jennifer M. Groh \\
    Department of Neurobiology, Department of Psychology \& Neuroscience,\\ Department of Biomedical Engineering, Department of Computer Science,\\ Duke University\\
    and\\
    Surya T. Tokdar\\
    Department of Statistical Science, Duke University}
  \maketitle
} \fi

\if0\blind
{
  \bigskip
  \bigskip
  \bigskip
  \begin{center}
    {\LARGE\bf Supplementary Materials for ``Modeling Neural Switching via Drift-Diffusion Models''}
\end{center}
  \medskip
} \fi

\bigskip
\begin{abstract}
Supplementary materials for the manuscript ``Modeling Neural Switching via Drift-Diffusion Models''. Section \ref{sec: Proofs} Contains the proof to Lemma 4.1 in the main manuscript, showing that the constructed Markov chain satisfies detailed balance with respect to the posterior distribution. Section \ref{appendix: MCMC} provides a comparison between our proposed method of sampling the labels and $\delta$ together compared to a simpler sampling scheme, and outlines details on how the MCMC is tuned. Section \ref{sec: WAIC_appendix} outlines how to calculate WAIC for the competition model and contains simulation studies demonstrating WAIC's effectiveness in model selection. Section \ref{Sec: Sim_convergence} contains information on how the simulation study in Section 5 of the main manuscript was conducted and information on coverage probability and credible interval widths. Lastly, Section \ref{sec: CS_appendix} contains additional information on the case study using data collected from the inferior colliculus.
\end{abstract}

\vfill

\newpage
\spacingset{1.9}

\section{Proofs}
\label{sec: Proofs}
\subsection{Lemma 4.1}
We would like to show that the Markov chain $\left(\{(\delta^{t}, \{\mathbf{l}_i^{t}\}_{i = 1}^{N^{AB}}) : t \in \mathbb{N}\}\right)$ generated from Algorithm 2 in the manuscript satisfies detailed balance with respect to $f(\delta, \{\mathbf{l}_i\}_{i = 1}^{N^{AB}} \mid \{\mathcal{S}^{AB}_i\}_{i =1}^{N^{AB}}, \boldsymbol{\theta}_{-\delta}, \alpha_\delta, \beta_\delta)$. Thus, we would like to show that
\begin{align}
    \nonumber f\left(\delta^t, \{\mathbf{l}_i^t\}_{i = 1}^{N^{AB}} \mid \{\mathcal{S}^{AB}_i\}_{i =1}^{N^{AB}}, \boldsymbol{\theta}_{-\delta}, \alpha_\delta, \beta_\delta\right)T\left((\delta^t, \{\mathbf{l}_i^t\}_{i = 1}^{N^{AB}}) \longrightarrow (\delta^{t+1}, \{\mathbf{l}_i^{t+1}\}_{i = 1}^{N^{AB}}) \right) = \\
    \nonumber f\left(\delta^{t+1}, \{\mathbf{l}_i^{t+1}\}_{i = 1}^{N^{AB}} \mid \{\mathcal{S}^{AB}_i\}_{i =1}^{N^{AB}}, \boldsymbol{\theta}_{-\delta}, \alpha_\delta, \beta_\delta\right)T\left((\delta^{t + 1}, \{\mathbf{l}_i^{t+1}\}_{i = 1}^{N^{AB}}) \longrightarrow (\delta^{t}, \{\mathbf{l}_i^{t}\}_{i = 1}^{N^{AB}}) \right),
\end{align}
where $T\left((\delta^t, \{\mathbf{l}_i^t\}_{i = 1}^{N^{AB}}) \longrightarrow (\delta^{t+1}, \{\mathbf{l}_i^{t+1}\}_{i = 1}^{N^{AB}}) \right)$ denotes the probability density of moving from $(\delta^t, \{\mathbf{l}_i^t\}_{i = 1}^{N^{AB}})$ to $(\delta^{t+1}, \{\mathbf{l}_i^{t+1}\}_{i = 1}^{N^{AB}})$.

Suppose that $\delta^{t} \ne \delta^{t+1}$. Then we have that
{\scriptsize
\begin{align}
    \nonumber & f\left(\delta^t, \{\mathbf{l}_i^t\}_{i = 1}^{N^{AB}} \mid \{\mathcal{S}^{AB}_i\}_{i =1}^{N^{AB}}, \boldsymbol{\theta}_{-\delta}, \alpha_\delta, \beta_\delta\right)T\left((\delta^t, \{\mathbf{l}_i^t\}_{i = 1}^{N^{AB}}) \longrightarrow (\delta^{t+1}, \{\mathbf{l}_i^{t+1}\}_{i = 1}^{N^{AB}}) \right)\\
    = \nonumber & f(\delta^t, \{\mathbf{l}_i^t\}_{i = 1}^{N^{AB}} \mid \{\mathcal{S}^{AB}_i\}_{i =1}^{N^{AB}}, \boldsymbol{\theta}_{-\delta}, \alpha_\delta, \beta_\delta) \left[M_\delta q_\delta(\delta^{t+1})\int \dots \int \frac{w_{\delta}(\delta^{t+1})}{w_{\delta}(\delta^{t+1}) + w_{\delta}(\delta^{t}) + \sum_{m=2}^{M_\delta} w_\delta(\hat{\delta}_m)} \prod_{m = 2}^{M_\delta} q_\delta(\hat{\delta}_m) \text{d}\hat{\delta}_m\right]\\ 
    \nonumber & \times T\left(\{\mathbf{l}_i^{t}\}_{i = 1}^{N^{AB}} \longrightarrow  \{\mathbf{l}_i^{t+1}\}_{i = 1}^{N^{AB}} \mid \delta^{t+1}\right),
\end{align}}
where the fact that $\hat{\delta}_m$ for $m = 1, \dots, M_\delta$ are exchangeable is used. Using the independence between sampling labels from different trials, we have 
$$T\left(\{\mathbf{l}_i^{t}\}_{i = 1}^{N^{AB}} \longrightarrow  \{\mathbf{l}_i^{t+1}\}_{i = 1}^{N^{AB}} \mid \delta^{t+1}\right) = \prod_{i=1}^{N^{AB}} T\left(\mathbf{l}_i^{t} \longrightarrow  \mathbf{l}_i^{t+1} \mid \delta^{t+1}\right).$$
Noticing that the transitions are independent of the previous label values, we have
\begin{align}
    \nonumber T\left(\mathbf{l}_i^{t} \longrightarrow  \mathbf{l}_i^{t+1} \mid \delta^{t+1}\right)  = &  P\left(L_{in_i^{AB}} = l_{in_i^{AB}}^{t+1}\mid \mathcal{S}_i^{AB}, \boldsymbol{\theta}_{-\delta}, \delta^{t+1}\right) \\
    \nonumber & P\left(L_{i(n_i^{AB}-1)} = l_{i(n_i^{AB}-1)}^{t+1}\mid \mathcal{S}_i^{AB}, l_{in_i^{AB}}^{t+1}, \boldsymbol{\theta}_{-\delta}, \delta^{t+1}\right) \\
    \nonumber& \times \prod_{j=1}^{n_i^{AB} -2} P\left(L_{ij} = l_{ij}^{t+1} \mid \{x^{AB}_{ik}\}_{k=1}^{j+1}, \{l^{t+1}_{ik}\}_{k=j+1}^{n_i^{AB}}, \boldsymbol{\theta}_{-\delta}, \delta^{t+1} \right)\\
    \nonumber = & P\left(\mathbf{L}_i = \mathbf{l}_i^{t+1} \mid \mathcal{S}_i^{AB}, \boldsymbol{\theta}_{-\delta}, \delta^{t+1}\right),
\end{align}
where the last equality can be seen since {\footnotesize$P\left(L_{ij} = l_{ij}^{t+1} \mid \{x^{AB}_{ik}\}_{k=1}^{j+1}, \{l^{t+1}_{ik}\}_{k=j+1}^{n_i^{AB}}, \boldsymbol{\theta}_{-\delta}, \delta^{t+1} \right) = P\left(L_{ij} = l_{ij}^{t+1} \mid \mathcal{S}_i^{AB}, \{l^{t+1}_{ik}\}_{k=j+1}^{n_i^{AB}}, \boldsymbol{\theta}_{-\delta}, \delta^{t+1} \right)$} for $ j = 1, \dots, n_i^{AB} -2$, due to conditional independence.
Thus we have that
$$T\left(\{\mathbf{l}_i^{t}\}_{i = 1}^{N^{AB}} \longrightarrow  \{\mathbf{l}_i^{t+1}\}_{i = 1}^{N^{AB}} \mid \delta^{t+1}\right) = \prod_{i=1}^{N^{AB}}P\left(\mathbf{L}_{i} = \mathbf{l}_{i}^{t+1}\mid \mathcal{S}^{AB}_i, \boldsymbol{\theta}_{-\delta}, \delta^{t + 1}\right).$$
Thus, using the definition of $w_\delta(\delta)$, we have
{\scriptsize
\begin{align}
    & f\left(\delta^t, \{\mathbf{l}_i^t\}_{i = 1}^{N^{AB}} \mid \{\mathcal{S}^{AB}_i\}_{i =1}^{N^{AB}}, \boldsymbol{\theta}_{-\delta}, \alpha_\delta, \beta_\delta\right)T\left((\delta^t, \{\mathbf{l}_i^t\}_{i = 1}^{N^{AB}}) \longrightarrow (\delta^{t+1}, \{\mathbf{l}_i^{t+1}\}_{i = 1}^{N^{AB}}) \right)\\
    =  & \left[M_\delta f\left(\delta^{t+1} \mid \{\mathcal{S}^{AB}_i\}_{i =1}^{N^{AB}}, \boldsymbol{\theta}_{-\delta}, \alpha_\delta, \beta_\delta \right)\int \dots \int \frac{f\left(\{\mathcal{S}^{AB}_i\}_{i =1}^{N^{AB}} \mid \boldsymbol{\theta}_{-\delta}, \alpha_\delta, \beta_\delta\right)}{w_{\delta}(\delta^{t+1}) + w_{\delta}(\delta^{t}) + \sum_{m=2}^{M_\delta} w_\delta(\hat{\delta}_m)} \prod_{m = 2}^{M_\delta} q_\delta(\hat{\delta}_m) \text{d}\hat{\delta}_m\right]\\ 
    \nonumber & \times f\left(\delta^t, \{\mathbf{l}_i^t\}_{i = 1}^{N^{AB}} \mid \{\mathcal{S}^{AB}_i\}_{i =1}^{N^{AB}}, \boldsymbol{\theta}_{-\delta}, \alpha_\delta, \beta_\delta\right) \left[\prod_{i=1}^{N^{AB}}P\left(\mathbf{L}_{i} = \mathbf{l}_{i}^{t+1}\mid \mathcal{S}^{AB}_i, \boldsymbol{\theta}_{-\delta}, \delta^{t + 1}\right)\right]\\
     \label{eq: symmetry_dif_delta} =  & \left[M_\delta \int \dots \int \frac{f\left(\{\mathbf{x}^{AB}_i\}_{i =1}^{N^{AB}} \mid \boldsymbol{\theta}_{-\delta}, \alpha_\delta, \beta_\delta\right)}{w_{\delta}(\delta^{t+1}) + w_{\delta}(\delta^{t}) + \sum_{m=2}^{M_\delta} w_\delta(\hat{\delta}_m)} \prod_{m = 2}^{M_\delta} q_\delta(\hat{\delta}_m) \text{d}\hat{\delta}_m\right]\\ 
    \nonumber & \times f\left(\delta^t, \{\mathbf{l}_i^t\}_{i = 1}^{N^{AB}} \mid \{\mathcal{S}^{AB}_i\}_{i =1}^{N^{AB}}, \boldsymbol{\theta}_{-\delta}, \alpha_\delta, \beta_\delta\right) f\left(\delta^{t+1}, \{\mathbf{l}_i^{t+1}\}_{i = 1}^{N^{AB}} \mid \{\mathcal{S}^{AB}_i\}_{i =1}^{N^{AB}}, \boldsymbol{\theta}_{-\delta}, \alpha_\delta, \beta_\delta\right)\\
    =  & f\left(\delta^{t+1}, \{\mathbf{l}_i^{t+1}\}_{i = 1}^{N^{AB}} \mid \{\mathcal{S}^{AB}_i\}_{i =1}^{N^{AB}}, \boldsymbol{\theta}_{-\delta}, \alpha_\delta, \beta_\delta\right)T\left((\delta^{t+1}, \{\mathbf{l}_i^{t+1}\}_{i = 1}^{N^{AB}}) \longrightarrow (\delta^{t}, \{\mathbf{l}_i^{t}\}_{i = 1}^{N^{AB}}) \right),
\end{align}}
where the last equality can be seen by the symmetry of $\left(\delta^{t}, \{\mathbf{l}_i^{t}\}_{i = 1}^{N^{AB}}\right)$ and $\left(\delta^{t+1}, \{\mathbf{l}_i^{t+1}\}_{i = 1}^{N^{AB}}\right)$ in Equation \ref{eq: symmetry_dif_delta}. 

If $\delta^t = \delta^{t+1}$, then we have
{\scriptsize
\begin{align}
    \nonumber & f\left(\delta^t, \{\mathbf{l}_i^t\}_{i = 1}^{N^{AB}} \mid \{\mathcal{S}^{AB}_i\}_{i =1}^{N^{AB}}, \boldsymbol{\theta}_{-\delta}, \alpha_\delta, \beta_\delta\right)T\left((\delta^t, \{\mathbf{l}_i^t\}_{i = 1}^{N^{AB}}) \longrightarrow (\delta^{t+1}, \{\mathbf{l}_i^{t+1}\}_{i = 1}^{N^{AB}}) \right)\\
    \nonumber = &  f\left(\delta^t, \{\mathbf{l}_i^t\}_{i = 1}^{N^{AB}} \mid \{\mathcal{S}^{AB}_i\}_{i =1}^{N^{AB}}, \boldsymbol{\theta}_{-\delta}, \alpha_\delta, \beta_\delta\right) T\left(\delta^t \longrightarrow \delta^{t+1}\right) T\left(\{\mathbf{l}_i^{t}\}_{i = 1}^{N^{AB}} \longrightarrow  \{\mathbf{l}_i^{t+1}\}_{i = 1}^{N^{AB}} \mid \delta^{t+1}\right) \\
    \nonumber = &  f\left(\delta^t, \{\mathbf{l}_i^t\}_{i = 1}^{N^{AB}} \mid \{\mathcal{S}^{AB}_i\}_{i =1}^{N^{AB}}, \boldsymbol{\theta}_{-\delta}, \alpha_\delta, \beta_\delta\right) T\left(\delta^t \longrightarrow \delta^{t+1}\right) \left[\prod_{i=1}^{N^{AB}}P\left(\mathbf{L}_{i} = \mathbf{l}_{i}^{t+1}\mid \mathcal{S}^{AB}_i, \boldsymbol{\theta}_{-\delta}, \delta^{t + 1}\right) \right] \\
    \nonumber = &  f\left(\delta^t, \{\mathbf{l}_i^t\}_{i = 1}^{N^{AB}} \mid \{\mathcal{S}^{AB}_i\}_{i =1}^{N^{AB}}, \boldsymbol{\theta}_{-\delta}, \alpha_\delta, \beta_\delta\right) T\left(\delta^t \longrightarrow \delta^{t+1}\right) \left[\frac{f\left(\delta^{t+1}, \{\mathbf{l}_i^{t+1}\}_{i = 1}^{N^{AB}} \mid \{\mathcal{S}^{AB}_i\}_{i =1}^{N^{AB}}, \boldsymbol{\theta}_{-\delta}, \alpha_\delta, \beta_\delta\right)}{f\left(\delta^{t+1}\mid \{\mathcal{S}^{AB}_i\}_{i =1}^{N^{AB}}, \boldsymbol{\theta}_{-\delta}, \alpha_\delta, \beta_\delta\right)} \right] \\
    \nonumber = & f\left(\delta^{t+1}, \{\mathbf{l}_i^{t+1}\}_{i = 1}^{N^{AB}} \mid \{\mathcal{S}^{AB}_i\}_{i =1}^{N^{AB}}, \boldsymbol{\theta}_{-\delta}, \alpha_\delta, \beta_\delta\right)T\left((\delta^{t+1}, \{\mathbf{l}_i^{t+1}\}_{i = 1}^{N^{AB}}) \longrightarrow (\delta^{t}, \{\mathbf{l}_i^{t}\}_{i = 1}^{N^{AB}})\right),
\end{align}}
where the last equality can be seen since $\delta^{t} = \delta^{t+1}$. Thus, we can see that the Markov chain generated by Algorithm 2 satisfies detailed balance with respect to $f(\delta, \{\mathbf{l}_i\}_{i = 1}^{N^{AB}} \mid \{\mathcal{S}^{AB}_i\}_{i =1}^{N^{AB}}, \boldsymbol{\theta}_{-\delta}, \alpha_\delta, \beta_\delta)$.

\section{MCMC for Posterior Inference}
\label{appendix: MCMC}
In this section, we will compare our MCMC algorithm with more standard MCMC algorithms to exemplify the advantages of our proposed MCMC scheme. In addition, we will discuss some of the adaptation schemes used to tune the MCMC algorithms. 

\subsection{Detailed Algorithms}

\begin{alg}
\label{alg1_appendix}
Let $P(L_{ij} = \mathscr{S} \mid \{x^{AB}_{ik}\}_{k=1}^j, \boldsymbol{\theta})$ and $P(L_{in_{i}^{AB}} =  \mathscr{S} \mid \mathcal{S}_i^{AB}, \boldsymbol{\theta})$ be calculated in the forward filtering step (Equation 10in the main manuscript) for $i = 1,\dots, N^{AB}$ and $j = 1, \dots, n_i^{AB}-1$. For $i = 1, \dots, N^{AB}$ repeat the following:
\begin{enumerate}
    \item Set $\tilde{l}_{in_i^{AB}} = \mathscr{S}$ with probability $P(L_{in_i^{AB}} = \mathscr{S} \mid \mathcal{S}_i^{AB}, \boldsymbol{\theta})$ for $\mathscr{S} \in \{A, B\}$,
    \label{BS1}
    \item For $ j = n_i^{AB} - 1, \dots, 1$, set $\tilde{l}_{ij} = \mathscr{S}$ ($\mathscr{S} \in \{A, B\}$) with probability $w^\mathscr{S}_{ij}$ , where
    \begin{footnotesize}
        \begin{align}
        \nonumber w^\mathscr{S}_{ij} =& \frac{P(L_{ij} = \mathscr{S} \mid \{x^{AB}_{ik}\}_{k=1}^j, \boldsymbol{\theta})f^{AB}_{j+1}\left(x_{i(j+1)}^{AB}, \tilde{l}_{i(j+1)} \mid L_{ij} = \mathscr{S}, \boldsymbol{\theta}, s^{AB}_{i(j-1)}\right)}{\sum_{\mathscr{S'} = A, B} P(L_{ij} = \mathscr{S}' \mid \{x^{AB}_{ik}\}_{k=1}^j, \boldsymbol{\theta})f^{AB}_{j+1}\left(x_{i(j+1)}^{AB}, \tilde{l}_{i(j+1)} \mid L_{ij} = \mathscr{S}', \boldsymbol{\theta}, s^{AB}_{i(j-1)}\right)}\\
        \nonumber = & P(L_{ij} = \mathscr{S} \mid \{x^{AB}_{ik}\}_{k=1}^{j+1}, \{\tilde{l}_{ik}\}_{k=j+1}^{n_i^{AB}}, \boldsymbol{\theta}).
    \end{align}
    \end{footnotesize}
    \label{BS_Appendix}
\end{enumerate}
\end{alg}

\begin{alg}
\label{alg2_appendix}
Given the data $\{\mathbf{x}_i^{AB}\}_{i =1}^{N^{AB}}$ and parameters $\boldsymbol{\theta}_{-\delta}$, generate a Markov chain $\left(\left\{\left(\delta^{s}, \{\mathbf{l}_i^{s}\}_{i = 1}^{N^{AB}}\right) : s \in \mathbb{N}\right\}\right)$ as follows:
    \begin{enumerate}
    \item Start with some $\delta^0$ and $\{\mathbf{l}_i^{0}\}_{i = 1}^{N^{AB}}$.
    \label{joint sampler 1}
    \item For $s = 1, 2, \dots$ repeat the following:
    \label{joint sampler 2}
        \begin{enumerate}
            \item Set $\hat{\delta}_0 = \delta^{s-1}$ and randomly generate $\hat{\delta}_1, \dots, \hat{\delta}_{M_\delta}$ from $q_\delta(\delta)$ ($M_\delta \ge 1$).
            \item Letting $\hat{\boldsymbol{\theta}}_m = \{I^A, I^B, \sigma^A, \sigma^B, \hat{\delta}_m, \boldsymbol{\phi}^A, \boldsymbol{\phi}^B\}$, calculate $P(L_{ij} = \mathscr{S} \mid \{x^{AB}_{ik}\}_{k=1}^j, \hat{\boldsymbol{\theta}}_m)$ and $P(L_{in_{i}^{AB}} =  \mathscr{S} \mid \mathcal{S}_i, \hat{\boldsymbol{\theta}}_m)$ using the forward filtering step (Equation 10 in the main manuscript) for $m = 0, \dots, M_\delta$, $j = 1, \dots, n_i^{AB}-1$, and $i = 1, \dots, N^{AB}$. Calculate the marginal likelihood
            \begin{footnotesize}
                \begin{equation}
                \label{eq: marginal_likelihood}
                \begin{split}
                \mathcal{L}_{Comp}^{AB}\left(\hat{\boldsymbol{\theta}}_m \mid \mathcal{S}_i^{AB}\right) = f\left(x_{i1}^{AB} \mid \hat{\boldsymbol{\theta}}_m \right)\left(\prod_{j=1}^{n_i^{AB} - 1} f\left( x_{ij}^{AB} \mid \{x_{ik}^{AB}\}_{k=1}^{j-1}, \hat{\boldsymbol{\theta}}_m\right)\right)\\
                     \times f\left( x^{AB}_{in^{AB}_i}, X^{AB}_{i(n_i^{AB} +1)} > T - s_{in_{i}^{AB}}^{AB} \mid \{x_{ik}^{AB}\}_{k=1}^{n_i^{AB}-1}, \hat{\boldsymbol{\theta}}_m \right)
                \end{split}
            \end{equation}
            \end{footnotesize}
            
             which can be obtained via the normalizing constants obtained in the forward filtration step, as discussed in the Section \ref{sec: marginal} of the Supplementary Materials.
            \item Set $\delta^s = \hat{\delta}_m$ with probability $w_\delta(\hat{\delta}_m)$ ($m = 0, \dots, M$), where 
            \begin{align}
                \nonumber w_\delta(\hat{\delta}_m) & \propto \frac{\prod_{i=1}^{N^{AB}} \mathcal{L}_{Comp}^{AB}\left(\hat{\boldsymbol{\theta}}_m \mid \mathcal{S}_i^{AB}\right)\pi_\delta(\hat{\delta}_m\mid \alpha_\delta, \beta_\delta)}{q_\delta(\hat{\delta}_m)}\\ 
                \nonumber & \propto \frac{f\left(\hat{\delta}_m\mid \left\{\mathcal{S}_i^{AB}\right\}_{i=1}^{N^{AB}}, \boldsymbol{\theta}_{-\delta}, \alpha_\delta, \beta_\delta\right)}{q_\delta(\hat{\delta}_m)},
            \end{align}
            where $\pi_\delta(\delta \mid \alpha_\delta, \beta_\delta)$ is the prior probability distribution function of $\delta$.
            \item Let $\tilde{m}$ denote the chosen index such that $\delta^s = \hat{\delta}_{\tilde{m}}$. Given $\delta^s$, $P(L_{ij} = \mathscr{S} \mid \{x^{AB}_{ik}\}_{k=1}^j, \hat{\boldsymbol{\theta}}_{\tilde{m}})$, and  $P(L_{in_{i}^{AB}} =  \mathscr{S} \mid \mathcal{S}_i^{AB}, \hat{\boldsymbol{\theta}}_{\tilde{m}})$  for $j = 1,
            \dots, n_i^{AB}-1$ and $i = 1, \dots, N^{AB}$, obtain $\{\mathbf{l}_i^s\}_{i=1}^{N^{AB}}$ through the backward sampling step (Algorithm \ref{alg1_appendix}).
        \end{enumerate}
\end{enumerate}
\end{alg}

\subsection{Comparison of MCMC Schemes}
\label{sec: Comparison_MCMC}
To efficiently generate samples from the posterior distribution of the proposed model, we proposed using Algorithm \ref{alg2_appendix} (Algorithm 2 in the main manuscript) which jointly samples the labels and $\delta$ parameter from the conditional posterior distribution. In this section, we compare the sampling method proposed in Algorithm \ref{alg2_appendix} to a simpler sampling algorithm where the $\delta$ parameter and the labels are sampled separately according to their respective conditional posterior distributions.

\begin{figure}[htbp]
    \centering
    \includegraphics[width = 0.99\textwidth]{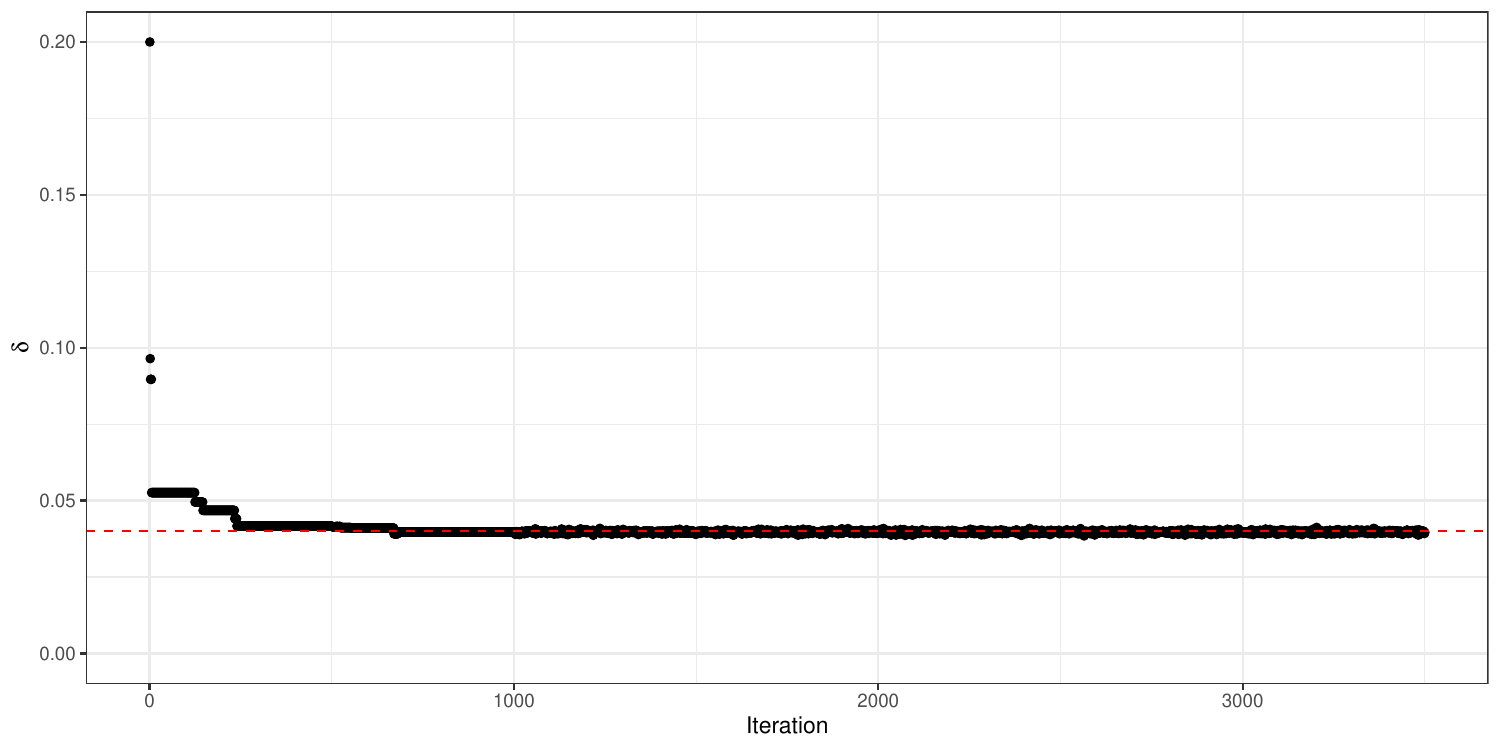}
    \caption{Trace plots of $\delta$ using the first sampling scheme (Algorithm \ref{alg2_appendix}). The dotted red line represents the value of $\delta$ used to generate to data.}
    \label{fig: trace_joint}
\end{figure}

In this section, the parameters $I^A, I^B, \sigma^A,$ and $\sigma^B$ will be sampled in one block using HMC, and the parameters $\phi^A$ and $\phi^B$ will be sampled in an additional block using HMC. The differences in the two sampling schemes compared in this section will be in how we sample from the conditional posterior distribution of the label and the $\delta$ parameter. In the first sampling scheme, we will use the sampling scheme in Algorithm \ref{alg2_appendix} to jointly sample the labels and the $\delta$ parameter. In the second sampling scheme, we will use Algorithm 1 to sample the labels from the conditional posterior distribution $f(\mathbf{l}_i\mid I^A, I^B, \sigma^A, \sigma^B, \delta, \phi^A, \phi^B, \mathcal{S}^{AB}_{i})$, and we will use random walk metropolis to generate samples from the conditional posterior of $\delta$, $f(\delta \mid I^A, I^B, \sigma^A, \sigma^B, \delta, \phi^A, \phi^B, \{\mathcal{S}_i^{AB}\}_{i=1}^{N^{AB}}, \{\mathbf{l}_i\}_{i=1}^{N^{AB}})$.

\begin{figure}
    \centering
    \includegraphics[width = 0.99\textwidth]{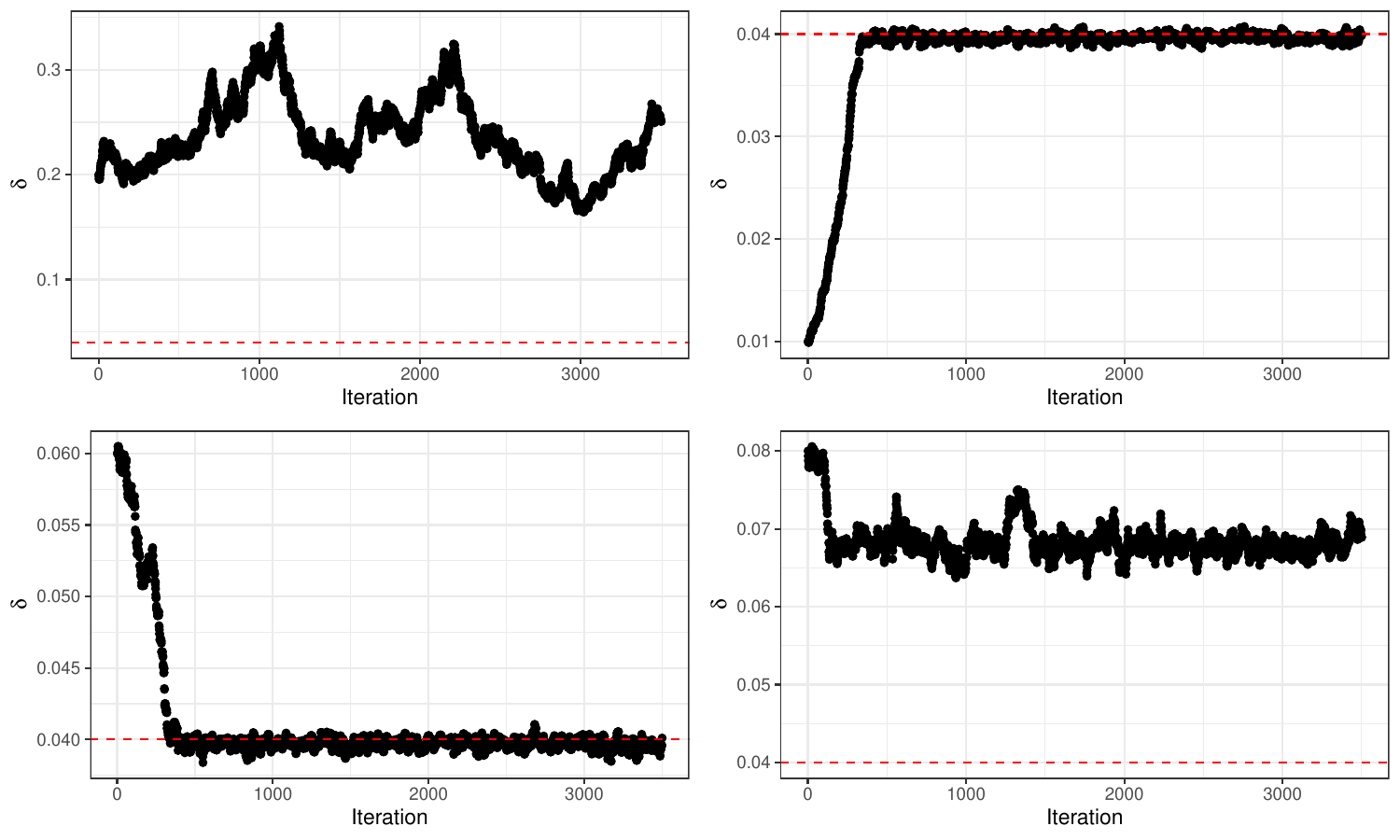}
    \caption{Trace plots of $\delta$ using the second sampling scheme ($\delta$ and the labels are sampled separately). The dotted red line represents the value of $\delta$ used to generate to data.}
    \label{fig: trace_seperate}
\end{figure}

\begin{figure}[htbp]
    \centering
    \includegraphics[width = 0.99\textwidth]{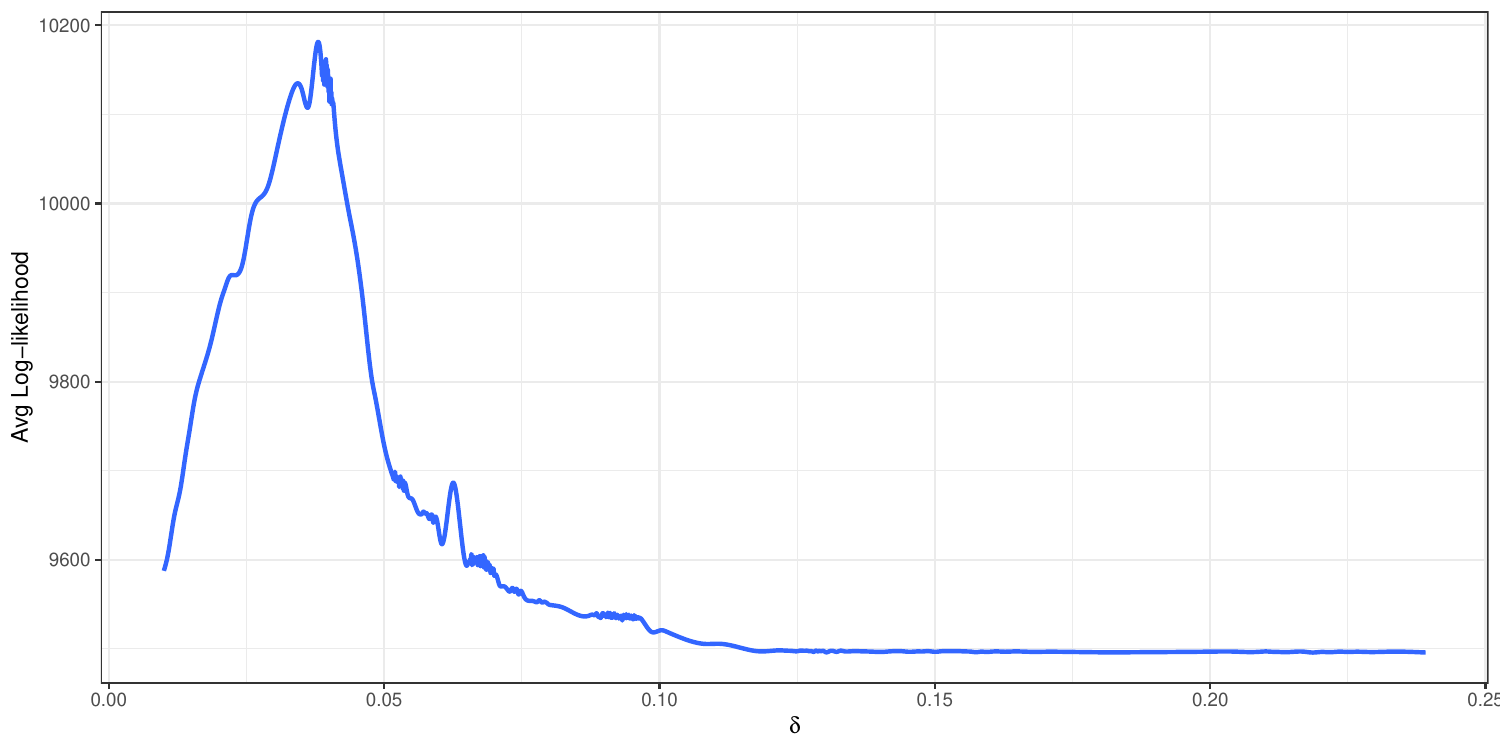}
    \caption{Plot of the log profile likelihood as a function of $\delta$.}
    \label{fig: llik_seperate}
\end{figure}

Figure \ref{fig: trace_joint} contains a trace plot of $\delta$ using the first sampling scheme (Algorithm \ref{alg2_appendix}). From this figure, we can see that the chain quickly converges to the conditional posterior distribution of $\delta$. Alternatively, from Figure \ref{fig: trace_seperate}, we can see that the convergence of the Markov chain depends greatly on the starting value of $\delta$. We can see that if we pick a value too large, the Markov chain does not converge in the 3500 iterations conducted. Looking at the bottom right panel of Figure \ref{fig: trace_seperate}, we can see that the Markov chain appears to have converged; however, looking at Figure \ref{fig: llik_seperate}, we can see that the Markov chain got stuck in a local mode and was unable to get out of the local mode. 
We note that adjusting the step size in the random walk did not affect the sampler's ability to leave the local mode, as the labels were not jointly sampled with $\delta$. Thus, we can see that Algorithm \ref{alg2_appendix} provides an efficient sampler that converges relatively fast, regardless of the starting position of the Markov chain.

\subsection{MCMC Implementation Details}
As discussed in the main manuscript, we use a combination of Hamiltonian Monte Carlo (HMC) and our proposed MCMC algorithm to conduct posterior inference. The samplers, implemented in RCPP, use automatic differentiation to obtain the gradients of the potential energy. To ensure stability while performing HMC, we jitter both the step size $\epsilon$, and the number of leapfrog steps $L$ \citep{mackenze1989improved, neal2011mcmc}. Specifically, we uniformly draw $\epsilon_s$ from the interval $[0.9\epsilon, 1.1\epsilon]$ ($\epsilon$ is adaptively learned) and uniformly draw $L_s$ from the set $\{l\in \mathbb{N}\mid 1 \le l \le 2L\}$, where $L$ is user-defined.

\begin{figure}[htbp]
    \centering
    \includegraphics[width = 0.99\textwidth]{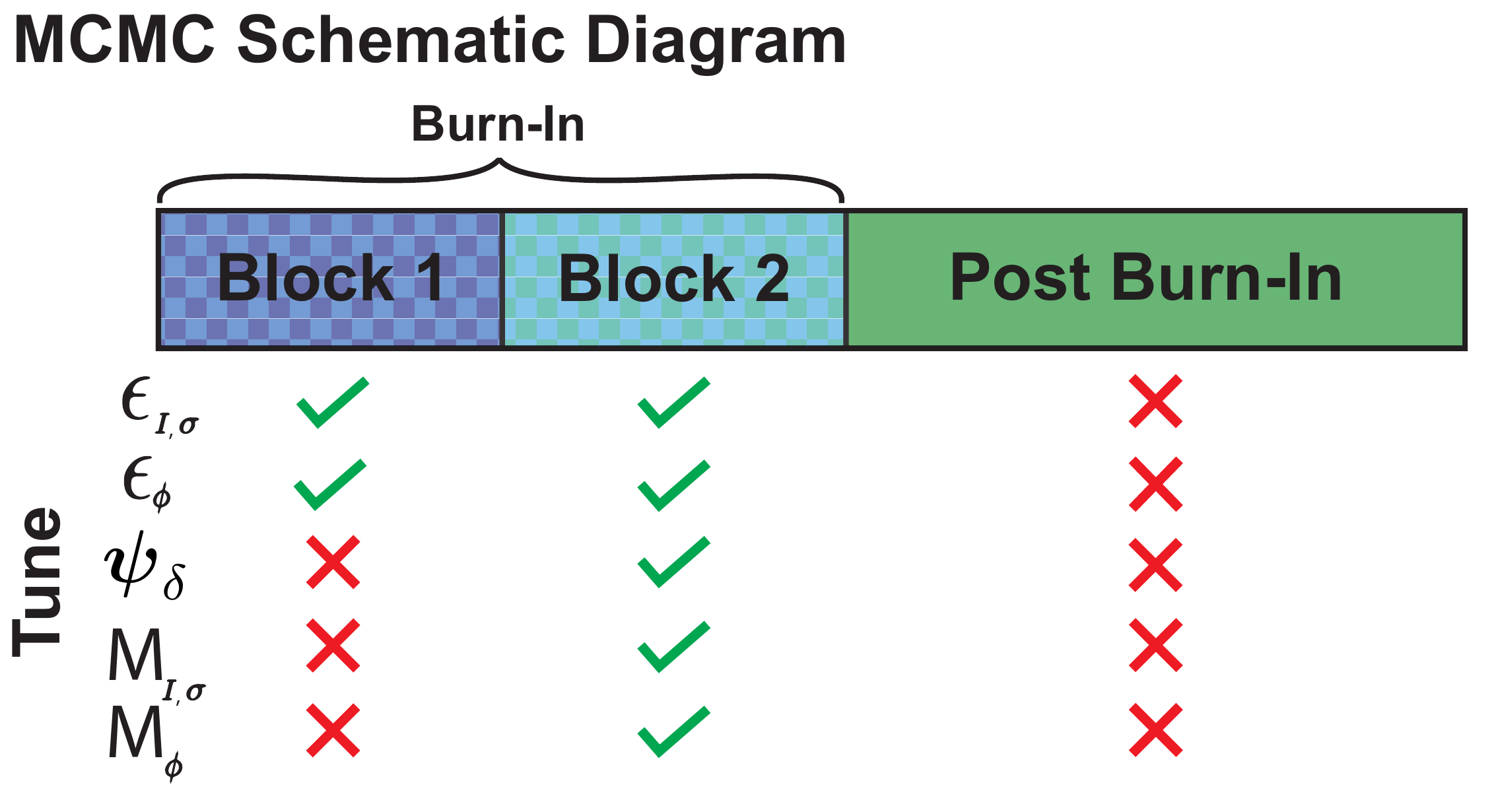}
    \caption{Schematic diagram of which parameters are adapted in each block of the MCMC algorithm.}
    \label{fig: MCMC_Scheme}
\end{figure}

To achieve efficient sampling from the posterior distribution, it is crucial to tune the MCMC algorithm by choosing the optimal parameters (i.e., mass matrices, step sizes, and proposal distribution for sampling $\delta$). The MCMC scheme can be broken down into three main blocks, as illustrated in Figure \ref{fig: MCMC_Scheme}. The first block starts by adapting the step sizes $\epsilon_{I, \sigma}$ and $\epsilon_\phi$, using mass matrices equal to the identity. In this block, the chain should ideally converge to the stationary distribution and (perhaps inefficiently) start sampling from the posterior distribution. The step sizes throughout this algorithm are adapted based on the acceptance rates of the HMC samplers. As we start the second block, we use our MCMC samples to start adaptation of the mass matrices, $M_{I, \sigma}$ and $M_\phi$. Specifically, we set the mass matrices to the sample precision matrix of the parameters using the MCMC samples. Before inverting the sample covariance, a small diagonal matrix ($0.001\mathbf{I}$) is added to the sample covariance for stability. The Mass matrix will be periodically updated throughout the second block, using the previous batch of MCMC samples. As discussed in the main manuscript, $\psi_\delta$ is chosen so that the first two moments of the proposal distribution match the first two moments of the posterior distribution of $\delta$. Thus, $\psi_\delta$ is periodically updated using the previous MCMC samples of $\delta$. After completion of the first two blocks, we should have an efficient sampler obtained by using the MCMC samples and acceptance rates to tune all of the hyperparameters. After the first two blocks, no further adaptation occurs, ensuring an ergodic Markov chain.

\section{Estimation of WAIC for the Competition Model}
\label{sec: WAIC_appendix}

To perform model comparison, we use the widely applicable information criterion (WAIC) \citep{watanabe2010asymptotic, watanabe2013widely} to compare the proposed competition model with a simpler inverse Gaussian process model. Following \citet{gelman2014understanding}, we have that
\begin{equation}
    \label{lppd_supplement}
    \text{lppd}_{\mathcal{M}} = \sum_{\mathscr{H} \in \{A, B, AB\}}\sum_{i=1}^{N^{\mathscr{H}}}\log \left(\frac{1}{S} \sum_{s = 1}^S \mathcal{L}^{\mathscr{H}}_{\mathcal{M}}\left(\boldsymbol{\theta}^s \mid \mathcal{S}_i^{\mathscr{H}}\right) \right),
\end{equation}
and the computed effective number of parameters is defined as
\begin{equation}
     \label{eff_num_param_supplement}
    p_{\mathcal{M}} = \sum_{\mathscr{H} \in \{A, B, AB\}}\sum_{i=1}^{N^{\mathscr{H}}}\text{Var}_{s=1}^S \left(\log \mathcal{L}^{\mathscr{H}}_{\mathcal{M}}\left(\boldsymbol{\theta}^s \mid \mathcal{S}_i^{\mathscr{H}}\right) \right),
\end{equation}
where $\boldsymbol{\theta}$ denotes the model parameters and $\boldsymbol{\theta}^s$ denotes the $s^{th}$ posterior draw, and $\text{Var}_{s=1}^S(x_s) = \frac{1}{S-1}\sum_{s=1}^L(x_s - \bar{x})^2$. From this we can calculate the WAIC estimate, which is defined as
\begin{equation}
    \label{WAIC_supplement}
    \text{WAIC}_\mathcal{M} = \text{lppd}_{\mathcal{M}} - p_{\mathcal{M}}.
\end{equation}

Since $f_{X^{AB}}\left(x_{ij}^{AB}\mid \boldsymbol{\theta}^{s}\right)$ is cannot be written in analytic form, approximation methods must be used. In this section, we outline four different methods for calculating WAIC.

\subsection{Marginal WAIC (Competition Model)}
\label{sec: marginal}
A marginal version of WAIC can be calculated by using the marginal likelihood, marginalizing out the labels ($L_{ij}$). As shown in \citet{watanabe2010asymptotic}, WAIC is asymptotically equivalent to Bayesian leave-one-out cross-validation. In this case, the marginal WAIC proposed in this section is similar to leave-one-trial-out cross-validation. The marginal WAIC can be calculated using the normalization constants of Equation 10 in the main manuscript (forward filtration step). Specifically, we have that 
\begin{align}
    \nonumber f(x_{ij}^{AB} \mid \{x_{ik}^{AB}\}_{k=1}^{j-1}, \boldsymbol{\theta}) = & \sum_{\mathscr{S} \in \{A, B\}}\sum_{\mathscr{S}' \in \{A, B\}}  \bigg[P\left(L_{i(j-1)} = \mathscr{S}'\mid \{x^{AB}_{ik}\}_{k=1}^{j-1}, \boldsymbol{\theta}\right) \\
    \nonumber & \times f^{AB}_{j}\left(x_{ij}^{AB}, L_{ij} = \mathscr{S}\mid L_{i(j-1)} = \mathscr{S}', \boldsymbol{\theta}, s_{i(j-1)}^{AB}\right)\bigg],
\end{align}
for $ j = 2, \dots, n_i^{AB} - 1$. Similarly, for the last spike, we have
\begin{footnotesize}
    \begin{equation}
    \begin{split}
    \nonumber
    f\left(x_{\tilde{n}_i}^{AB}, X_{\tilde{n}_i + 1} > T - s_{\tilde{n}_i}^{AB}\mid \{x^{AB}_{ik}\}_{k=1}^{n_{i}^{AB}-1}, \boldsymbol{\theta}\right)  = \\     \sum_{\mathscr{S} \in \{A, B\}}\Bigg\{\bigg[\sum_{\mathscr{S}' = \{ A, B\}}  P\left(L_{i(n_{i}^{AB}-1)} = \mathscr{S}'\mid \{x^{AB}_{ik}\}_{k=1}^{n_{i}^{AB}-1}, \boldsymbol{\theta}\right) \\
      \times f^{AB}_{\tilde{n}_i}\left(x_{\tilde{n}_i}^{AB},  L_{\tilde{n}_i} = \mathscr{S}\mid L_{\tilde{n}_i-1} = \mathscr{S}', \boldsymbol{\theta}, s^{AB}_{\tilde{n}_i - 1}\right)\bigg] \\
      \times \left[1 - F^{\mathscr{S}^C}_{\tilde{n}_i +1}\left(T - s_{\tilde{n}_i}^{AB} - \delta \mid \boldsymbol{\theta}, s_{\tilde{n}_i}^{AB}\right) \right] \\
     \times \left[1 - F^{\mathscr{S}}_{\tilde{n}_i + 1}\left(T - s_{\tilde{n}_i}^{AB} \mid \boldsymbol{\theta}, s_{\tilde{n}_i}^{AB}\right) \right]\Bigg\},
\end{split}
\end{equation}
\end{footnotesize}
From this we have 
\begin{footnotesize}
    \begin{equation}
    \label{eq: marginal_likelihood_supplement}
    \begin{split}
    \mathcal{L}_{Comp}^{AB}\left(\boldsymbol{\theta} \mid \mathcal{S}_i^{AB}\right) = f\left(x_{i1}^{AB} \mid \boldsymbol{\theta} \right)\left(\prod_{j=1}^{n_i^{AB} - 1} f\left( x_{ij}^{AB} \mid \{x_{ik}^{AB}\}_{k=1}^{j-1}, \boldsymbol{\theta}\right)\right)\\
         \times f\left( x^{AB}_{\tilde{n}_i}, X^{AB}_{\tilde{n}_i + 1} > T - s_{\tilde{n}_i}^{AB} \mid \{x_{ik}^{AB}\}_{k=1}^{n_i^{AB}-1}, \boldsymbol{\theta} \right).
    \end{split}
\end{equation}
\end{footnotesize}
Thus, we can calculate the marginal WAIC using Algorithm \hyperref[alg 3]{3}.

\begin{alg3}
\label{alg 3}
Given the observed data $\{\mathbf{x}_i^{A}\}_{i=1}^{N^A}, \{\mathbf{x}_i^{B}\}_{i=1}^{N^B}, \{\mathbf{x}_i^{AB}\}_{i=1}^{N^{AB}}$, and posterior samples $\boldsymbol{\theta}^{s}$ and $\mathbf{l}_i^s$ for $i = 1, \dots, N^{AB}$ and $s = 1, \dots, S$, the marginal WAIC can be obtained as follows:
\begin{enumerate}
    \item Calculate $\mathcal{L}^{\mathscr{S}}(\boldsymbol{\theta}^s \mid \mathcal{S}_i^{\mathscr{S}} )$ using Equation 9 in the main manuscript for $\mathscr{S} = A, B$, $s = 1, \dots, S$ and $i = 1, \dots, N^{\mathscr{S}}$.
    \item For $s = 1, \dots, S$ and $i = 1, \dots, N^{AB}$, repeat the following: 
    \begin{enumerate}
        \item Calculate $$f_{X^{AB}}(x_{i1}^{AB} \mid \boldsymbol{\theta}^s) = \sum_{\mathscr{S} \in \{A,B\}}f_{X^{AB},L}\left(x_{i1}^{AB}, L_{i1} = \mathscr{S}\mid \boldsymbol{\theta}^s\right).$$
        \item Run the forward filtration step, specified in Equation 10 of the main manuscript, to obtain $f_{X^{AB}}(x_{ij}^{AB} \mid \{x_{ik}^{AB}\}_{k=1}^{j-1}, \boldsymbol{\theta}^s)$ ($j = 2, \dots, n_i^{AB}-1$) and \\
        $f_{X^{AB}}\left(x_{\tilde{n}_i}^{AB}, X_{\tilde{n}_i + 1} > T - s_{\tilde{n}_i}^{AB}\mid \{x^{AB}_{ik}\}_{k=1}^{n_{i}^{AB}-1}, \boldsymbol{\theta}\right)$.
        \item Calculate $\mathcal{L}_{Comp}^{AB}\left(\boldsymbol{\theta}^s \mid \mathcal{S}_i^{AB}\right)$ using Equation \ref{eq: marginal_likelihood_supplement}.
    \end{enumerate}
    \item Calculate the marginal log pointwise predictive density by
    \begin{align}
        \nonumber \text{lppd}_{comp} = & \sum_{\mathscr{S} \in \{A, B\}}\sum_{i=1}^{N^{\mathscr{S}}}\log \left(\frac{1}{S} \sum_{s = 1}^S\mathcal{L}^{\mathscr{S}}\left(\boldsymbol{\theta}^s \mid \mathcal{S}_i^{\mathscr{S}}\right) \right) \\
        \nonumber & + \sum_{i=1}^{N^{AB}}\log \left(\frac{1}{S} \sum_{s = 1}^S \mathcal{L}_{Comp}^{AB}\left(\boldsymbol{\theta}^s \mid \mathcal{S}_i^{AB}\right)\right).
    \end{align}
    \item Calculate the effective number of parameters by
    \begin{align}
        \nonumber p_{comp} = &  \sum_{\mathscr{S} \in \{A, B\}}\sum_{i=1}^{N^{\mathscr{S}}}\text{Var}_{s=1}^S\left(\log \mathcal{L}^{\mathscr{S}}\left(\boldsymbol{\theta}^s \mid \mathcal{S}_i^{\mathscr{S}}\right) \right) \\
        \nonumber & + \sum_{i=1}^{N^{AB}} \text{Var}_{s=1}^S\left(\log \mathcal{L}_{Comp}^{AB}\left(\boldsymbol{\theta}^s \mid \mathcal{S}_i^{AB}\right)\right).
    \end{align}
    \item Calculate the marginal WAIC by $\text{WAIC}_{comp} = \text{lppd}_{comp} - p_{comp}$.
\end{enumerate}

\end{alg3}

\subsection{WAIC Simulation Study 1: Effectiveness of WAIC in Distinguishing Between the Competition Model and IIGPP Model}
\label{sec: marginal_vs_conditional}
In this simulation study, we study how informative WAIC is in model selection. To evaluate the performance, we will generate data from both the competition model and the IIGPP model and see if WAIC can recover the correct generating model. $\text{WAIC}_{IIGPP}$ can be calculated using Equation \ref{WAIC_supplement}, as all quantities are easily computable. $\text{WAIC}_{comp}$ can be calculated using Algorithm \hyperref[alg 3]{3}, which requires calculating the normalizing constants obtained from the forward filtration step of FFBS.  In this simulation study, we generated 100 data sets from the competition model, as well as 100 data sets where the spike trains are generated from the IIGPP model, each with 25 spike trains per condition ($N^A = N^B = N^{AB} = 25$).

The 100 data sets generated from the proposed model were created by randomly sampling the model parameters in the following way:
\begin{align}
    \nonumber I^A & \sim \mathcal{N}^+(40, 16), \\
    \nonumber I^B & \sim \mathcal{N}^+(80, 16), \\
    \nonumber \sigma^A & \sim \mathcal \mathcal{N}^+(\sqrt{40}, 4),\\
    \nonumber \sigma^B & \sim \mathcal \mathcal{N}^+(\sqrt{80}, 4), \\
    \nonumber \delta & \sim \text{LogNormal}(-2.5, 0.25), \text{(for 80\% of the data sets)}\\
    \nonumber \delta & = 0, \text{(for 20\% of the data sets)}\\
    \nonumber \boldsymbol{\phi}^A & \sim \mathcal{N}_6(\mathbf{0}, 0.09\mathbf{I}_6) ,\\
    \nonumber \boldsymbol{\phi}^B & \sim \mathcal{N}_6(\mathbf{0}, 0.09\mathbf{I}_6),
\end{align}
where $\mathcal{N}^+$ denotes a truncated normal distribution with support on $[0, \infty)$. In this simulation study, we considered $\mathcal{T} = [0,1]$ and used B splines to capture the inhomogeneity of the firing rates over time. Specifically, we used B-splines of degree 3, with boundary knots $(0,1)$ and internal knots $(0.25, 0.5, 0.75)$. The 100 data sets generated from the three inverse Gaussian processes were created by randomly sampling the model parameters in the following way:
\begin{align}
    \nonumber I^A & \sim \mathcal{N}^+(40, 16), \\
    \nonumber I^B & \sim \mathcal{N}^+(80, 16), \\
    \nonumber I^{AB} & \sim \mathcal{N}^+(80, 64), \\
    \nonumber \sigma^A & \sim \mathcal \mathcal{N}^+(\sqrt{40}, 4),\\
    \nonumber \sigma^B & \sim \mathcal \mathcal{N}^+(\sqrt{80}, 4), \\
    \nonumber \sigma^{AB} & \sim \mathcal \mathcal{N}^+(\sqrt{60}, 16), \\
    \nonumber \boldsymbol{\phi}^A & \sim \mathcal{N}_6(\mathbf{0}, 0.09\mathbf{I}_6),\\
    \nonumber \boldsymbol{\phi}^B & \sim \mathcal{N}_6(\mathbf{0}, 0.09\mathbf{I}_6),\\
    \nonumber \boldsymbol{\phi}^{AB} & \sim \mathcal{N}_6(\mathbf{0}, 0.09\mathbf{I}_6).
\end{align}

\begin{figure}
    \centering
    \includegraphics[width = 0.99\textwidth]{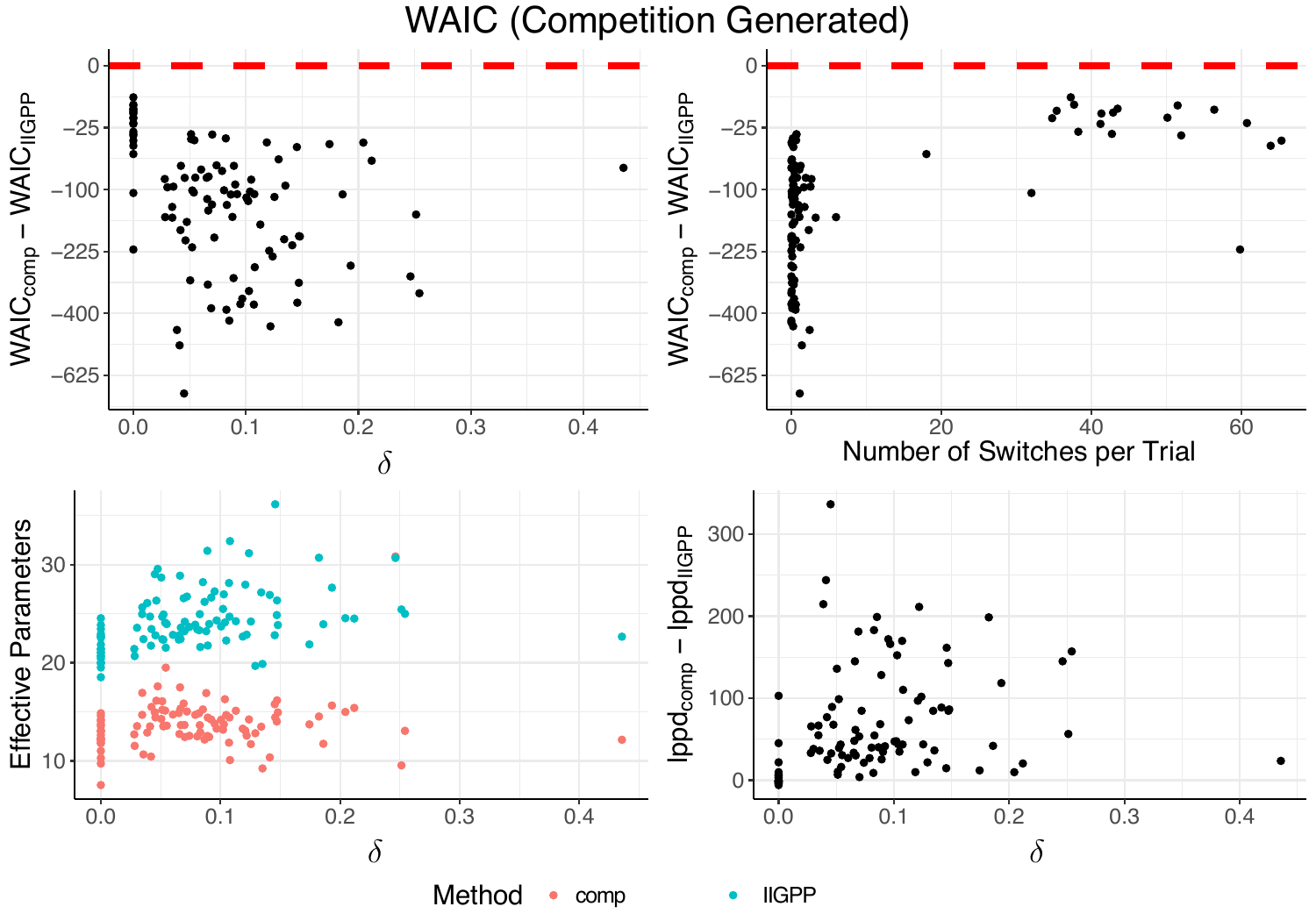}
    \caption{Visualization of the performance of $WAIC_{comp}$ and $WAIC_{IIGPP}$ from data generated from the proposed competition model. The top-left panel visualizes the WAIC estimates as a function of the delta used to generate the data. The top-right panel visualizes the WAIC estimates as a function of the average true number of switches from the $A$ process to the $B$ process (or vice versa) in an $AB$ condition spike train. The two bottom panels visualize the effective number of parameters and lppd as a function of the true delta used to generate the data.}
    \label{fig: sim_WAIC_comp}
\end{figure}

\begin{figure}
    \centering
    \includegraphics[width = 0.99\textwidth]{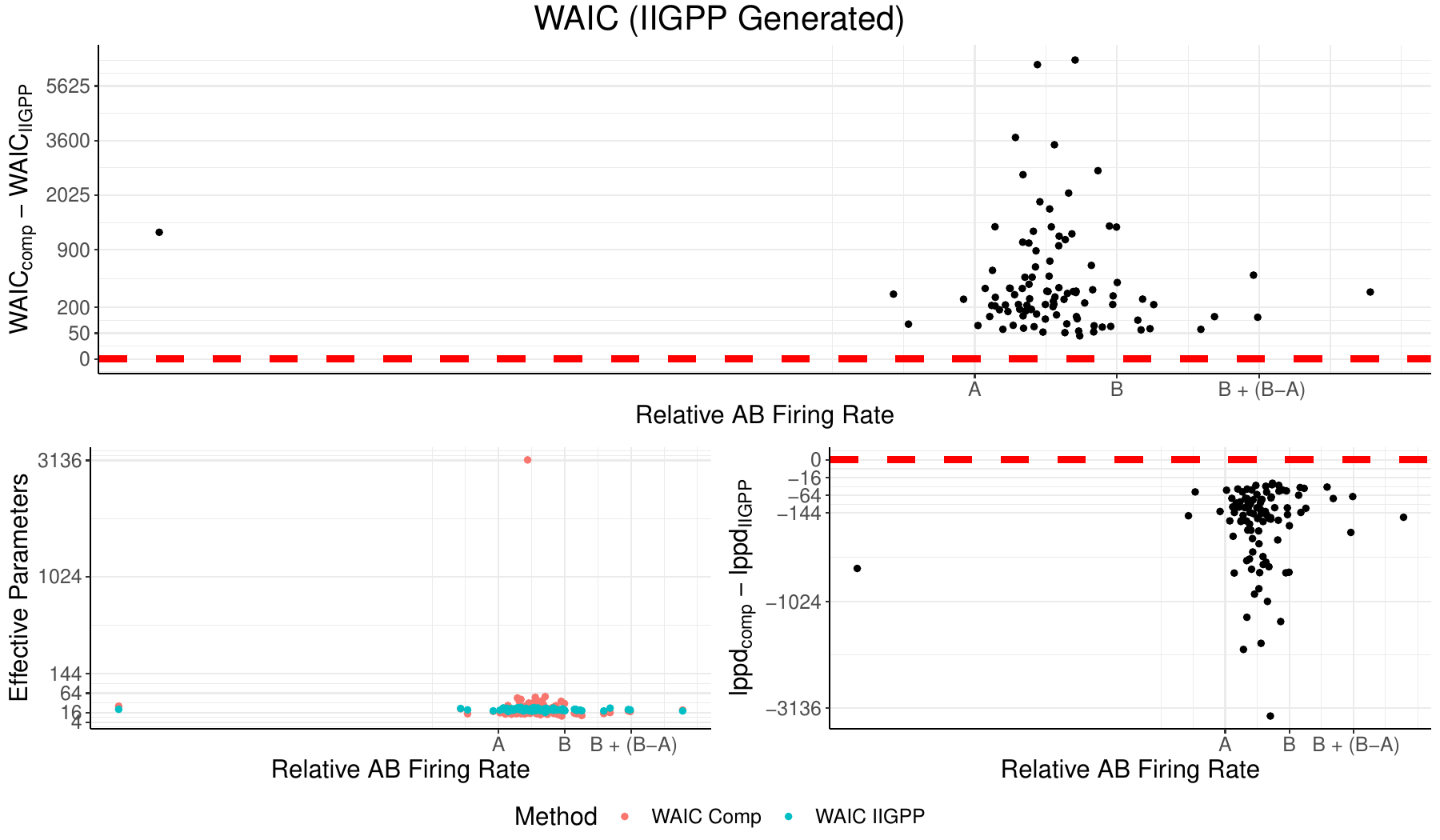}
    \caption{Visualization of the performance of $WAIC_{comp}$ and $WAIC_{IIGPP}$ from data generated from the IIGPP model. The top panel illustrates the difference between $WAIC_{comp}$ and $WAIC_{IIGPP}$, with positive values suggesting the IIGPP model over the competition model. The bottom panels show the effective number of parameters and the difference in the log pointwise predictive distributions under the two models.}
    \label{fig: sim_WAIC_IIGPP}
\end{figure}

From Figures \ref{fig: sim_WAIC_comp} and \ref{fig: sim_WAIC_IIGPP}, we can see that WAIC is highly informative in distinguishing between data generated from the competition model and data generated from the IIGPP model. In particular, WAIC suggested the competition model for all 100 occasions when the datasets were generated from the competition framework, and suggested the IIGPP model for all 100 instances when the datasets were generated from the IIGPP framework.  This simulation study also shows that when $\delta = 0$, WAIC can be informative in determining whether the $AB$-stimulus spike trains were generated from the competition framework or IIGPP framework. We note that if we assumed a Poisson process, the ISIs would follow an exponential distribution. Thus, the competition process would assume that the ISIs of the $AB$-stimulus spike trains, which would be the minimum of two random variables that are exponentially distributed, also follow an exponential distribution. Thus, the resulting $AB$ condition spike trains could be modeled as a Poisson process when $\delta = 0$. Although this result does not hold for ISIs that have an inverse Gaussian distribution, it was unclear whether WAIC could differentiate between the competition model and the IIGPP model when $\delta = 0$. However, from this simulation, we can see that WAIC can distinguish between a competition model with $\delta = 0$ and a IIGPP model.

\subsection{WAIC Simulation Study 2: WAIC Performance under Partial Switching Behavior}
\label{sec: appendix_sim3}
 In this section, we are interested in exploring the performance of conditional and marginal WAIC when only a proportion $\alpha \in (0,1)$ of the spike trains come from the competition framework and $(1 - \alpha)$ come from the IIGPP framework.  To study the performance of the information criteria under partial switching behavior, we generated 100 datasets from a mixture of the IIGPP model and the competition model and calculated the information criteria of interest ($WAIC_{comp}$ and $WAIC_{IIGPP}$). The 100 datasets, with 25 spike trains per condition ($N^A = N^B = N^{AB} = 25$), were generated as follows:
\begin{align}
    \nonumber I^A & \sim \mathcal{N}^+(40, 16), & I^B & \sim \mathcal{N}^+(80, 16), \\
    \nonumber I^{AB} & \sim \mathcal{N}^+(80, 64) &  \sigma^A & \sim \mathcal \mathcal{N}^+(\sqrt{40}, 4),\\
    \nonumber \sigma^B & \sim \mathcal \mathcal{N}^+(\sqrt{80}, 4), & \sigma^{AB} & \sim \mathcal \mathcal{N}^+(\sqrt{60}, 16), \\
    \nonumber \delta & \sim \text{LogNormal}(-2.5, 0.25), & \boldsymbol{\phi}^A & \sim \mathcal{N}_6(\mathbf{0}, 0.09\mathbf{I}_6) ,\\
    \nonumber \boldsymbol{\phi}^B & \sim \mathcal{N}_6(\mathbf{0}, 0.09\mathbf{I}_6), & \boldsymbol{\phi}^{AB} & \sim \mathcal{N}_6(\mathbf{0}, 0.09\mathbf{I}_6),\\
\end{align}
\begin{align}
    \nonumber N^{AB}_{Comp} &\sim Unif(\{n \in \mathbb{N}| 2 \le n \le 24 \}),\\
    \nonumber \mathcal{S}^{AB}_1, \dots, \mathcal{S}^{AB}_{N^{AB}_{Comp}} &\sim \text{Competition Framework} (I^A, I^B, \boldsymbol{\phi}^{A}, \boldsymbol{\phi}^{B}, \sigma^A, \sigma^B, \delta),\\
    \nonumber \mathcal{S}^{AB}_{N^{AB}_{Comp} + 1}, \dots, \mathcal{S}^{AB}_{25} & \sim \text{IIGPP} (I^{AB}, \boldsymbol{\phi}^{AB}, \sigma^{AB}),\\
    \nonumber \mathcal{S}^A_1, \dots, \mathcal{S}^A_{25} & \sim \text{IIGPP} (I^{A}, \boldsymbol{\phi}^{A}, \sigma^{A}), \\
    \nonumber \mathcal{S}^B_1, \dots, \mathcal{S}^B_{25} & \sim \text{IIGPP} (I^{B}, \boldsymbol{\phi}^{B}, \sigma^{B}).
\end{align}

\begin{figure}
    \centering
    \includegraphics[width=0.99\linewidth]{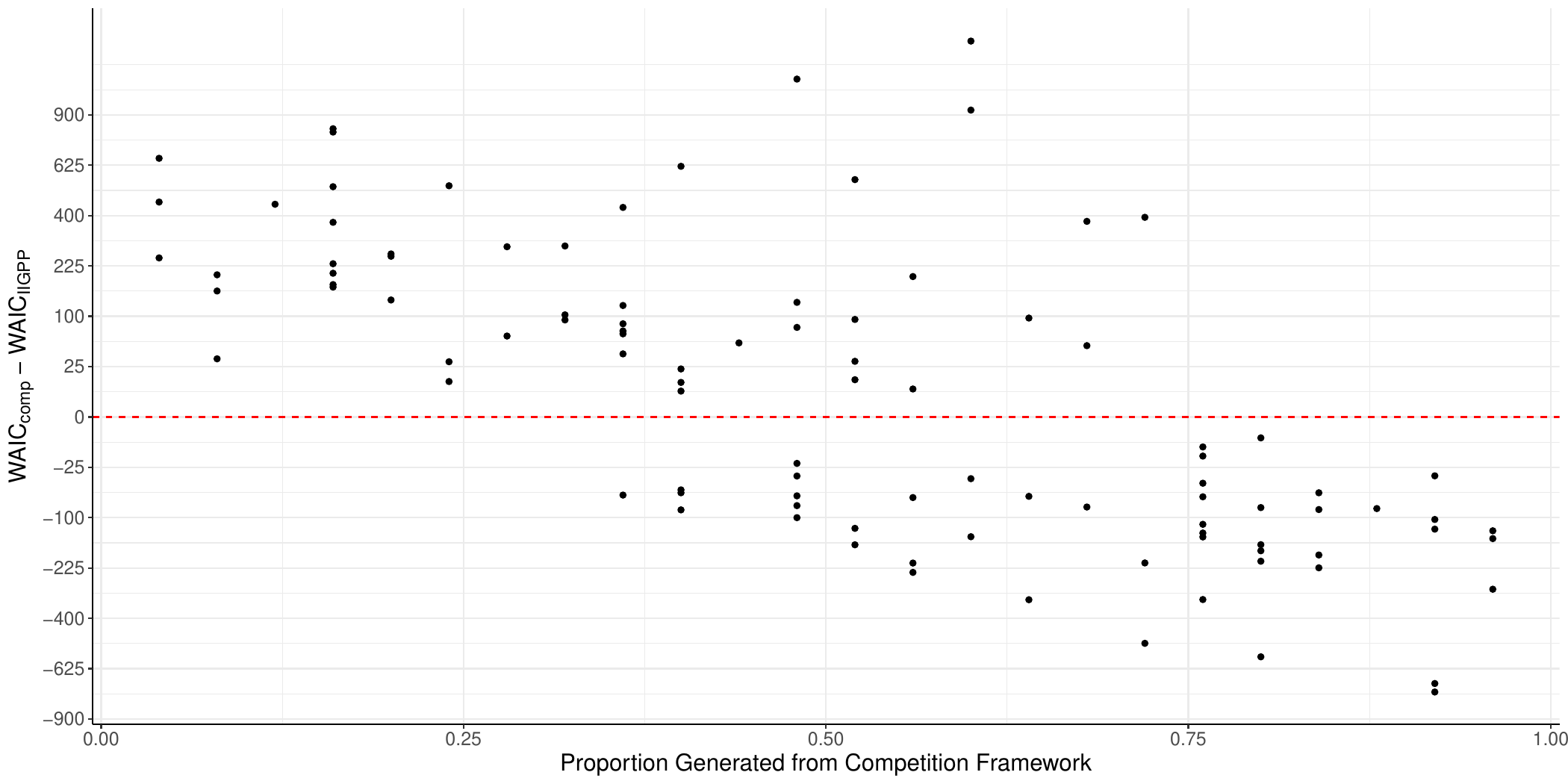}
    \caption{Performance of WAIC when only a subset of the $AB$ trials are generated from the competition framework.}
    \label{fig: Partial_Competition_WAIC}
\end{figure}

Figure \ref{fig: Partial_Competition_WAIC} contains the results of the simulation study. When less than 25\% of the trials contain multiplexing behavior, WAIC is likely to suggest the IIGPP model over the competition model. Alternatively, when over 75\% of the trials contain multiplexing behavior, WAIC is likely to suggest the competition framework. When roughly half of the trials contain multiplexing behavior, WAIC is roughly equally likely to suggest the multiplexing model and the IIGPP model. Specifically, the model chosen in these cases depends on the values of $\boldsymbol{\phi}^{AB}$, $I^{AB}$, and $\sigma^{AB}$. If the chosen parameters lead to behavior similar to the multiplexing behavior, then WAIC is more likely to choose the competition framework over the IIGPP framework.

\subsection{WAIC Simulation Study 3: WAIC Performance under Spike Trains Generated from a HMM}
\label{sec: WAIC_HMM}

In this simulation study, we are interested in how marginal and conditional WAIC perform when the spike trains are generated from a hidden Markov model (HMM). Under our modeling framework, only the competition model assumes a fluctuating firing rate, leading to questions of whether data generated from a HMM will cause WAIC to incorrectly suggest that multiplexing is occurring. 

To explore the performance of WAIC under model misspecification, we will first generate the $A$ and $B$ condition spike trains using a time-inhomogeneous inverse Gaussian point process ($X^A_{ij} \sim IG\left(\frac{1}{I^A \exp\left\{\left(\boldsymbol{\phi}^A\right)^\intercal \mathbf{b}\left(s^{A}_{i(j-1)}\right)\right\}}, \left(\frac{1}{\sigma^A}\right)^2 \right)$ and $X^B_{ij} \sim IG\left(\frac{1}{I^B \exp\left\{\left(\boldsymbol{\phi}^B\right)^\intercal \mathbf{b}\left(s^{B}_{i(j-1)}\right)\right\}}, \left(\frac{1}{\sigma^B}\right)^2 \right)$). Next, we assume that $X_{i1}^{AB}$ has a mixture distribution with equal weight between the two single-stimuli distributions. The following ISIs are generated conditionally on the previous encoding state. Specifically, letting $l_{ij} \in \{A, B\}$ be the variable denoting which stimulus is encoded in the $j^{th}$ spike and $i^{th}$ trial, we have $P(L_{ij} = l_{i(j-1)}) = p_s$. Thus, we will continue encoding stimulus $\mathscr{S}$ with probability $p_s$ in the next spike, and switch encodings with probability $1 - p_s$. 

To evaluate the performance of the information criteria, we generated 100 datasets with $N^A = N^B = N^{AB} = 25$. The parameters were generated randomly as follows: 
\begin{align}
    \nonumber I^A & \sim \mathcal{N}^+(40, 16), & \nonumber I^B & \sim \mathcal{N}^+(80, 16), \\
    \nonumber \sigma^A & \sim \mathcal \mathcal{N}^+(\sqrt{40}, 4), & \sigma^B & \sim \mathcal \mathcal{N}^+(\sqrt{80}, 4), \\
    \nonumber p_s & \sim \text{Beta}(10, 2), & \boldsymbol{\phi}^A & \sim \mathcal{N}_6(\mathbf{0}, 0.09\mathbf{I}_6) ,\\
    \nonumber \boldsymbol{\phi}^B & \sim \mathcal{N}_6(\mathbf{0}, 0.09\mathbf{I}_6), & &
\end{align}

\begin{figure}
    \centering
    \includegraphics[width=1\linewidth]{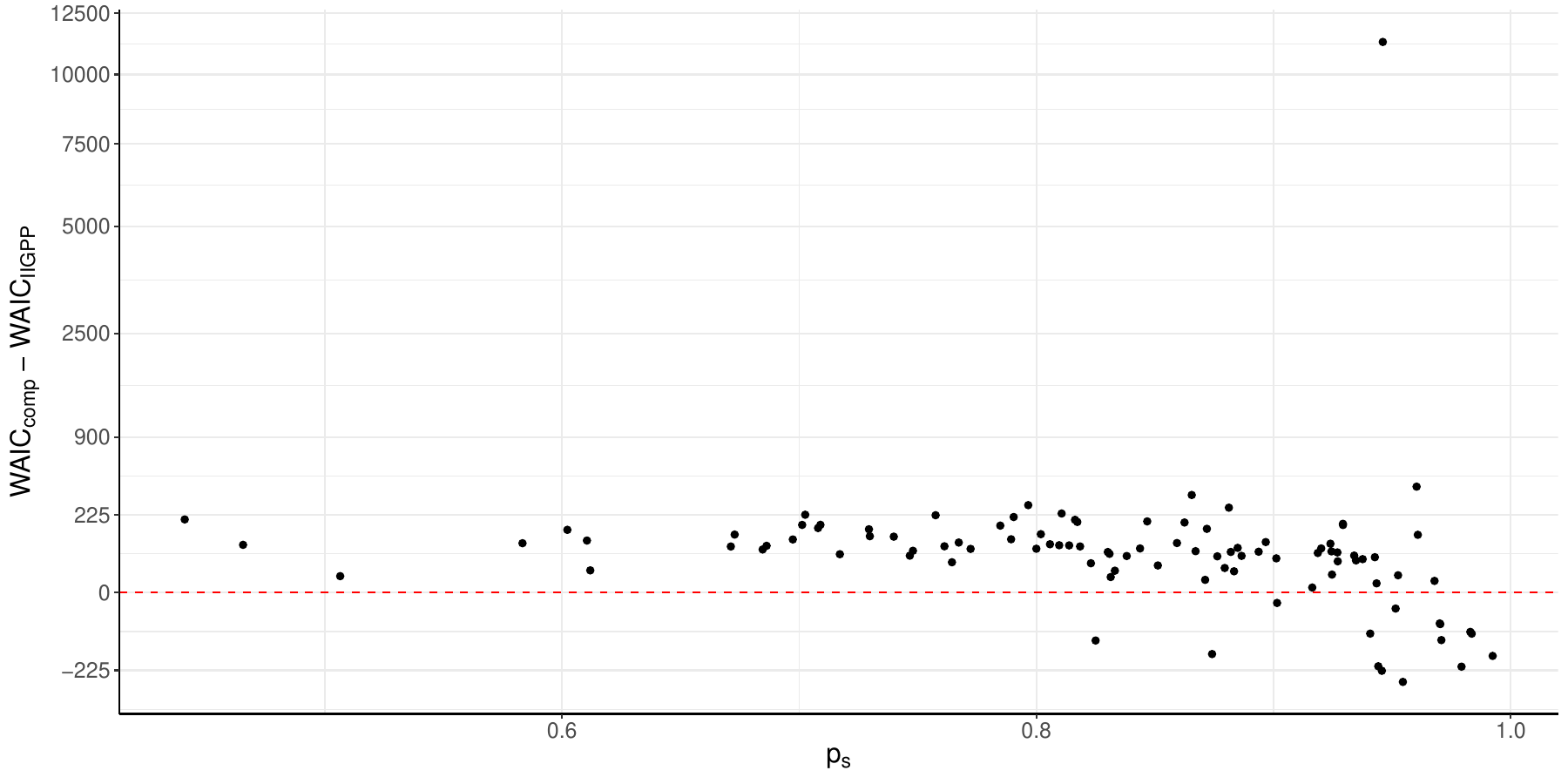}
    \caption{Performance of WAIC over the 100 datasets. As $p_s$ moves towards 1, we get a model very similar to \textsc{Slow Juggling}, with the major difference being that the first spike is not a competition between two diffusion processes but randomly chosen with probability 0.5.}
    \label{fig:HMM_Sim_results}
\end{figure}

Figure \ref{fig:HMM_Sim_results} contains the results of this simulation. When the data were generated from a HMM model, we can see that WAIC suggests the competition model 15\% of the time.  As $p_s$ gets closer to 1,  WAIC is more likely to suggest the competition model, which is not surprising, since a HMM model with a very low probability of switching encodings is very similar to our competition model under a \textsc{Slow Juggling} (large $\delta$) scenario. Specifically, when $p_s$ is close to 1, we are unlikely to switch encodings within a trial and are equally likely to encode stimulus $A$ and $B$ for each trial under the HMM framework (since $P(L_{i1} = \mathscr{S}) = 0.5$ for $\mathscr{S} \in \{A, B\}$). Similarly, under the \textsc{Slow Juggling} scenario, we are unlikely to switch encodings within a trial, but the probability of encoding stimulus $\mathscr{S}$ controlled by the competition process.

\begin{figure}
    \centering
    \includegraphics[width=1\linewidth]{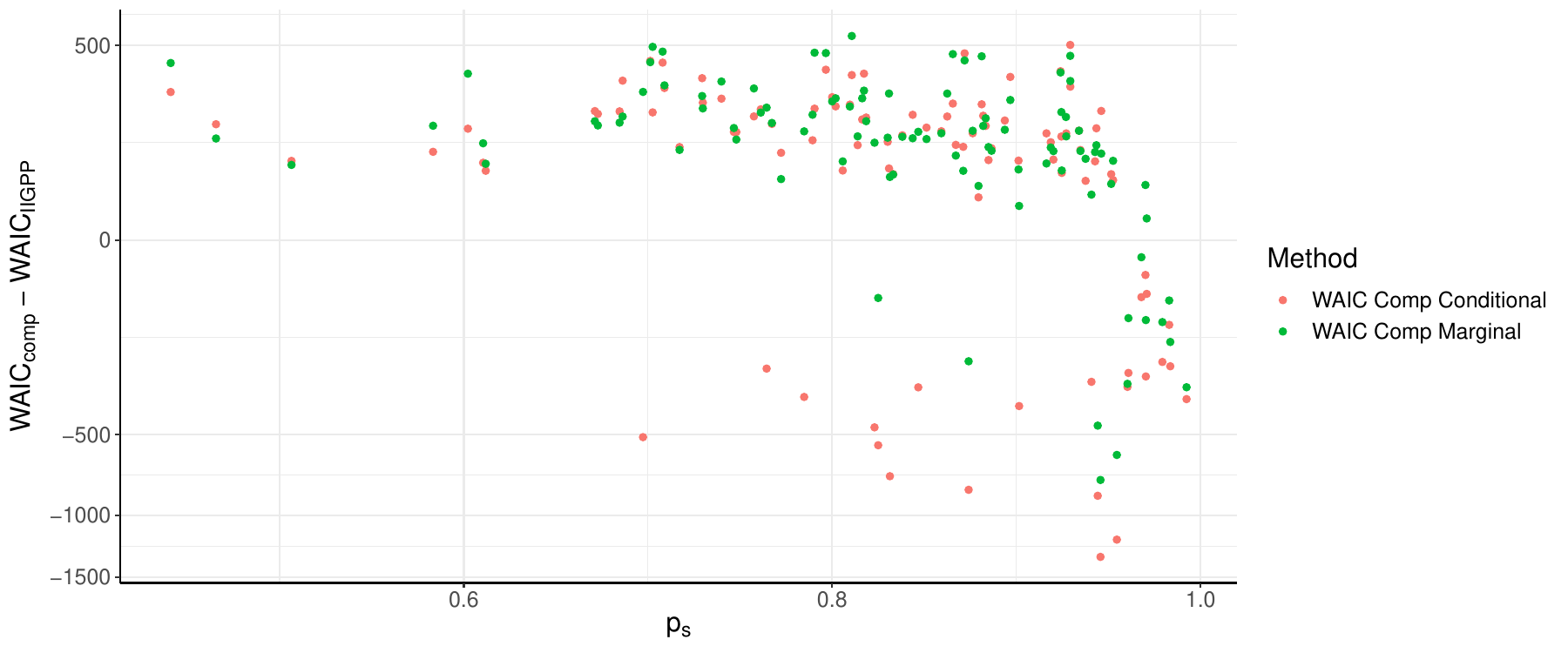}
    \caption{Performance of marginal WAIC versus conditional WAIC when considering data drawn from a simple HMM.}
    \label{fig: marginal_conditional_waic}
\end{figure}

\begin{remark}[Marginal vs Conditional WAIC]
When considering WAIC for latent variable models, and in this case state-space models, one consideration is whether to use the marginal likelihood (states marginalized out) or the conditional likelihood when calculating WAIC. While performance was relatively similar in the correctly specified models, we noticed that marginal WAIC had better performance when there was model misspecification. In particular, conditional WAIC did not penalize improbable switches between states to a great enough extent. Thus, if considering model selection when using state-space models, we advocate the use of marginal WAIC over conditional WAIC. The simulation results are shown in Figure \ref{fig: marginal_conditional_waic}.
\end{remark}

\subsection{WAIC Simulation Study 4: WAIC Performance under Spike Trains Generated from a Winner-Take-All Model}
\label{sec: WTA_WAIC}
Winner-take-all coding schemes posit that each neuron will encode only one of the two stimuli ($A$ or $B$). To represent these models, we assume that the $AB$-condition spike trains can be represented by the corresponding single-stimulus model. In this simulation study, we aim to study the performance of marginal and conditional WAIC under data generated from a winner-take-all scenario and alternatively when the spike trains are generated from our competition process. We will first focus on the case where the data were generated from a winner-take-all scenario. The spike trains were generated as follows:

\begin{align}
    \nonumber I^A & \sim \mathcal{N}^+(40, 36), & I^B & \sim \mathcal{N}^+(80, 36), \\
    \nonumber \sigma^A & \sim \mathcal \mathcal{N}^+(\sqrt{40}, 4), & \sigma^B & \sim \mathcal \mathcal{N}^+(\sqrt{80}, 4), \\
    \nonumber \boldsymbol{\phi}^A & \sim \mathcal{N}_6(\mathbf{0}, 0.09\mathbf{I}_6) , & \boldsymbol{\phi}^B & \sim \mathcal{N}_6(\mathbf{0}, 0.09\mathbf{I}_6).
\end{align}
Once the single-stimuli parameters were generated, the $AB$ condition parameters were chosen such that the neuron encoded $A$ with probability 0.5 and $B$ with probability 0.5:
\begin{align}
        \left\{
\begin{array}{ll}
      I^{AB} = I^A, \sigma^{AB} = \sigma^{A}, \phi^{AB} = \phi^A & \text{with probability 0.5}\\
      I^{AB} = I^B, \sigma^{AB} = \sigma^{B}, \phi^{AB} = \phi^B & \text{with probability 0.5}\\
\end{array} 
\right. \nonumber .
\end{align}
We generated 25 spike trains for each of the three conditions ($A$, $B$, and $AB$), fit all of the proposed models (Competition, IIGPP, Winner-take-all (A), and Winner-take-all (B)) and computed each of the WAIC ($WAIC_{comp}$, $WAIC_{IIGPP}$, $WAIC_{WTA(A)}$, and $WAIC_{WTA(B)}$). 

\begin{table}
    \centering
    \begin{tabular}{|p{0.3cm}|p{2.2cm}|p{1.3cm}p{1.3cm}p{1.3cm} p{1.6cm}|}
    \hline
     \multicolumn{6}{|c|}{WAIC Selection} \\
     \hline
      & \multicolumn{1}{c|}{} & \multicolumn{1}{c}{IIGPP} &  \multicolumn{1}{c}{WTA (A)} & \multicolumn{1}{c}{WTA (B)} & \multicolumn{1}{c|}{Competition} \\
     \hline
     \parbox[c]{2mm}{\multirow{2}{*}{\rotatebox[origin=c]{90}{Truth}}} & WTA (A) & 3 & 39 & 0 & 0 \\
    & WTA (B) & 3 & 0 & 55 & 0 \\
     \hline
     \hline
     & Total & 6 & 39 & 55 & 0 \\
     \hline
    \end{tabular}
    \caption{Table showing the chosen model using WAIC when the datasets were generated from a winner-take-all scenario. \textbf{Abbreviations:} WTA (A) -- winner-take-all (encoding stimulus $A$), WTA (B) -- winner-take-all (encoding stimulus $B$).}
    \label{tab: WTA_WAIC}
\end{table}

From Tables \ref{tab: WTA_WAIC} we can see that WAIC reliably chooses the winner-take-all scenario when the data are generated from a winner-take-all scenario. We can see that WAIC sometimes suggests that the data come from the IIGPP model, which is understandable because the winner-take-all model can be considered a nested model within the class of IIGPP models. However, these few cases are not scientifically concerning, as both the IIGPP model and the winner-take-all models represent alternative encoding scenarios to multiplexing. Crucially, WAIC does not incorrectly suggest the competition model, supporting the notion that WAIC is a reliable technique to conduct model comparison with the purpose of collecting evidence of multiplexing.

Alternatively, we want to ensure that the WAIC suggests the competition model under data generated from the competition model. While Simulation Study 2 (Section \ref{sec: marginal_vs_conditional}) covers a similar scenario, we study the performance under the larger class of alternative models (including winner-take-all scenarios). The data were generated from the competition model, with the parameters selected as follows:

\begin{align}
    \nonumber I^A & \sim \mathcal{N}^+(40, 36), & I^B & \sim \mathcal{N}^+(80, 36), \\
    \nonumber \sigma^A & \sim \mathcal \mathcal{N}^+(\sqrt{40}, 4), & \sigma^B & \sim \mathcal \mathcal{N}^+(\sqrt{80}, 4), \\
    \nonumber \delta & \sim \text{LogNormal}(-2.5, 0.25), & \boldsymbol{\phi}^A & \sim \mathcal{N}_6(\mathbf{0}, 0.09\mathbf{I}_6) ,\\
    \nonumber \boldsymbol{\phi}^B & \sim \mathcal{N}_6(\mathbf{0}, 0.09\mathbf{I}_6). & & 
\end{align}
Using these parameters, we generated 25 spike trains for each condition from our competition model and evaluated the performance of WAIC. As illustrated by the results in Table \ref{tab: comp_WAIC}, WAIC reliably picks the competition model when the spike trains were generated from the competition model (99\% of the time). Although WAIC chose the winner-take-all model once, we note that in this scenario only $1.03\%$ (22 out of 2139 spikes) of the $AB$ condition spikes were generated as a result of the $A$ process winning. Thus, we can see that this dataset generated from the competition framework is very similar to a dataset generated from a winner-take-all (B) framework. Therefore, in this case, WAIC chose the simpler winner-take-all (B) framework over the competition framework. Overall, we can see that WAIC can reliably differentiate between the different classes of models, and is relatively robust to model misspecification as illustrated in Section \ref{sec: WAIC_HMM}.

\begin{table}
    \centering
    \begin{tabular}{|p{1.3cm}p{1.3cm}p{1.3cm} p{1.6cm}|}
    \hline
     \multicolumn{4}{|c|}{WAIC Selection} \\
     \hline
     \multicolumn{1}{|c}{IIGPP} &  \multicolumn{1}{c}{WTA (A)} & \multicolumn{1}{c}{WTA (B)} & \multicolumn{1}{c|}{Competition}  \\
     \hline
      0 & 0 & 1 & 99\\
     \hline
    \end{tabular}
    \caption{Table showing the chosen model using WAIC. \textbf{Abbreviations:} WTA (A) -- winner-take-all (encoding stimulus $A$), WTA (B) -- winner-take-all (encoding stimulus $B$)}
    \label{tab: comp_WAIC}
\end{table}

\section{Simulation Study: Recovery of Scientific Quantities of Interest}
\label{Sec: Sim_convergence}

 In this simulation study, our objective was to study the convergence properties of our model. To evaluate this, we generated 100 datasets under four different sample sizes. We evaluated how well we could recover the model parameters and posterior predictive distributions of interest by calculating the relative squared error (RSE) and the relative Wasserstein distance (RWD), respectively. For a parameter $\theta$, the relative squared error was calculated as $\text{RSE} = \frac{\Vert \hat{\theta} - \theta \Vert_2^2}{\Vert \theta \Vert_2^2}$, where $\hat{\theta}$ is defined as the posterior median estimate. The RSE for functional parameters, such as $I^\mathscr{S} \exp\left(\left(\boldsymbol{\phi}^\mathscr{S}\right)^\intercal \mathbf{b}(t) \right)$, were evaluated on a dense finite-dimensional grid over $\mathcal{T} = [0,1]$ prior to the calculation of RSE. The Wasserstein distance between two probability measures $\mu$ and $\nu$ characterizes the minimum amount of ``mass'' that must be moved to reconfigure $\mu$ into $\nu$ \citep{panaretos2019statistical}; leading to the alternate name of ``Earth mover's distance''. We characterized the relative Wasserstein distances in terms of empirical measures $\mu_n$ and $\nu_n$. Letting $X_1, \dots, X_n$ be the posterior predictive samples with empirical measure $\mu_n$ and $Y_1, \dots, Y_n$ be the samples generated under the true model with empirical measure $\nu_n$, the relative Wasserstein distance (RWD) was calculated as $\text{RWD}(\mu_n, \nu_n) = \frac{\sum_{i=1}^n\mid X_{(i)} - Y_{(i)}\mid}{\sum_{i=1}^n \mid Y_{(i)}\mid}$, where $X_{(i)}$ and $Y_{(i)}$ are the respective order statistics; thereby scaling the Wasserstein distance in relation to the expected value of $Y$ in our case. For this simulation study, $n = 100,000$ was used to calculate the RWD.
 
 In this simulation study, we considered the following number of triplets: $N^A = N^B = N^{AB} = 5, 10, 25, 50$. The spike trains used in this simulation were generated in the following way:
 
 \begin{align}
    \nonumber I^A & \sim \mathcal{N}^{10+}(40, 400), & I^B & \sim \mathcal{N}^{10+}(80, 400), \\
    \nonumber \sigma^A & \sim \mathcal \mathcal{N}^{3+}(\sqrt{40}, 25), & \sigma^B & \sim \mathcal \mathcal{N}^{3+}(\sqrt{80}, 25), \\
    \nonumber \delta & \sim \text{LogNormal}(-3.5, 1), \text{(for 80\% of the data sets)} &\delta & = 0, \text{(for 20\% of the data sets)}\\
    \nonumber \boldsymbol{\phi}^A & \sim \mathcal{N}_6(\mathbf{0}, 0.09\mathbf{I}_6) , & \boldsymbol{\phi}^B & \sim \mathcal{N}_6(\mathbf{0}, 0.09\mathbf{I}_6),
\end{align}
 where $\mathcal{N}^{x+}(\mu,\sigma^2)$ is a truncated normal distribution with support on $[x, \infty)$.
 
 Table \ref{tab: sim_coverage} provides the coverage obtained from the simulation study conducted in Section 5 of the main manuscript. We can see that the coverage of the credible intervals was nominal, covering the true parameter value roughly 95\% of the time as the sample size increased. We can also see that as we gained more information (more spike trains), the width of the credible intervals decreased, as expected. 
 
\begin{table}
    \centering
    \begin{tabular}{|c|| c |c| c|} 
     \hline
     $N^A=N^B = N^{AB}$ & $I^\mathscr{S} \exp\left(\left(\boldsymbol{\phi}^\mathscr{S}\right)^\intercal \mathbf{b}(t) \right)$ & $\sigma^{\mathscr{S}}$ & $\delta$ \\
     \hline\hline
     5 & 92.0\% (1) & 97.0\% (1) & 90.4\% (1) \\
     \hline
     10 & 91.9\% (0.599) & 94.0\% (0.691)& 87.6\% (0.653)\\
     \hline
     25 & 93.1\% (0.309) & 96.5\% (0.427) & 92.0\% (0.221) \\ 
     \hline
     50 & 92.5\% (0.179) & 94.5\% (0.316) & 96.0\% (0.015)\\
     \hline
    \end{tabular}
    \caption{Observed frequentist coverage for 95\% credible intervals of the model parameters, with average relative credible interval width (shown in parenthesis). The coverage for $I^\mathscr{S} \exp\left(\left(\boldsymbol{\phi}^\mathscr{S}\right)^\intercal \mathbf{b}(s) \right)$ is pointwise-coverage, evaluated on a dense finite-dimensional grid over $\mathcal{T}$.}
    \label{tab: sim_coverage}
\end{table}

\section{Case Study: Caruso}
\label{sec: CS_appendix}
This section of the Supplementary Materials contains additional information on the analysis of the IC spike trains collected in \citet{caruso2018single}.

\subsection{Experimental Setup}

\begin{figure}
    \centering
    \includegraphics[width=0.99\linewidth]{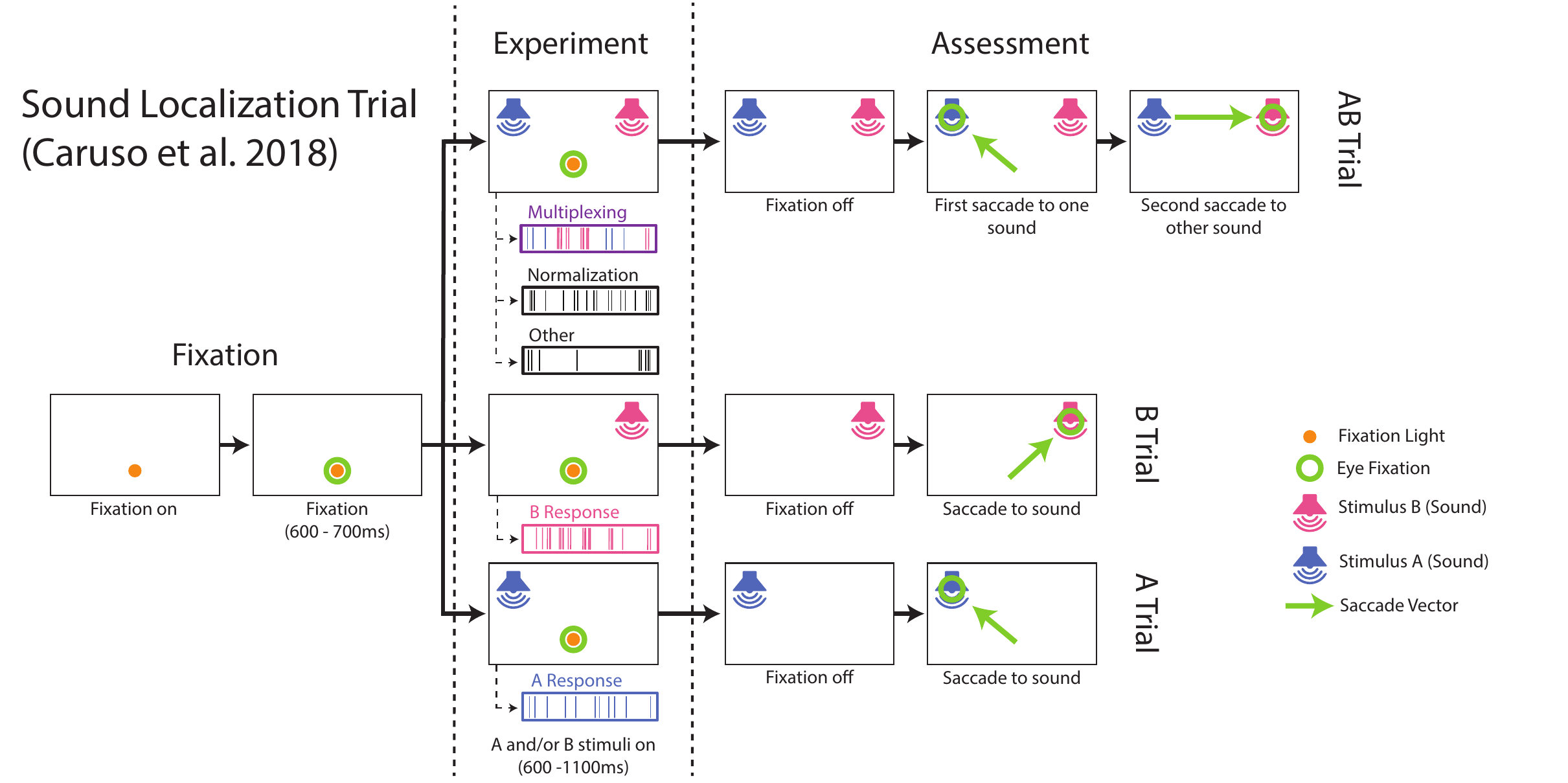}
    \caption{Visualization of the sound localization task conducted in \citet{caruso2018single}.}
    \label{fig:caruso_detailed}
\end{figure}

The experimental setup consists of recordings under various triplets of conditions (Figure \ref{fig:caruso_detailed}); consisting of trials under an $A$ condition ($A$ stimulus only), trials under a $B$ condition ($B$ stimulus only), and trials under an $AB$ condition ($A$ and $B$ stimulus simultaneously). For each triplet, one of the stimuli ($A$ or $B$) consisted of a bandpass noise with a 742 Hz center frequency at -24, -6, 6, or 24 degrees horizontally. The other stimulus consisted of a bandpass noise with center frequency of 500 Hz, 609 Hz, 903 Hz, 1100 Hz, 1340 Hz, 1632 Hz, or 1988 Hz located 30 degrees from the 742 Hz stimulus (-24, -6, 6, or 24 degrees horizontally). The frequency differences were important because IC neurons are sensitive to sound frequency as well as location; and because perceptually, simultaneous sounds of the same frequency will typically fuse to be perceived at a single location \citep{yin1994physiological}. A trial would start with a visual fixation period of 600--700 ms (Figure \ref{fig:caruso_detailed} -- Fixation), after which either the single-sound ($A$ or $B$ stimulus) or the dual-sound ($A$ and $B$ stimulus simultaneously), along with the visual stimulus, would be present for 600--1100 ms (Figure \ref{fig:caruso_detailed} -- Experiment). After this, the visual fixation light would extinguish, and the macaque would have to visually identify where the sound or sounds were originating from (Figure \ref{fig:caruso_detailed} -- Assessment); ensuring that the auditory stimuli were correctly perceived. The data analyzed in this section were recorded when the sound(s) and fixation light were both present (Figure \ref{fig:caruso_detailed} -- Experiment). To avoid potential interdependence between trials, the conditions ($A$, $B$, and $AB$) and the stimuli (frequency and location) of the various triplets were randomly interleaved.

\subsection{Inclusion Criteria}
\begin{figure}
    \centering
    \includegraphics[width=\linewidth]{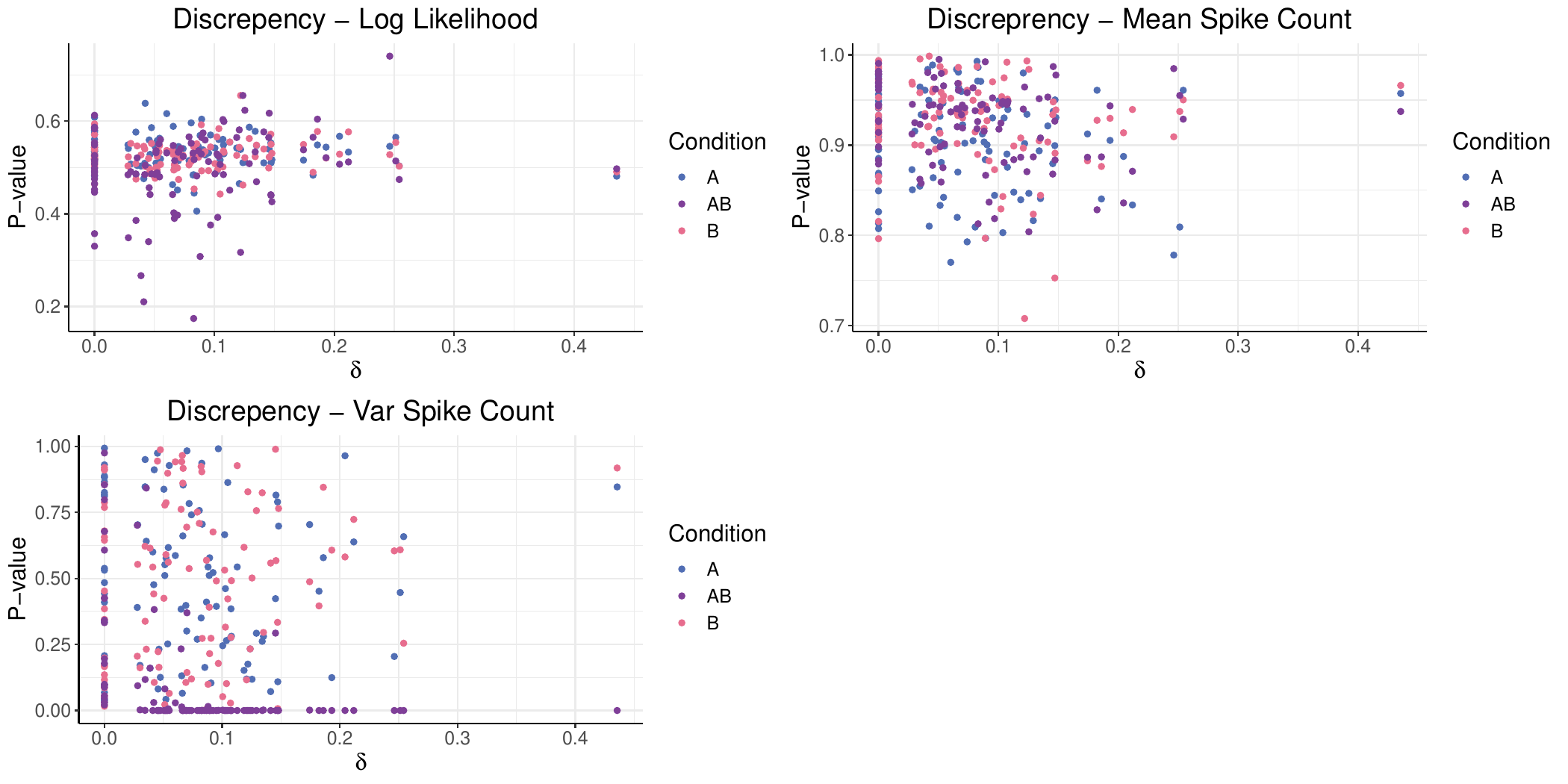}
    \caption{Performance results from the 3 discrepancy variables. The data was generated from our competition model, where the $A$ and $B$ are true inhomogeneous inverse Gaussian Point Processes, and $AB$ is generated from the competition framework.}
    \label{fig: posterior_pval}
\end{figure}

To be included in the analysis, the triplet must satisfy the following conditions:
\begin{enumerate}
    \item At least five trials for each of the three conditions ($A$, $B$, and $AB$),
    \item single-stimulus spike trains can be represented by a time-inhomogeneous inverse Gaussian point process,
    \item Distinguishably different distributions of spike trains for the $A$ and $B$ conditions.
\end{enumerate}
The first one ensures that we have a sufficient amount of data for each of the three conditions and is straightforward to implement. The second condition will remove any triplets where the single-stimuli conditions ($A$ or $B$ conditions) are a mixture of processes or multiplexing itself. The screening is done by calculating posterior p-values \citep{meng1994posterior, gelman1996posterior} and discarding any triplets where the p-values are less than 0.05 for either the $A$ or $B$ conditions for any of the discrepancy variables.  Specifically, the three discrepancy variables we use when calculating the posterior p-values are 
\begin{enumerate}
    \item Average Log-Likelihood -- measure of average discrepancy between observed ISIs and the posited distributions of the ISIs under the modeling assumptions. $$D_{avg-LL}(\{\mathcal{S}_i^{\mathscr{S}}\}_{i=1}^{N^{\mathscr{S}}}, \boldsymbol{\theta}):= \int \left(\prod_{i=1}^{N^{\mathscr{S}}} \mathcal{L}^{\mathscr{S}}(\boldsymbol{\theta}\mid \mathcal{S}_{i}^{\mathscr{S}})\right)f\left(\boldsymbol{\theta}\mid \{\mathcal{S}_i^{\mathscr{S}}\}_{i=1}^{N^{\mathscr{S}}}\right)\text{d}\boldsymbol{\theta},$$
    \item Mean of Spike Counts -- measure of average discrepancy between the observed mean spike counts and the posited mean spike count under the modeling assumptions.
    $$D_{mean-SC}(\{\mathcal{S}_i^{\mathscr{S}}\}_{i=1}^{N^{\mathscr{S}}}, \boldsymbol{\theta}):= \left|\left(\frac{1}{N^{\mathscr{S}}} \sum_{i=1}^{N^{\mathscr{S}}}n_i^{\mathscr{S}}\right) - \mathbb{E}_{\boldsymbol{\theta}}(n^{\mathscr{S}})\right|,$$
    \item Variance of Spike Counts -- measure of average discrepancy between the sample variance of the spike counts and the posited variance of spike counts under the modeling assumptions.
    $$D_{var-SC}(\{\mathcal{S}_i^{\mathscr{S}}\}_{i=1}^{N^{\mathscr{S}}}, \boldsymbol{\theta}):= \left| \left(\frac{1}{N^{\mathscr{S}}-1} \sum_{i=1}^{N^{\mathscr{S}}}(n_i^{\mathscr{S}} -\overline{n_i^{\mathscr{S}}})^2\right) - \text{Var}_{\boldsymbol{\theta}}(n^{\mathscr{S}}) \right|,$$
\end{enumerate}
where $\mathbb{E}_{\boldsymbol{\theta}}(n^{\mathscr{S}})$ and $\text{Var}_{\boldsymbol{\theta}}(n^{\mathscr{S}})$ are the expectation and variance of the trial-wise spike count with respect to the fitted model, respectively. Using the defined discrepancy metrics, we can define the tail area to get the corresponding posterior predictive p-value as follows:
$$p_\mathcal{D}\left(\{\mathcal{S}_i^{\mathscr{S}}\}_{i=1}^{N^{\mathscr{S}}}\right) = P\left(\{D_\mathcal{D}(\{\mathcal{\tilde{S}}_i^{\mathscr{S}}\}_{i=1}^{N^{\mathscr{S}}},\boldsymbol{\theta}) \ge D_\mathcal{D}(\{\mathcal{S}_i^{\mathscr{S}}\}_{i=1}^{N^{\mathscr{S}}},\boldsymbol{\theta}) \mid \{\mathcal{S}_i^{\mathscr{S}}\}_{i=1}^{N^{\mathscr{S}}}, H\right),$$
where $H$ denotes the model and $\mathcal{\tilde{S}}_i^{\mathscr{S}}$ denotes a spike train generated from the posterior predictive distribution. Performance results can be seen in Figure \ref{fig: posterior_pval}. In this simulation, we generated triplets from our competition model, which means that the triplets generated under the $A$ and $B$ conditions are generated from a time-inhomogeneous inverse Gaussian point process (IIGPP), and the spike trains generated under the $AB$ process are not. Under these simulation settings, we can see that $D_{var-SC}$ is able to determine that the $AB$ condition spike trains were not generated from a time-inhomogeneous inverse Gaussian point process, especially when $\delta$ is moderate or large. Although $D_{avg-LL}$ and $D_{mean-SC}$ were not very informative in this simulation study, they may be informative in other simulation settings and will be included in the screening procedure. Specifically, we will remove any triplet where $p_{avg-LL}$, $p_{mean-SC}$, or $p_{var-SC}$ is greater than 0.05. Since posterior p-values are known to not be uniformly distributed under the null \cite{gelman2013two} this can be considered a relatively conservative screening process -- only removing triplets where the $A$ or $B$ condition spike trains greatly violate our second assumption (single-stimulus spike trains can be represented by a time-inhomogeneous inverse Gaussian point process). 

To verify the triplet has distinguishably different distributions of spike trains for the $A$ and $B$ conditions, we fit a joint model and compare the pointwise predictive distributions of the joint model and separate model. Specifically, for the joint model, we assume that

\begin{equation}
    f_{j}^{joint}(x^{\mathscr{S}}_{ij}\mid I, \sigma, \phi, s^{\mathscr{S}}_{i(j-1)}) = \frac{1}{\sigma\sqrt{2\pi (x^{\mathscr{S}}_{ij})^3}}\text{exp}\left( -\frac{\left(1 - I \exp\left\{\left(\boldsymbol{\phi}\right)^\intercal \mathbf{b}\left(s^{\mathscr{S}}_{i(j-1)}\right) \right\}x^{\mathscr{S}}_{ij}\right)^2}{2(\sigma)^2x^{\mathscr{S}}_{ij}}\right),
\end{equation}
for $\mathscr{S} = A, B$. Thus we can see that both the $A$ and $B$ distribution of spike trains share the same parameters ($I, \sigma, \phi$). For the separate models, we have
\begin{equation}
    f^{A}_j(x^{A}_{ij}\mid \boldsymbol{\theta}, s^{A}_{i(j-1)}) = \frac{1}{\sigma^{A}\sqrt{2\pi (x^{A}_{ij})^3}}\text{exp}\left( -\frac{\left(1 - I^A \exp\left\{\left(\boldsymbol{\phi}^A\right)^\intercal \mathbf{b}\left(s^{A}_{i(j-1)}\right) \right\}x^{A}_{ij}\right)^2}{2(\sigma^{A})^2x^{A}_{ij}}\right),
\end{equation}
\begin{equation}
    f^{B}_j(x^{B}_{ij}\mid \boldsymbol{\theta}, s^{B}_{i(j-1)}) = \frac{1}{\sigma^{B}\sqrt{2\pi (x^{B}_{ij})^3}}\text{exp}\left( -\frac{\left(1 - I^B \exp\left\{\left(\boldsymbol{\phi}^B\right)^\intercal \mathbf{b}\left(s^{B}_{i(j-1)}\right) \right\}x^{B}_{ij}\right)^2}{2(\sigma^{B})^2x^{B}_{ij}}\right).
\end{equation}
We can see that the $A$ and $B$ condition spike trains have there own parameters ($I^{\mathscr{S}}, \sigma^{\mathscr{S}}, \phi^{\mathscr{S}})$. We will conclude that the triplet has distinguishably different distributions of spike trains for the $A$ and $B$ conditions if the following holds:
$$\text{lppd}_{seperate} - \text{lppd}_{joint} > \log(3),$$
where
\begin{align}
    \nonumber \text{lppd}_{joint} & =  \sum_{\mathscr{S} \in \{A, B\}}\sum_{i=1}^{N^{\mathscr{S}}}\log \left(\frac{1}{S} \sum_{s = 1}^S\mathcal{L}^{\mathscr{S}}\left(I^s, \sigma^s, \phi^s \mid \mathcal{S}_i^{\mathscr{S}}\right) \right),\\
    \nonumber \text{lppd}_{seperate} & =  \sum_{\mathscr{S} \in \{A, B\}}\sum_{i=1}^{N^{\mathscr{S}}}\log \left(\frac{1}{S} \sum_{s = 1}^S\mathcal{L}^{\mathscr{S}}\left(\boldsymbol{\theta}^s \mid \mathcal{S}_i^{\mathscr{S}}\right) \right),
\end{align}
where $S$ is the number of MCMC samples. From the 2225 triplets recorded in \citet{caruso2018single}, 1231 triplets passed the first criterion of at least five trials per condition. From there, 645 of the 1241 triplets passed the second criterion. Lastly, we arrive at the final number of 571 triplets that fit all three criteria for inclusion.

\subsection{A Comparison Between SCAMPI and the Proposed Spike Train Analysis}
As illustrated throughout the main manuscript, this work builds on the SCAMPI framework (spike count analysis) proposed by \citet{chen2024} by providing a more granular spike train framework for analysis. Table \ref{tab: SCAMPI_vs_proposed} shows an overview of the similarities between the results obtained from the two types of analysis. Crucially, we can see that many of the triplets thought to be \textsc{Fast Juggling} under the SCAMPI framework were estimated to not be multiplexing under our proposed framework (IIGPP or Winner-take-all (preferred)). Although the posited model allows multiplexing to achieve overall higher firing rates, we note that this does not account for many of the differences between the two models. Rather, it is postulated that the more granular analysis allows us to capture ISI-variability and time-inhomogeneous firing rates, which are crucial to (1) defining the single-stimulus encoding signature and (2) inferring switches between encoding the two stimuli at a sub-trial level (i.e., Fast Juggling).

\begin{table}
    \centering
    \begin{tabular}{|p{0.3cm}|p{2.2cm}|p{1.3cm}p{1.3cm}p{1.3cm} p{1.3cm} p{1cm}|}
    \hline
     \multicolumn{7}{|c|}{Proposed Framework} \\
     \hline
      & \multicolumn{1}{c|}{} & \multicolumn{1}{c}{IIGPP} &  \multicolumn{1}{c}{WTA (P)} & \multicolumn{1}{c}{WTA (NP)} & \multicolumn{1}{c}{Fast Jug} & \multicolumn{1}{c|}{Slow Jug}\\
     \hline
     \parbox[c]{2mm}{\multirow{5}{*}{\rotatebox[origin=c]{90}{SCAMPI}}} & Alt & 45 & 4 & 3 & 6 & 4\\
    & WTA (P) & 24 & 27 & 1 & 8  & 2\\
     &WTA (NP) & 29 & 0 & 21 & 6  & 3\\
     & Fast Jug& 20 & 10 & 1 & 7 &  1\\
     & Slow Jug & 2 & 1 & 2 & 2 & 0\\
     \hline
    \end{tabular}
    \caption{Comparison of the results obtained from SCAMPI and the proposed framework. The total number of overlapping triplets are 229. \textbf{Abbreviations:} WTA (P) -- winner-take-all (Preferred), WTA (NP) -- winner-take-all (Non-preferred),  Alt (SCAMPI) -- Overreaching, Fixed -- Middle, and Fixed -- Outside, Fast Jug -- Fast Juggling, Slow Jug -- Slow Juggling.}
    \label{tab: SCAMPI_vs_proposed}
\end{table}

\subsection{Simultaneously Recorded Cells}
A subset of the 166 neurons recorded in the \citet{caruso2018single} dataset was recorded simultaneously with another neuron (two neurons recorded at the same time). Thus, we have access to pairs of simultaneously recorded spike trains from two different neurons in the IC. The recording of these spike trains is a very time-intensive and skill-intensive task, which can lead to potential concerns of shifting of the probes or other artifacts that can influence the spike trains. Thus, in this subsection, we will explore whether the results of one neuron seem to be dependent on the results of the other simultaneously recorded neuron. Although dependence could indicate that perhaps a large population of neurons are simultaneously multiplexing and not a result of artificial artifacts, independent results would suggest that the findings are not due to movement of the probe, movement of the monkey, or other artifacts. Table \ref{tab: dual_recoring_WAIC} contains the results of the preferred model (competition vs. IIGPP) using WAIC. When performing Fisher's exact test, we obtain a p-value of 0.428, indicating that we did not observe a significant amount of dependence in the model selection results between the two simultaneously recorded neurons. 

\begin{table}
    \centering
    \begin{tabular}{|p{0.3cm}|p{2.2cm}|p{1.3cm}p{1.3cm}p{1.3cm} p{1.3cm}|| p{1cm}|}
    \hline
     \multicolumn{7}{|c|}{Neuron 1} \\
     \hline
      & \multicolumn{1}{c|}{} & \multicolumn{1}{c}{IIGPP} &  \multicolumn{1}{c}{WTA (P)} & \multicolumn{1}{c}{WTA (NP)} & \multicolumn{1}{c||}{Comp} & \multicolumn{1}{c|}{Total}\\
     \hline
     \parbox[c]{2mm}{\multirow{4}{*}{\rotatebox[origin=c]{90}{Neuron 2}}} & IIGPP & 0 & 3 & 2 & 3 & 8\\
    & WTA (P) & 2 & 2 & 2 & 3  & 9\\
     &WTA (NP) & 1 & 4 & 0 & 2  & 7\\
     & Comp & 6 & 3 & 1 & 3 &  13\\
     \hline
     \hline
     & Total & 9 & 12 & 5 & 11 & 37\\
     \hline
    \end{tabular}
    \caption{Contingency table showing the results of WAIC from the subset of simultaneously recorded neurons (two cells simultaneously recorded). The p-value obtained from performing Fisher's exact test is 0.428. \textbf{Abbreviations:} WTA (P) -- winner-take-all (Preferred), WTA (NP) -- winner-take-all (Non-preferred),  Comp -- Competition.}
    \label{tab: dual_recoring_WAIC}
\end{table}

\subsection{Can We Recover Which Stimulus is Encoded in a Spike?}
\begin{figure}
    \centering
    \includegraphics[width=\linewidth]{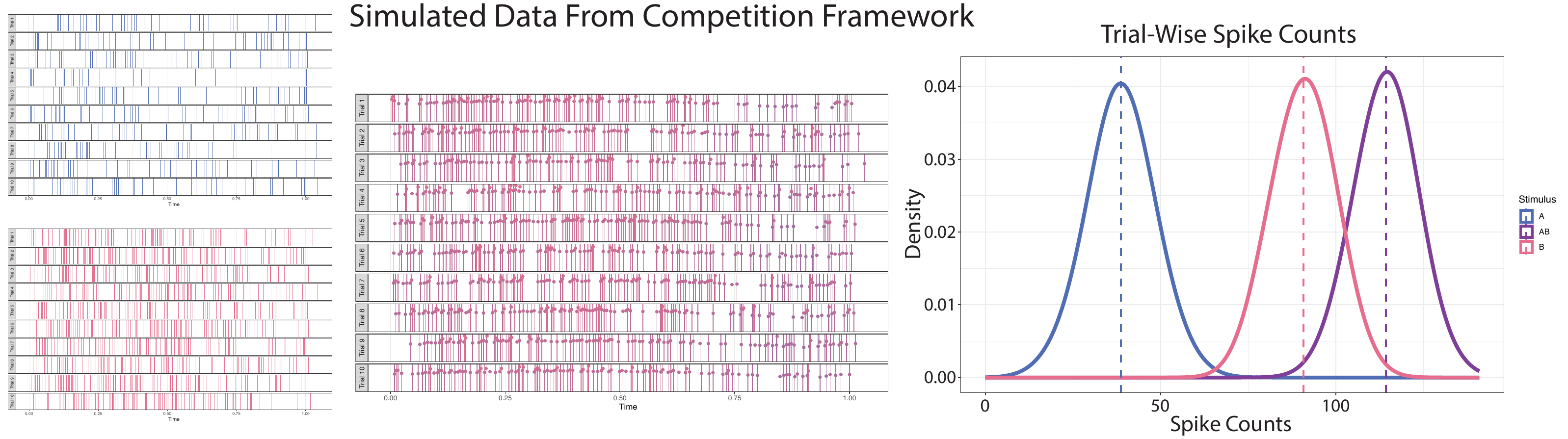}
    \caption{Results from a triplet generated from the competition framework with $\delta = 0$.}
    \label{fig:Simulated_Triplet_res_delta_0}
\end{figure}

As evident from the case study, we are able to recover which stimulus a set of spikes is encoding when $\delta$ is large or moderate. However, it was evident that when $\delta$ was small, we were unable to determine the specific stimulus encoded by an individual spike.
 As $\delta$ becomes smaller, recovering the stimulus for which a spike is encoding for becomes a more challenging task. However, as shown in Section \ref{sec: marginal_vs_conditional} of the Supplementary Materials, WAIC was able to determine whether the data were generated from the competition model or the IIGPP model. From Figure \ref{fig:Simulated_Triplet_res_delta_0}, we can see that even under simulated data from the competition model, we cannot recover which stimulus a set of spikes encodes. Therefore, the fact that we cannot ascertain which stimulus is encoded in each spike in the small $\delta$ scenario is not inherently an indication of model mispecification. In fact, our simulation studies show that WAIC was relatively robust to model mispecification in frequent switching scenarios, picking the IIGPP framework when the datasets were generated from a HMM with $p_s$ close to 0.5 (Section \ref{sec: WAIC_HMM} of the Supplementary Materials). Therefore, based on the simulations conducted, we are confident that the triplets classified as fast-switching (small $\delta$) exhibit behavior consistent with multiplexing, despite not being able to recover which stimulus an individual spike encodes. 
 
\bibliographystyle{abbrvnat}
\bibliography{Neural_Switching}